\documentclass[iop]{emulateapj}
\usepackage{natbib,xspace,color}
\usepackage{verbatim}
\usepackage{rotating}
\usepackage{gensymb }
\usepackage{mathrsfs }  
\usepackage{amsmath}
\usepackage{amssymb }
\usepackage[backref,breaklinks,colorlinks,citecolor=blue]{hyperref}
\usepackage[all]{hypcap} 

\usepackage{url} 
\newcommand{\ang}{\mbox{\AA} }

\shorttitle{Characterizing Dust Attenuation in Local Star Forming Galaxies}
\shortauthors{Battisti et al.}

\begin{document}

\title{Characterizing Dust Attenuation in Local Star Forming Galaxies: UV and Optical Reddening}
\author{A. J. Battisti\altaffilmark{1}, 
D. Calzetti\altaffilmark{1}
R.-R. Chary\altaffilmark{2}
}

\email{abattist@astro.umass.edu}

\altaffiltext{1}{Department of Astronomy, University of Massachusetts, Amherst, MA 01003, USA}
\altaffiltext{2}{MS314-6, U.S. Planck Data Center, California Institute of Technology, 1200 East California Boulevard, Pasadena, CA 91125, USA}

\begin{abstract}
The dust attenuation for a sample of $\sim$10000 local ($z\lesssim0.1$) star forming galaxies is constrained as a function of their physical properties. We utilize aperture-matched multi-wavelength data available from the \textit{Galaxy Evolution Explorer} (\textit{GALEX}) and the Sloan Digital Sky Survey (SDSS) to ensure that regions of comparable size in each galaxy are being analyzed. We follow the method of \citet{calzetti94} and characterize the dust attenuation through the UV power-law index, $\beta$, and the dust optical depth, which is quantified using the difference in Balmer emission line optical depth, $\tau_B^l=\tau_{\mathrm{H}\beta} - \tau_{\mathrm{H}\alpha}$. The observed linear relationship between $\beta$ and $\tau_B^l$ is similar to the local starburst relation, but the large scatter ($\sigma_{\mathrm{int}}=0.44$) suggests there is significant variation in the local Universe. We derive a selective attenuation curve over the range $1250\mathrm{\AA}<\lambda<8320\mathrm{\AA}$ and find that a single attenuation curve is effective for characterizing the majority of galaxies in our sample. This curve has a slightly lower selective attenuation in the UV compared to previously determined curves.  We do not see evidence to suggest that a 2175~\AA\ feature is significant in the average attenuation curve. Significant positive correlations are seen between the amount of UV and optical reddening and galaxy metallicity, mass, star formation rate (SFR), and SFR surface density. This provides a potential tool for gauging attenuation where the stellar population is unresolved, such as at high-$z$.
\end{abstract} 

\section{Introduction}
The presence of dust in a galaxy causes its spectral energy distribution (SED) to experience reddening, a consequence of the highest attenuation occurring in the ultraviolet (UV) and decreasing towards longer wavelengths out to the infrared (IR) \citep[see review by][]{draine03}. The nature of this reddening is dependent on both the dust properties and its geometry within the galaxy \citep[see review by][]{calzetti01}. Taken together, these effects limit the interpretation of galaxy SEDs to determine fundamental quantities such as the stellar population age, stellar mass, and star formation rate (SFR). In the Milky Way \citep[MW;][]{cardelli89,fitzpatrick99} and Magellanic Clouds \citep{gordon03}, where dust extinction\footnote{We define extinction to be the combination of absorption and scattering of light out of the line of sight by dust (no dependence on geometry).} can be directly measured using individual stars, these effects have been well studied and are routinely corrected using extinction curves. The size distribution and composition of dust grains in these galaxies is directly related to the observed extinction \citep{weingartner&draine01}.

When observing galaxies beyond the MW, the Magellanic Clouds, and other very nearby galaxies \citep[e.g., M31,][]{bianchi96,clayton15}, it is usually no longer feasible to utilize point sources to derive extinction curves. Exceptions include using quasars \citep[e.g.,][]{gallerani10} or gamma ray burst \citep[e.g.,][]{perley11}. Instead, we must rely on using unresolved stellar populations, which have more complicated SEDs as a result of being composed of many stars of varying spectral types and will depend on the star formation history and the initial mass function (IMF). In addition, using collections of stars, as opposed to single stars, introduces a wide range of possible geometries for the light sources with respect to the dust. The nature of this geometry determines the importance of light scattering into the line of sight \citep{calzetti01}, and is the reason that it is important to distinguish between extinction and attenuation\footnote{We define attenuation to be a combination of extinction and scattering of light into the line of sight by dust (strong dependence on geometry).}. This additional component has the effect of flattening or ``graying'' the overall effective extinction, due to bluer light experiencing more efficient scattering. These added complications hinder a general prescription for correcting for dust attenuation in external galaxies.

One exception to the aforementioned problem has been for starburst galaxies, for which a tight positive correlation exists between the slope of the UV flux density, $\beta$, and the color excess of the nebular gas, $E(B-V)_{\mathrm{gas}}$, which are both related to the wavelength-dependent attenuation \citep{calzetti94}. The color excess can be inferred using the \textit{differential} optical depth of the dust from the Balmer decrement, $\tau_B^l$, which we will term ``Balmer optical depth'' for sake of brevity. Tight positive correlations also exist between $\beta$ and the ratio of IR to UV luminosity, termed the ``infrared excess'' ($IRX=L_{\rm{IR}}/L_{\rm{UV}}$), which is a proxy for the total dust attenuation \citep{meurer99}. The simple interpretation of the correlation between $\beta$ and $IRX$ is that dust attenuation increases with higher dust-to-gas ratios. Unfortunately, the $IRX-\beta$ correlation has been shown to break down as one moves from starburst galaxies to more ``normal'' star forming galaxies (SFGs) \citep{kong04,buat05,hao11}. This breakdown has been attributed to effects of evolved stellar populations, different star formation histories, and variations in the dust/star geometry \citep[e.g.,][]{boquien09,grasha13}, all of which can impact the $IRX$ and/or $\beta$ values and introduce large scatter.

The influence of dust appears to increase at intermediate redshifts ($z\sim1-3$), as the SED of galaxies are more heavily attenuated by dust than in the local Universe, corresponding to a larger fraction of the star formation within these galaxies being enshrouded by dust \citep{lefloch05,magnelli09,elbaz11,murphy11,reddy12}. These effects can result in imprecise values of SFRs, stellar mass, extinction corrections, and photometric redshifts of individual galaxies. Uncertainties in the latter two quantities are among the biggest factors limiting current precision dark energy studies, affecting both weak-lensing and supernovae measurements. Therefore, it is imperative to accurately characterize the dust attenuation in galaxies in order to reduce the uncertainties in the interpretation of data in future missions that seek to measure cosmological quantities with unprecedented precision. In addition, these quantities are crucial for studies of galaxy formation and evolution. Understanding the dust attenuation in local SFGs as a function of their properties (e.g., metallicity, stellar mass, SFR) creates a baseline with which to determine appropriate corrections for higher redshift galaxies. This will allow for accurate determination of galaxy properties when limited measurements are available, as is typically the case for higher redshift systems.

For their sample of 39 local starburst galaxies, \citet{calzetti94} find a linear correlation between $\beta$ and $\tau_B^l$. This result implies that the dust behaves as a foreground distribution to the ionized gas, because in this scenario the reddening of the stellar continuum linearly correlates with the reddening of nebular regions. They derive the selective attenuation for this sample by comparing the average SEDs of galaxies binned according to $\tau_B^l$. Virtually all studies of SFGs, both local and distant, make use of the attenuation curve derived from this sample \citep{calzetti00} to correct for effects of dust and determine properties of those galaxies. The appeal of this method stems from its simple approach; determining rest-frame UV colors ($\beta$), which correspond to observer-frame optical colors for high-$z$ galaxies, allows for the dust-free luminosity to be recovered. However, since the \citet{calzetti00} attenuation curve was calibrated with a relatively small number of local starburst galaxies, it is not clear how accurate such generalizations are to more typical SFGs with lower specific star formation rates (sSFRs; $\mathrm{SFR}/M_*$). More importantly, the extent to which the relation holds true as a function of redshift has not been conclusively determined. Recent results by \citet{reddy15} suggest that the relationship between $\beta$ and $\tau_B^l$ is shallower for galaxies at $z\sim2$ and is dependent on the sSFR. They also find that the attenuation curve for these galaxies is lower by about 20\% in the UV. A similar study of a large number of local SFGs will provide a strong foundation with which to compare and address why such differences exist. 

Other studies have examined the nature of attenuation for large samples of local galaxies \citep[e.g.,][]{johnson07b,wild11} in order to address the degree to which it can vary. However, these studies have used different techniques than those described in \citet{calzetti94,calzetti00}, which can introduce different biases and make direct comparison unclear. More specifically, \citet{johnson07b} use average SEDs of galaxies separated according to $IRX$ and \citet{wild11} utilize a galaxy pair-matching technique (matched in gas-phase metallicity, sSFR, axial ratio, and redshift) to compare the SEDs of more dusty and less dusty galaxies as determined by $\tau_B^l$. Despite the different methods for constructing attenuation curves among these studies, a common picture has developed in which the dust content of galaxies appear to have two components \citep[e.g.,][]{calzetti94,charlot&fall00,wild11}; one associated with short-lived dense clouds where massive stars form HII regions and another associated with the diffuse interstellar medium. 

For this study, we follow the methodology used in \citet{calzetti94} to determine the behavior of dust attenuation in a large sample of SFGs with a wide range of properties in order to test the extent to which the dust attenuation and geometry found in that work might hold for a more diverse sample of galaxies. Understanding the geometry of the dust in these systems gives important information on where dust is located in galaxies. The large sample size will also allow us to examine sources of scatter in the attenuation properties of individual galaxies. Furthermore, our results will allow for more detailed future analysis into the nature of the $IRX-\beta$ relationship breakdown for normal SFGs at $z\sim0$.

Throughout this work we adopt a $\Lambda$-CDM concordance cosmological model, $H_0=70$~km/s/Mpc, $\Omega_M=0.3$, $\Omega_{vac}=0.7$. We also assume a Kroupa IMF \citep{kroupa01} when making comparisons with stellar population models. To avoid confusion, we make explicit distinction between the color excess of the stellar continuum $E(B-V)_{\mathrm{star}}$, which traces the reddening of the bulk of the galaxy stellar population, and the color excess seen in the nebular gas emission $E(B-V)_{\mathrm{gas}}$, which traces the reddening of the ionized gas around massive stars located within HII regions. In principle these two parameters need not be related because $E(B-V)_{\mathrm{star}}$ and $E(B-V)_{\mathrm{gas}}$ are a result of attenuation and extinction, respectively (i.e., they use different obscuration curves; see \S~\ref{method_tau}), but they have been found to be correlated in starburst galaxies, with $\langle E(B-V)_{\mathrm{star}}\rangle=(0.44\pm0.03)\langle E(B-V)_{\mathrm{gas}}\rangle$ \citep{calzetti00}, and in star forming regions within local galaxies, with $\langle E(B-V)_{\mathrm{star}}\rangle=(0.470\pm0.006)\langle E(B-V)_{\mathrm{gas}}\rangle$ \citep{kreckel13}.

\section{Data and Measurements}
\subsection{Sample Selection}\label{data}
Our sample is constructed using the \textit{Galaxy Evolution Explorer} \citep[\textit{GALEX}; ][]{martin05,morrissey07} catalogs of \citet{bianchi14}. These catalogs represent unique sources in the \textit{GALEX} data release 6/7 (GR6/7) and are separated for the All-sky Imaging Survey (AIS; depth $m_{\mathrm{AB}}\sim20.5$~mag in FUV/NUV), containing $\sim$71 million sources, and the Medium Imaging Survey (MIS; depth $m_{\mathrm{AB}}\sim22.7$~mag), containing $\sim$16.6 million sources \citep{bianchi14}. We only make use of sources that can be cross-matched to the Sloan Digital Sky Survey (SDSS) data release 7 \citep[DR7;][]{abazajian09} sources and have available SDSS spectroscopy, as the latter is required to determine the Balmer decrement. These cross-matched cases are determined using the Mikulski Archive for Space Telescopes (MAST) CasJobs website\footnote{\url{http://galex.stsci.edu/casjobs/}}, with the requirement that the separation between objects be within $3\arcsec$ to be a match. Since our analysis requires detection in both the \textit{GALEX} FUV ($1344-1786$~\AA) and NUV ($1771-2831$~\AA) bands, we have chosen to only consider galaxies within the MIS catalog because spectroscopic SDSS galaxies detected within both bands in the shallower AIS are found to be highly biased towards very blue galaxies (see \S~\ref{sample_compare} for details). Within the MIS catalog, this restricts the parent sample to 63691 galaxies. 

For the purpose of this study, we further constrain the sample to only star forming galaxies (SFGs) and exclude cases in which a significant fraction of the flux density is produced from an active galactic nucleus (AGN). The galaxy type is determined using the traditional optical emission line diagnostics $[\rm{N II}]\lambda6583/H\alpha$, a proxy for gas phase metallicity, and $[\rm{O III}]\lambda5007/H\beta$, a measure of the hardness of the radiation field \citep[i.e., the Baldwin-Phillips-Terlevich (BPT) diagram,][]{baldwin81,veilleux&osterbrock87,kewley01,kauffmann03c}. The optical spectroscopic measurements for these galaxies are from the Max Planck Institute for Astrophysics and Johns Hopkins University (MPA/JHU) group\footnote{\url{http://www.mpa-garching.mpg.de/SDSS/DR7/}}, which is based on the method presented in \citet{tremonti04}. To summarize, line fluxes are corrected for stellar absorption by fitting a non-negative combination of stellar population synthesis models from \citet{bruzual&charlot03} for the SDSS DR4 and updated in DR7 using a newer version of these models (unpublished). As recommended by the MPA/JHU group starting with DR4, we increase the uncertainties associated with each emission line. We adopt the values listed in \citet{juneau14}, which are updated for the DR7 dataset. It has been found by \citet{groves12} that the equivalent width of the H$\beta$ emission line of galaxies in the MPA/JHU catalog appear to be underestimated by $\sim$$0.35$~\AA\ due to their method of correcting Balmer absorption. However, this correction value was derived assuming these galaxies follow the \citet{calzetti00} attenuation curve and because we are seeking to determine if there are significant departures from that relation, we choose not to adopt it. We have examined the effect of including this correction and found that it causes a systematic decrease in the estimated Balmer optical depth of $\Delta\tau_B^l\sim-0.1$. As was seen in \citet{groves12}, we find this shift to be uniform across the entire range of Balmer emission line strengths (i.e., $\tau_B^l$ values). For this reason, we expect that this will not influence the attenuation curve derived later because only the difference in Balmer optical depth is used and not the absolute value. 

We require that all emission lines have a signal-to-noise ratio ($S/N$) greater than 5 for classification, with the additional constraint that the FUV and NUV measurements have $S/N>5$ and that the redshift of the galaxy be $z\le0.105$. The redshift restriction is required to ensure that the FUV passband ($\lambda_{\rm{FUV}}=$1516~\AA, FWHM=269~\AA) lies above the numerous stellar absorption features that occur below 1250~\AA. Using this selection criteria gives a final sample of 9813 SFGs. 

All photometry and spectroscopy has been corrected for foreground Milky Way extinction using the \textit{GALEX} provided $E(B-V)_{\mathrm{MW}}$ with the extinction curve of \citet{fitzpatrick99}. For the \textit{GALEX} bands, we adopt the values of $k_{\mathrm{FUV}}=8.06$\footnote{$k(\lambda)\equiv A_\lambda /E(B-V)$ is the total-to-selective extinction.} and $k_{\mathrm{NUV}}=8.05$, which represent the average value of the MW extinction curve convolved with each filter on SEDs with UV slopes $-2.5 <\beta<0.5$, the typical range for our SFG sample.

All measurements of galaxy properties utilized in this work are those provided by the MPA/JHU group and correspond only to the 3\arcsec\ SDSS fiber, which is typically centered on the nuclear region and represents only a fraction of the total galaxy. The stellar masses are based on fits to the photometric data following the methodology of \citet{kauffmann03a} and \citet{salim07}. The SFRs are based on the method presented in \citet{brinchmann04}. The gas phase metallicities are estimated using \citet{charlot&longhetti01} models as outlined in \citet{tremonti04}. The fiber regions of the 9813 SFGs in this sample span the following range in properties: $5.99<\log[M_* (M_{\odot})]<10.67$, $-3.66<\log[SFR (M_{\odot}\mathrm{yr}^{-1})]<1.60$, and $7.67<12+\log(O/H)<9.37$. For comparison with \citet{calzetti94}, our sample consists of galaxies with lower sSFRs (average H$\alpha$ emission equivalent width, $\langle \mathrm{EW(H}\alpha)\rangle \sim-40$~\AA; a proxy for sSFR) relative to their starburst sample \citep[$\langle \mathrm{EW(H}\alpha)\rangle \sim-110$~\AA;][]{mcquade95,storchi-bergmann95}.

\subsection{UV-Optical Aperture Matching}
In order to test the existence of any relation between the UV flux density measured by \textit{GALEX} and the optical flux density measured by SDSS it is crucial that the apertures be closely matched in order to ensure the they arise from regions of the galaxy that are comparable in size. This is also essential in order to utilize the SEDs of these galaxies to derive the underlying attenuation curve. The limiting factor in this respect is the $3\arcsec$ diameter of the SDSS spectroscopic fiber. This aperture is smaller than the point-spread function (PSF) of \textit{GALEX} at both FUV ($\mathrm{FWHM}=4.2\arcsec$) and NUV ($\mathrm{FWHM}=4.9\arcsec$). Using an aperture which is smaller than \textit{GALEX} PSF would add positional uncertainty that could introduce UV emission unassociated with the fiber location. However, using an aperture much larger than the PSF would require large aperture corrections for our fiber measurements, which would also introduce uncertainty. In addition, the large PSF of \textit{GALEX} can lead to nearby objects contributing to the observed flux density within a given area and using a large aperture would increase the likelihood of this happening. As a compromise between these issues we choose to adopt a $4.5\arcsec$ diameter aperture for our analysis.  The UV photometry and associated uncertainties at this aperture were retrieved directly from the MAST database using CasJobs. We note that by adopting a fixed aperture, the physical sizes being probed will vary from sub-kpc at the lowest redshifts up to several kpc at the higher redshifts ($0.002\le z\le0.105$).  We address the impact of this effect on our results in \S~\ref{atten_vs_param} \& \ref{curve_vs_param} when we separate the sample according to redshift.

\begin{figure}
\begin{center}
\includegraphics[width=3.5in]{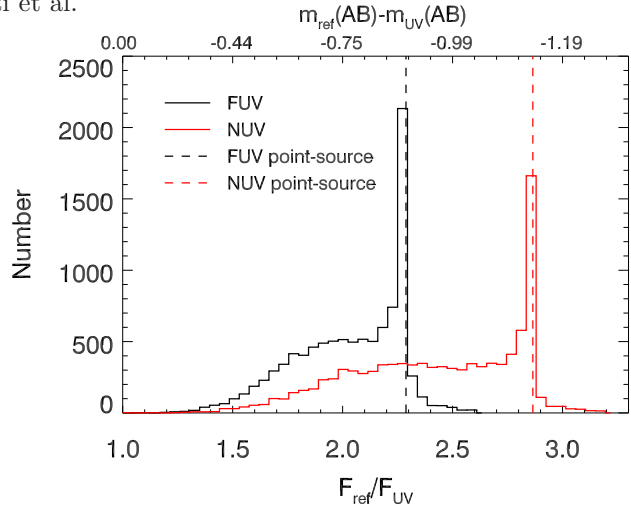}
\end{center}
\caption{Histogram of the flux in $4.5\arcsec$ apertures based on the SDSS $u$-band S\'ersic light profile, $F_{\mathrm{ref}}$, relative to the flux within the same aperture after convolving $u$-band light profile by the \textit{GALEX} PSFs, $F_{\mathrm{UV}}$, for the 9813 SFGs in our sample. This ratio corresponds to the aperture correction, $F_{\mathrm{ref}}/F_{\mathrm{UV}}$, under the assumption that the UV light follows a similar behavior to that of the $u$-band. Roughly half of the sources are well approximated as point-sources, which corresponds to the peaks in the distributions (vertical dashed lines). \label{fig:galex_aper_corr}}
\end{figure}

Given that the chosen aperture is roughly the size of the PSF for the \textit{GALEX} bands, determining the appropriate aperture corrections for the UV is non-trivial as a significant amount of light from outside of the aperture region can be spread within it. To distinguish these effects we utilize the light profile models of SDSS galaxies from the NYU Value-Added Galaxy Catalog\footnote{\url{http://sdss.physics.nyu.edu/vagc/}} \citep[NYU-VAGC;][]{blanton05}. These profiles are one-component S\'ersic fits of each individual band of SDSS,
\begin{equation}
I(r) = A \exp \left[ -(r/r_0)^{1/n} \right] \,,
\end{equation}
where $A$ is an amplitude in nanomaggies/arcsec$^2$, $r_0$ is an effective radius in arcsec, and $n$ is the S\'ersic index. A nanomaggie is a flux density unit defined such that 1~nanomaggie has a magnitude of 22.5 in any band, $m_{AB}=22.5-2.5\log[f(\mathrm{nMgy})]$, which for SDSS as a near AB magnitude system is $3.631$~$\mu$Jy. If we assume that the UV colors follow the $u$-band light profile, which is the shortest wavelength SDSS band, then it is possible to determine appropriate aperture corrections for each individual galaxy. This is done by measuring the amount of light within $4.5\arcsec$ in the modeled $u$-band profile, $F_{\mathrm{ref}}$, and comparing this to the same measurement after the light profile has been convolved with the \textit{GALEX} PSFs, $F_{\mathrm{UV}}$. These PSFs are obtained from the \textit{GALEX} data analysis website\footnote{\url{http://www.galex.caltech.edu/researcher/techdoc-ch5.html}}. The aperture correction is taken as the ratio of these values, $F_{\mathrm{ref}}/F_{\mathrm{UV}}$. The distribution of aperture corrections for the entire parent sample of galaxies is shown in Figure~\ref{fig:galex_aper_corr}. Roughly one-third of the sources in our sample are well approximated as point sources in \textit{GALEX} (corresponding to being near the peak in the distribution). The aperture correction in flux density for a point-source is a factor of 2.29 and 2.86 for FUV and NUV, respectively. 

\begin{figure}
\begin{center}$
\begin{array}{c}
\includegraphics[width=3.5in]{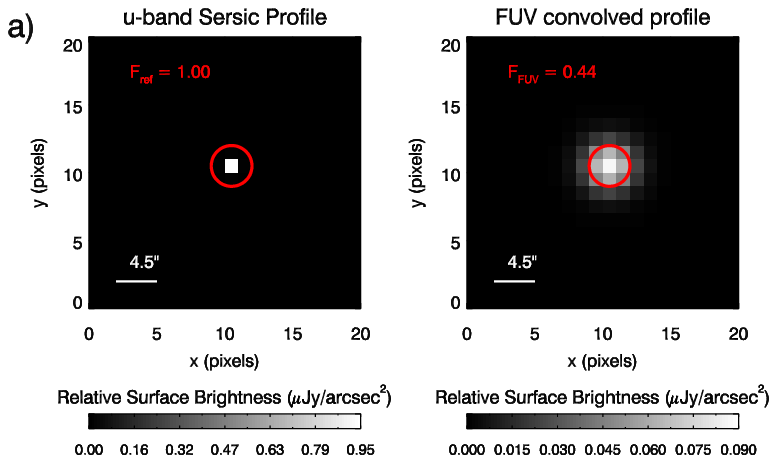} \\
\includegraphics[width=3.5in]{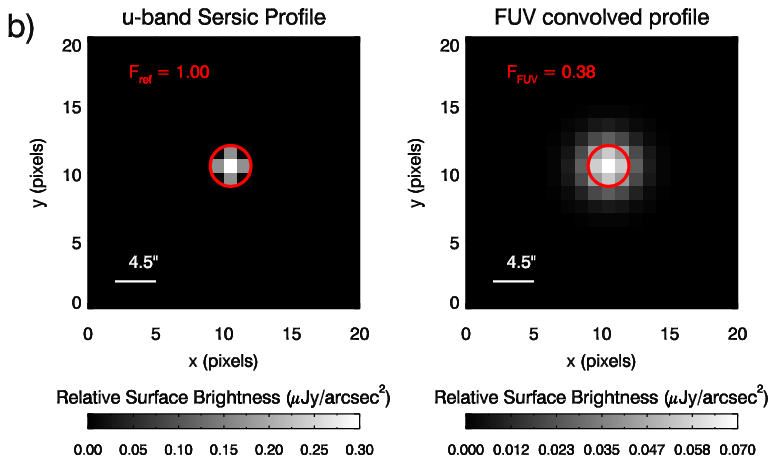} \\
\includegraphics[width=3.5in]{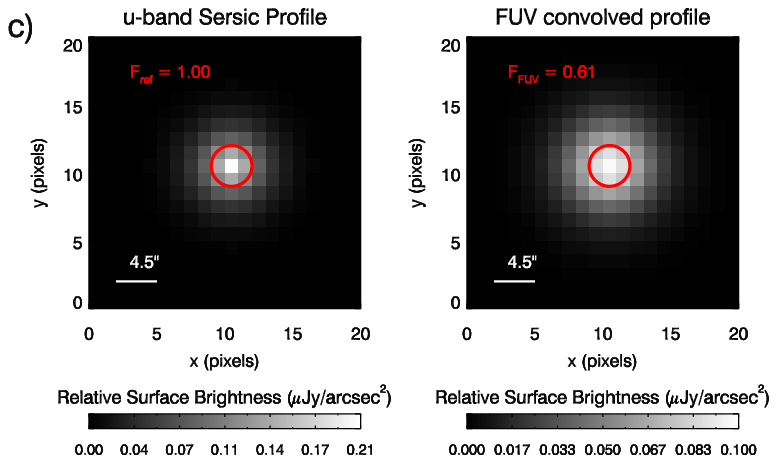} \\
\end{array}$
\end{center}
\caption{Demonstration of how different galaxy light profiles affect the aperture correction ($F_{\mathrm{ref}}/F_{\mathrm{UV}}$). The red circle represents the $4.5\arcsec$ aperture. The surface brightness profiles have been normalized such that the aperture flux density (shown in red) in the $u$-band value is $1~\mu$Jy. (a) Galaxies with radii of $r_{90,u}\lesssim1\arcsec$ or $r_{90,u}\gtrsim1\arcsec$ and large S\'ersic indices (i.e., steep light profile) are well described as point-sources. (b) Galaxies with $1\arcsec\lesssim r_{90,u}\lesssim3\arcsec$ and small S\'ersic indices (i.e., shallow light profile) have the light within $4.5\arcsec$ being spread out more than would be the case for a point-source and this leads to smaller $F_{\mathrm{FUV}}$ relative to case (a). (c) Galaxies with $r_{90,u}\gtrsim3\arcsec$ and small S\'ersic indices have significant light outside of the $4.5\arcsec$ aperture that can be spread into it and we find larger $F_{\mathrm{FUV}}$ relative to case (a). The NUV aperture corrections behave in a similar manner. \label{fig:galex_aper_corr_demo}}
\end{figure}

It can be seen in Figure~\ref{fig:galex_aper_corr} that there are some sources with aperture corrections that are larger than what is expected for a point source (vertical dashed lines). To determine the cause of this interesting behavior we examined the relationship between the aperture corrections with the effective radius of the galaxy and its S\'ersic index in the $u$-band. We choose to use the $90\%$ light radius from the NYU-VAGC light profile, $r_{90,u}$, instead of the variable $r_0$ in the S\'ersic fit, as it is a better representation of the size of the galaxy. We find that the behavior of the aperture correction is directly related to $r_{90,u}$ and the S\'ersic index, which we illustrate in Figure~\ref{fig:galex_aper_corr_demo} and discuss below.   

We find that sources with radii of $r_{90,u}\lesssim1\arcsec$ or with $r_{90,u}\gtrsim1\arcsec$ and large S\'ersic indices (i.e., steep light profile) are well described by a point-source correction (see panel (a) in Figure~\ref{fig:galex_aper_corr_demo}). For the regime of galaxies with $1\arcsec\lesssim r_{90,u}\lesssim3\arcsec$ and small S\'ersic indices (i.e., shallow light profile), there is little galaxy flux density outside of the $4.5\arcsec$ aperture to spread inside of it, due to the smearing effect of the PSF, but the light within it is being spread out more than would be the case for a point-source. Together these effects result in a larger aperture correction being necessary than would be the case for a point-source (see panel (b) in Figure~\ref{fig:galex_aper_corr_demo}). For galaxies with $r_{90,u}\gtrsim3\arcsec$ and small S\'ersic indices, there is significant light outside of the $4.5\arcsec$ aperture that can be spread into it and we find an aperture correction that is smaller than that for a point-source (see panel (c) in Figure~\ref{fig:galex_aper_corr_demo}).

Since we have chosen to adopt a $4.5\arcsec$ aperture for our galaxies, we also need to apply an aperture correction to the SDSS spectroscopy ($3\arcsec$). Following a similar methodology to before, we determine the correction for the observed photometry using the light profile models from the NYU-VAGC. The correction is taken to be the ratio of light within $4.5\arcsec$ and $3.0\arcsec$ in the modeled band profile. We then perform a chi-squared minimization to match the optical spectrum to the $4.5\arcsec$ photometry. 

All nebular line diagnostics used in this study are ratios of emission lines ($[\rm{N II}]/H\alpha$, $[\rm{O III}]/H\beta$, and H$\alpha$/H$\beta$), and the effects of the aperture corrections are very small. The reason for this is because the relative difference in the correction terms across the SDSS bands is small ($\sim6\%$, $1\sigma$ dispersion of $11\%$). As a check, we estimated a correction for each the optical emission lines by taking a linear interpolation between the corrections of the two closest bands. The largest aperture effects for the line ratios will occur for lines which are separated the furthest in wavelength; which in this study is the ratio of the H$\alpha$ to H$\beta$. We find that the distribution of the ratio of the aperture corrections for H$\alpha$ to H$\beta$ has a mean of $0.98$, with a $1\sigma$ dispersion of 0.03. This implies these corrections will only change the ratio measurements at the level of a few percent and will be even smaller for the other line ratios ($\ll1\%$). As these are minor changes we have chosen not to apply any aperture corrections to the emission lines. 

As a check on the accuracy of the aperture matching, we compare the corrected UV flux densities to the corrected optical spectra and inspected if the shape of the UV (inferred from $\beta_{\rm{GLX}}$) agrees with the shortest wavelength data available in the optical spectrum ($\lambda\sim3600$~\AA, for $z\sim0.05$). In other words, we examine whether the region between $2600-3600$~\AA, corresponding to the gap in our data, that would be extrapolated from the UV slope and the optical data shortward of the 4000~\AA\ break feature are in agreement. If our aperture corrections are inaccurate, then we expect to see systematic offsets between the flux densities in the UV and optical. As will be shown with our average templates in \S~\ref{attenuation curve}, we find that on average the UV data agrees well with the optical data, with no significant offsets between them. 

\section{Methodology for Characterizing Attenuation}\label{method}
\subsection{Balmer Optical Depth}\label{method_tau}
The dust attenuation in a galaxy can be measured from the optical depth, $\tau(\lambda)$. For the simple case of a uniform layer of dust between a source of intensity, $I_\lambda^0$, and the observer, the optical depth is defined as
\begin{equation}\label{eq:intensity}
I_\lambda = I_\lambda^0e^{-\tau(\lambda)} \,,
\end{equation}
where $I_\lambda$ is the observed intensity and all quantities are dependent on the wavelength. In the case of a point source such as a star with a well-characterized SED, the optical depth can easily be determined by comparing the observed and intrinsic SEDs. However, in the case of entire galaxies for which the underlying SED is strongly affected by many factors, including the underlying stellar population, star formation history, and IMF, this becomes much more difficult. To mitigate these problems the flux ratio of H$\alpha$ and H$\beta$ is often utilized, as the intrinsic flux ratio is set by quantum mechanics and is only affected by the electron temperature, $T_{\mathrm{e}}$, and density, $n_{\mathrm{e}}$,  at the $\sim5-10\%$ level \citep{osterbrock06}. This ratio is also relatively insensitive to the underlying stellar population and IMF \citep{calzetti01}. Therefore, large variations from the intrinsic ratio can be directly attributed to the reddening of dust. 

For our work we will use the Balmer decrement, $F(\mathrm{H}\alpha)/F(\mathrm{H}\beta)$, as a tracer of dust attenuation, with the assumption that this dust acts as a foreground screen for these lines. Since the ionized gas from which these lines arise is primarily located in HII regions, which have small angular extent relative to the rest of the galaxy (and presumably the dust), and are usually distributed within a short height of the galaxy's midplane, this is a reasonable assumption. Following from equation~(\ref{eq:intensity}) and \citet{calzetti94}, we define the Balmer optical depth as
\begin{equation}\label{eq:tau}
\tau_B^l = \tau_{\mathrm{H}\beta} - \tau_{\mathrm{H}\alpha} = \ln \left(\frac{F(\mathrm{H}\alpha)/F(\mathrm{H}\beta)}{2.86}\right)\,,
\end{equation}
where $F($H$\alpha)$ and $F($H$\beta)$ are the flux of the nebular emission lines located at 6562.8~\AA and 4861.4~\AA, respectively, and the value of 2.86 comes from the theoretical value expected for the unreddened ratio H$\alpha/$H$\beta$ undergoing Case B recombination with $T_{\mathrm{e}}=10^4$~K and $n_{\mathrm{e}}=100$~cm$^{-3}$ \citep{osterbrock89,osterbrock06}. The superscript \textit{l} is used to emphasize that this quantity is coming from emission lines and should be distinguished from optical depths associated with the stellar continuum. If one assumes knowledge of the total-to-selective extinction, $k(\lambda)\equiv A_\lambda /E(B-V)$, then $\tau_B^l$ can be directly related to the color excess of the nebular gas, $E(B-V)_{\mathrm{gas}}$, through
\begin{equation}\label{eq:EBV_gas}
E(B-V)_{\mathrm{gas}}=\frac{A(\mathrm{H}\beta)-A(\mathrm{H}\alpha)}{k(\mathrm{H}\beta)-k(\mathrm{H}\alpha)}=\frac{1.086\tau_B^l}{k(\mathrm{H}\beta)-k(\mathrm{H}\alpha)} \,,
\end{equation}
where $A(\lambda)$ is the total extinction at a given wavelength. If the extinction in other galaxies at these wavelengths were to be identical to the MW, which is unlikely to always be the case, then we can use $k(\mathrm{H}\alpha)-k(\mathrm{H}\beta)=1.257$ \citep{fitzpatrick99}.

Since the hydrogen recombination lines are primarily produced within the HII regions of massive O and B stars, this implies that they are only sensitive to the attenuation of ionized gas around these massive stars. In general these same stars will also contribute greatly to the UV and optical light of the total stellar population, leading one to expect the reddening of the ionized gas to be related to the reddening seen in the stellar population (traced by $\beta$ or $E(B-V)_{\mathrm{star}}$). However, the distribution of these massive stars and the rest of the stellar population relative the distribution of dust, in addition to the age distribution of the stellar population (i.e., the relative contribution of older stars to the UV-optical flux density), will strongly influence how these quantities are related. For both starburst galaxies \citep{calzetti00} and star forming regions within local galaxies \citep{kreckel13} it appears that the stellar continuum suffers roughly one-half of the reddening of the ionized gas. This result suggests a scenario in which massive stars are more deeply embedded in molecular clouds than the long-lived stars (e.g., see descriptions in \citealt{charlot&fall00}, \citealt{calzetti01}, or \citealt{wild11}). Another concern is that the total amount of dust within a galaxy may only weakly relate to attenuation, as the regions of lowest optical depth ($\tau_B^l\lesssim 1$) are providing the majority of the flux density \citep{calzetti94}. Large amounts of the dust might exist in the regions of high optical depth ($\tau_B^l>>1$), which would provide negligible amounts of flux density and thus not be represented by our tracers. Fortunately, the presence of dust attenuation beyond the level of what can be measured in the optical has not been shown to be an issue for local starburst galaxies \citep{meurer99}, and therefore we do not expect this to be a major concern. 

\subsection{UV Slope}\label{method_beta}
For actively SFGs, where the UV is dominated by recent star formation (i.e., massive stars) and there is little contamination from earlier generations of stars, the dust attenuation can also be measured from the UV flux density spectral slope, $\beta$, which is defined as
\begin{equation}
F(\lambda)\propto\lambda^\beta\,,
\end{equation}
where $F(\lambda)$ is the flux density in the range $1250\le\lambda\le2600$~\AA. For reference, $\beta$ values cover the range between $-2.70$ and $-2.20$ for constant star formation \citep[i.e., they have a fairly small intrinsic variation;][]{calzetti01}. This makes this parameter ideal for measuring the wavelength-dependence of attenuation in the UV. In Appendix~\ref{opt_slope}, we examine the use of the optical slope as dust tracer when UV data is unavailable.

For this study, we use the UV power-law index $\beta_{\rm{GLX}}$ measured from observed \textit{GALEX} FUV and NUV photometry,
\begin{equation}
\beta_{\rm{GLX}} = \frac{\log[F_\lambda(\mathrm{FUV})/F_\lambda(\mathrm{NUV})]}{\log[\lambda_{\mathrm{FUV}}/\lambda_{\mathrm{NUV}}]}\,,
\end{equation}
where the flux density is in erg~s$^{-1}$~cm$^{-2}$~\AA$^{-1}$, $\lambda_{\rm{FUV}}=1516\ang/(1+z)$, and $\lambda_{\rm{NUV}}=2267\ang/(1+z)$. If one assumes a power-law fit to the region described above, there is no need to perform $k$-corrections on the flux density. We examine the possibility of a 2175~\AA\ absorption feature biasing this UV slope measurement in \S~\ref{influence_of_2175}. As is shown in \citet{calzetti94}, the quantities $\tau_B^l$ and $\beta$ can be used to \textit{derive} a dust attenuation curve independent of any prior knowledge of its shape. Such an analysis will be performed in \S~\ref{attenuation curve}.

\section{Dust Attenuation in Star Forming Galaxies}
\subsection{Relating Attenuation of UV Continuum to the Balmer Optical Depth}\label{beta_vs_tau}
\begin{figure}
\plotone{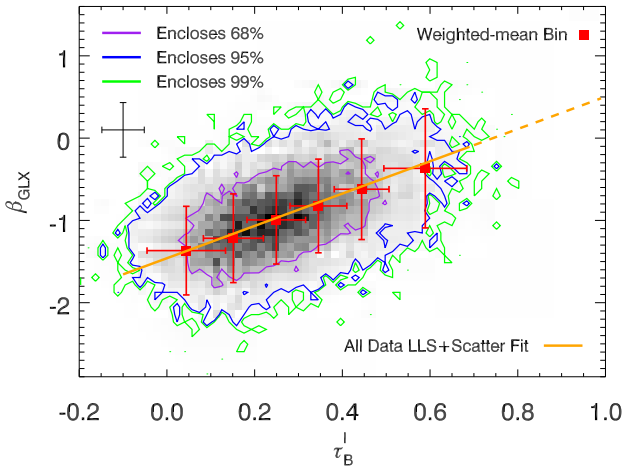}
\caption{The UV power-law index, $\beta_{\rm{GLX}}$, as a function of the Balmer optical depth, $\tau_B^l$, for our sample of SFGs. A representative error bar of the median measurement uncertainties for our sample is shown in the top left. A linear least-squares fit with error in both variables while also including a term to account for intrinsic scatter in the data is shown (orange line). Our fit at $\tau_B^l>0.7$ is shown with a dashed line to denote that there are limited data in this range. For comparison, the data are separated into six  bins of $\tau_B^l$ (red squares; see Table~\ref{Tab:tau_bin}). \label{fig:beta_tau}}
\end{figure}

In order to characterize the dust attenuation in SFGs, we first examine how the reddening in the UV stellar continuum (measured through $\beta_{\rm{GLX}}$) is related to the optical reddening of the ionized gas (measured through $\tau_B^l$). In Figure~\ref{fig:beta_tau} we show the $\beta_{\rm{GLX}}$ and $\tau_B^l$ values for our sample of 9813 SFGs. There does appear to be a significant correlation present, but with a large degree of scatter. Spearman and Kendall nonparametric correlation tests give $\rho_S=0.48$ and $\tau_K=0.33$, respectively, which indicate that this correlation is not particularly strong. Given the large sample size we are dealing with, it is not possible to report the significance of these correlation coefficients because the probability of no correlation existing is found to be very close to zero.

Looking at Figure~\ref{fig:beta_tau}, it can be seen that there are a number of cases for which $\tau_B^l<0$. This corresponds to cases where the observed flux ratio of the Balmer lines is below the assumed intrinsic value of 2.86. This can result from the uncertainties in the measured values of these lines (see representative error bar) or from variations in the intrinsic line ratio \citep[e.g., for $T_{\mathrm{e}}>10^4$~K and/or $n_{\mathrm{e}}>100$~cm$^{-3}$ the intrinsic ratio decreases;][]{osterbrock06}. Another interesting feature is the lack of data points at $\tau_B^l\gtrsim0.7$, which corresponds to galaxies experiencing the largest attenuation. It is worth determining if this is being driven by the selection criteria imposed for our sample, which requires the UV flux density and emission line fluxes to have $S/N>5$, and if it could result in any biases. We postpone the analysis of a UV selection bias until \S~\ref{sample_compare}, where we will compare sources from \textit{GALEX} surveys of different depths. To test for an emission line selection bias, we examine the effect that requiring a lower $S/N$ threshold for the weakest line in this study ([OIII]$\lambda5007$) has on the sample being selected. Imposing a threshold of $S/N>3$ for [OIII]$\lambda5007$, while still requiring $S/N>5$ for the other lines, increases the sample size by 2160 galaxies. Imposing lower thresholds on the other lines does not increase the sample size significantly. We do find that the weaker [OIII] systems are slightly more attenuated on average than the original sample (i.e., larger $\beta_{\rm{GLX}}$ \& $\tau_B^l$), but the vast majority are still $\tau_B^l<0.7$. Including these galaxies has no effect on the relationship between $\beta_{\rm{GLX}}$ and $\tau_B^l$. Therefore, we believe that we are not preferentially excluding objects located at $\tau_B^l\gtrsim0.7$. The lack of objects at these values may result from galaxies at this level of attenuation being relatively rare in the local Universe, which is consistent with the results of \citet{kauffmann03a}. 

Similar to the findings of \citet{calzetti94}, the observed relationship between $\beta_{\rm{GLX}}$ and $\tau_B^l$ is linear, which indicates that the dust behaves as a foreground-like screen to ionized gas regions. We fit a linear relationship to the data in Figure~\ref{fig:beta_tau} using the MPFITEXY routine \citep{williams10}, which utilizes the MPFIT package \citep{markwardt09}. This routine performs a linear least-squares fit with error in both variables while also including a term to account for intrinsic scatter in the data. A scatter within the data is expected for the y-axis given that $\beta_{\rm{GLX}}$ is likely to be dependent on variations in the age of the stellar population, the star formation history, and/or the metallicity of each galaxy \citep{calzetti94}. A fit to the $\beta_{\rm{GLX}}$ and $\tau_B^l$ values for all of the data gives 
\begin{equation}
\beta_{\mathrm{GLX}} = (1.96\pm0.03) \tau_B^l -(1.46\pm0.01) \,,
\end{equation}
with an intrinsic dispersion of $\sigma_{\mathrm{int}}=0.43$.  We denote the region of $\tau_B^l>0.7$ in our fit with a dashed line to indicate that there are limited data in this range. Interestingly, the width of the scatter in $\beta_{\rm{GLX}}$ does not appear to change with $\tau_B^l$. As a result of this behavior, we do not expect the scatter to be strongly driven by possible variations in the dust geometry. This is because the scatter around the value of $\beta_{\rm{GLX}}$ should decrease for ``low-dust'' systems ($\tau_B^l\sim0$), regardless of geometry, and approach the intrinsic value of $\beta_{\rm{GLX}}$ for each galaxy \citep[e.g., see][]{calzetti94,calzetti00}. We will examine this scatter in more detail in \S~\ref{atten_vs_param}. 

We also examine the data after binning it into 6 bins of $\tau_B^l$, which is useful for comparison in our derivation of the attenuation curves in \S~\ref{attenuation curve}. The ranges of the bins along with their weighted-mean values, which account for uncertainties in both variables, are shown in Table~\ref{Tab:tau_bin}. The error bar shown for each bin correspond to the measurement uncertainty and the sample dispersion added in quadrature. It can be seen that these bins are in good agreement with the previous fit, given the uncertainties. We postpone a comparison of our $\beta_{\rm{GLX}}-\tau_B^l$ relation to those in the literature until \S~\ref{beta_GLX_vs_z}, as we will make use of our derived attenuation curve to determine an appropriate way to compare different techniques for measuring the UV slope.  

\begin{table}
\begin{center}
\caption{Values of $\tau_B^l$ Bins \label{Tab:tau_bin}}
\begin{tabular}{ccccc}
\hline\hline 
 Bin & N & $\langle\tau_B^l\rangle$ & $\langle\beta_{\rm{GLX}}\rangle$ & $\langle\beta\rangle$ \\ \hline
  $-0.26\le\tau_B^l<0.10$& 1303 & $0.04\pm0.09$ & $-1.37\pm0.54$ & $-1.52\pm0.55$ \\
  $0.10\le\tau_B^l<0.20$ & 2244 & $0.15\pm0.07$ & $-1.22\pm0.54$ & $-1.37\pm0.55$ \\
  $0.20\le\tau_B^l<0.30$ & 2533 & $0.25\pm0.07$ & $-1.00\pm0.54$ & $-1.15\pm0.54$ \\
  $0.30\le\tau_B^l<0.40$ & 1913 & $0.35\pm0.06$ & $-0.83\pm0.57$ & $-0.98\pm0.58$ \\
  $0.40\le\tau_B^l<0.50$ & 1101 & $0.44\pm0.06$ & $-0.62\pm0.61$ & $-0.77\pm0.62$ \\
  $0.50\le\tau_B^l<1.01$ & 719  & $0.59\pm0.09$ & $-0.37\pm0.72$ & $-0.52\pm0.74$ \\ \hline
\end{tabular}
\end{center}
\textbf{Notes.} Columns list the (1) range in $\tau_B^l$ spanned, (2) number of objects, (3) weighted-mean Balmer optical depth (see \S~\ref{method_tau}), (4) weighted-mean UV slope using the \textit{GALEX} passbands (see \S~\ref{method_beta}), (5) weighted-mean UV slope after correcting for stellar absorption features in the \textit{GALEX} passbands (see \S~\ref{beta_GLX_vs_z}). The uncertainties shown correspond to the measurement uncertainty and the sample dispersion added in quadrature. 
\end{table}

\subsection{Tests on Possible UV Selection Effects} \label{sample_compare}
As mentioned in \S~\ref{data}, we did not use the \textit{GALEX} AIS sample for our analysis as it was realized that sources detected in both FUV and NUV in this survey are biased towards galaxies with blue UV slopes. We attribute this bias to the shallowness of the AIS. This effect is identical to selection biases which occur for high redshift galaxies \citep[e.g.,][]{dunlop12,bouwens12} and is a result of sources at the detection threshold being preferentially identified if they have bluer colors. In this section we demonstrate the bias in the AIS sample and also show that no such bias is evident for the MIS.

\begin{figure}
\plotone{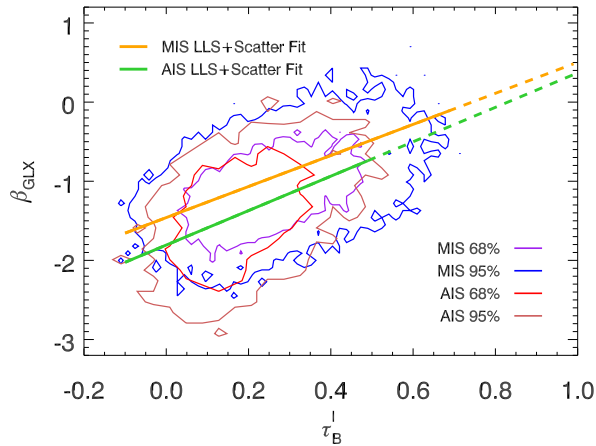}
\caption{Comparison of the $\beta_{\rm{GLX}}$-$\tau_B^l$ relation for galaxies in the \textit{GALEX} AIS and MIS surveys (depths of $m_{\mathrm{AB}}\sim20.5$~mag and $m_{\mathrm{AB}}\sim22.7$~mag, respectively) that satisfy our selection criteria. It is apparent that the sample in the shallower AIS survey is biased towards bluer galaxies (lower $\beta_{\rm{GLX}}$) with relatively little dust (lower $\tau_B^l$) compared to the sample from the deeper MIS survey. For this reason the AIS sample was excluded from our analysis. \label{fig:beta_tau_AIS_compare}}
\end{figure}

Following the same approach outlined for the MIS sample, we select sources in the AIS that have FUV and NUV data with $S/N>5$ and are designated as SFGs using the BPT diagram. This gives a sample of 3190 galaxies in the AIS. We plot the values of $\beta_{\rm{GLX}}$ vs. $\tau_B^l$ for the galaxies detected in these two survey in Figure~\ref{fig:beta_tau_AIS_compare}. Looking at this Figure, it is apparent that galaxies in the AIS are bluer with relatively small Balmer decrements (indicative of less dust) relative to MIS. This result suggests that the shallowness of the AIS is such that only galaxies with minimal attenuation can be detected in both UV bands. It is worth pointing out that \citet{wild11} use \textit{GALEX} AIS data for their analysis and this may account for differences in the UV region of the attenuation curve derived later in this study and theirs.

In light of the previous issue, we feel it necessary to check whether or not similar effects could be biasing the sample of galaxies selected in the MIS. To perform this test we examine galaxies within the Galaxy Multiwavelength Atlas from Combined Surveys (GMACS) dataset, which was observed with \textit{GALEX} as part of its deep imaging survey (DIS). The regions within the GMACS sample consist of the Lockman Hole, the Spitzer First Look Survey (FLS), and the SWIRE ELAIS-N1 and N2 fields. Unlike the MIS and AIS, the exposure times for different fields in the DIS can have significantly different exposure time but the typical depth is $m_{\mathrm{AB}}\sim25$~mag. This subsample of the DIS was chosen because the catalog of the cross-matched SDSS spectroscopic sources is publicly available\footnote{\url{http://user.astro.columbia.edu/~bjohnson/GMACS/catalogs.html}}.

\begin{figure}
\plotone{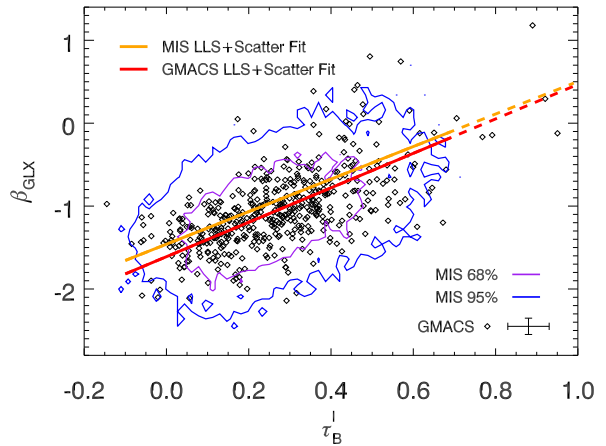}
\caption{Comparison of the $\beta_{\rm{GLX}}$-$\tau_B^l$ relation for galaxies in the \textit{GALEX} GMACS and MIS surveys (depths of $m_{\mathrm{AB}}\sim25$~mag and $m_{\mathrm{AB}}\sim22.7$~mag, respectively) that satisfy our selection criteria. These two samples occupy similar regions of $\beta_{\rm{GLX}}$-$\tau_B^l$ parameter space, despite having different depths. Therefore, we suspect that the MIS sample does not suffer from a UV-selection bias. \label{fig:beta_tau_GMACS_compare}}
\end{figure}

Following the same approach outlined for the MIS sample, we select sources in the GMACS sample that have FUV and NUV data with $S/N>5$ and are designated as SFGs using the BPT diagram. This gives a sample of 476 galaxies. We plot the values of $\beta_{\rm{GLX}}$ vs $\tau_B^l$ for this sample of galaxies compared to the MIS sample in Figure~\ref{fig:beta_tau_GMACS_compare}. A visual comparison suggests that this population appears to occupy a similar region of parameter space compared to the MIS sample. We take a LLS fit, with an added term for scatter, and find the relation,
\begin{equation}
\beta_{\mathrm{GLX,GMACS}} = (2.08\pm0.11) \tau_B^l -(1.61\pm0.03) \,,
\end{equation}
with an intrinsic dispersion of $\sigma_{\mathrm{int}}=0.36$. This relation is consistent with our original sample given the uncertainties and scatter. This implies that the UV depth of the MIS sample does not significantly bias our sample towards galaxies of a particular attenuation.  

As a final check, we also examined the distribution of $\tau_B^l$ values for the entire SDSS spectroscopic sample identified as a SFG using the BPT diagnostics with emission line strengths of $S/N>5$ and $z\le0.105$ but \textit{without} requiring any UV detection. This selection gives a sample of $\sim$150000 galaxies with a distribution of $\tau_B^l$ values which is nearly Gaussian with a mean of $\mu=0.26$ and a dispersion of $\sigma=0.17$. The range of values observed in the MIS selected sample is also nearly Gaussian with $\mu=0.25$ and a dispersion of $\sigma=0.15$. This would further argue that there is not a significant fraction of the full population that are missed as a result of the UV flux requirement.

\subsection{Deriving the Dust Attenuation Curve} \label{attenuation curve}
The main drawback in utilizing entire galaxies for deriving attenuation curves is that their spectra result from the contributions of many stellar populations of different ages, and therefore we have no knowledge of the underlying intrinsic spectra with which to directly compare (in contrast to using individual stars for extinction curves). However, given the large dataset on hand, we can take a statistical approach to determine the attenuation curve if we assume that the effects of different stellar populations, which should only significantly affect $\beta_{\rm{GLX}}$, can be averaged out within similar values of $\tau_B^l$. This is the same approach taken in \citet{calzetti94} and \citet{reddy15} to derive their attenuation curves. 

\begin{figure*}
\begin{center}$
\begin{array}{lr}
\includegraphics[width=3.3in]{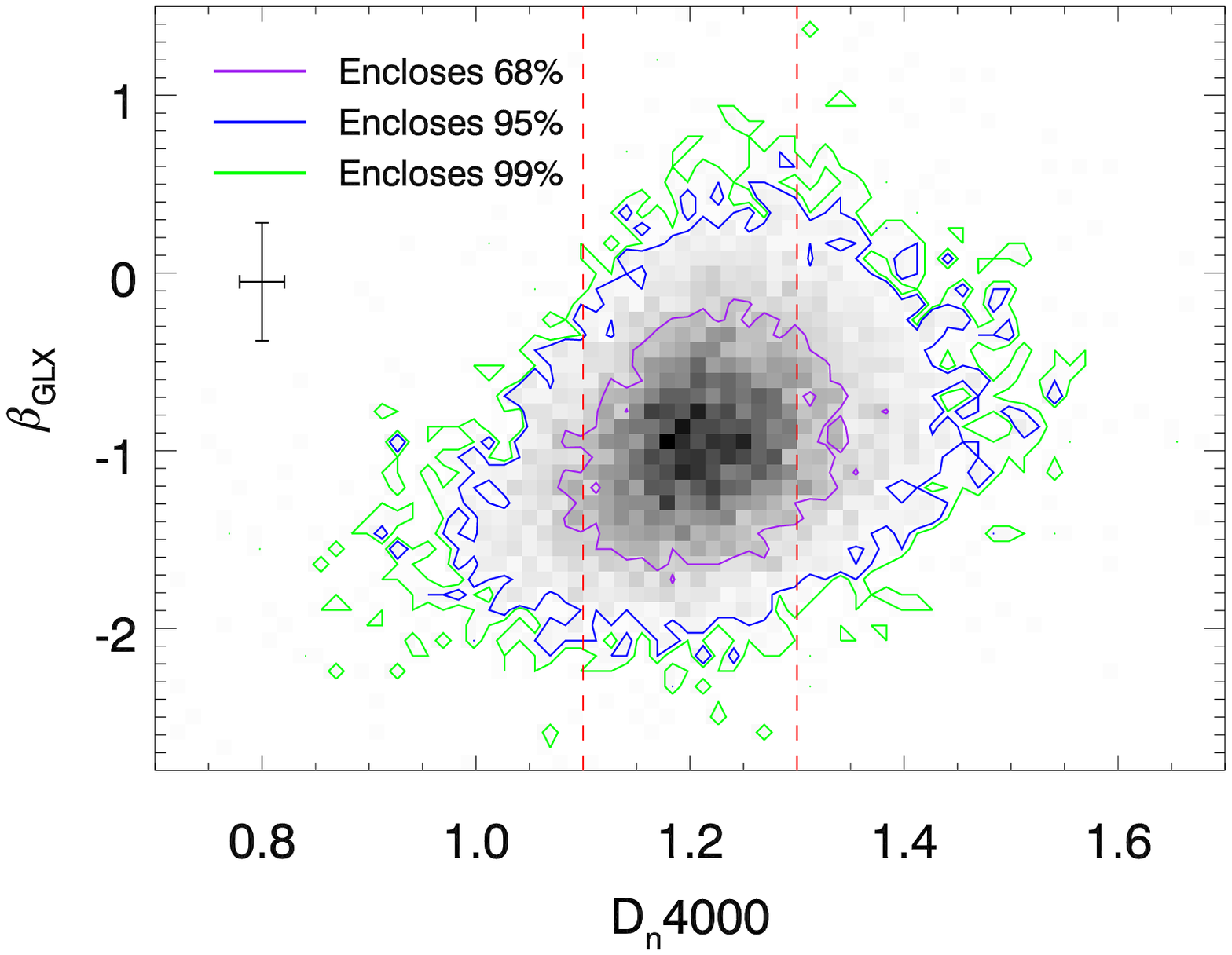}
\includegraphics[width=3.3in]{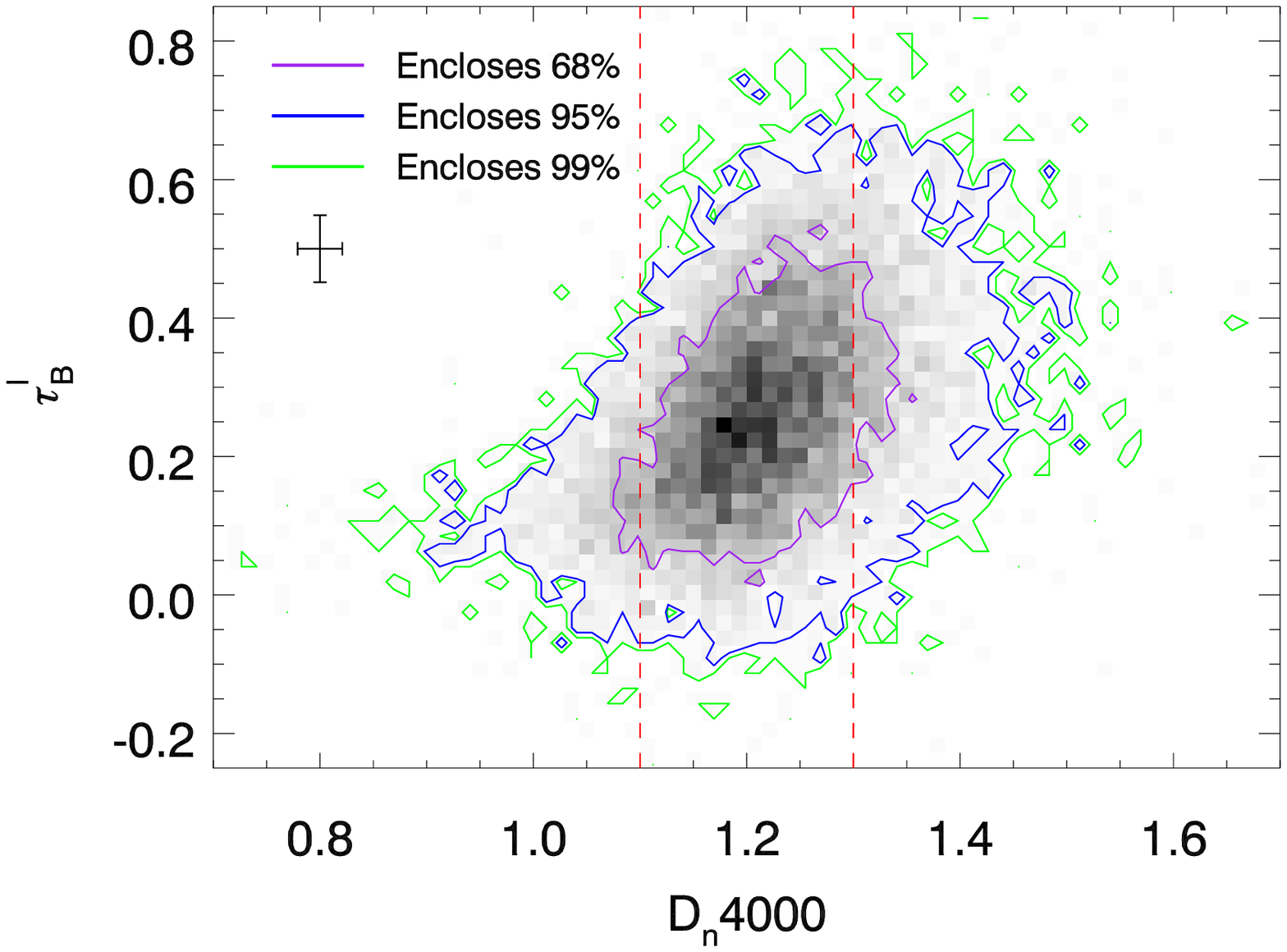} \\
\includegraphics[width=3.3in]{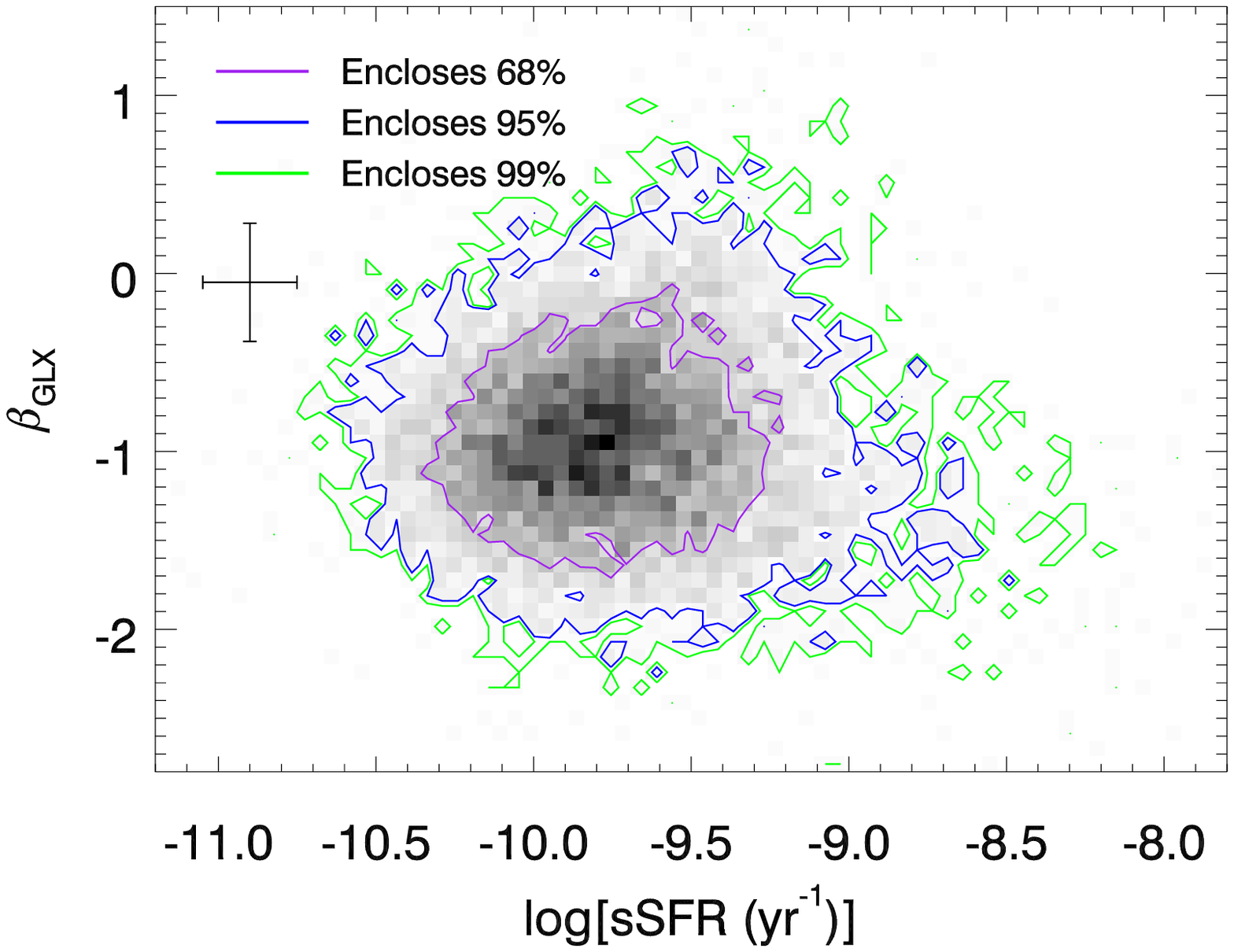}
\includegraphics[width=3.3in]{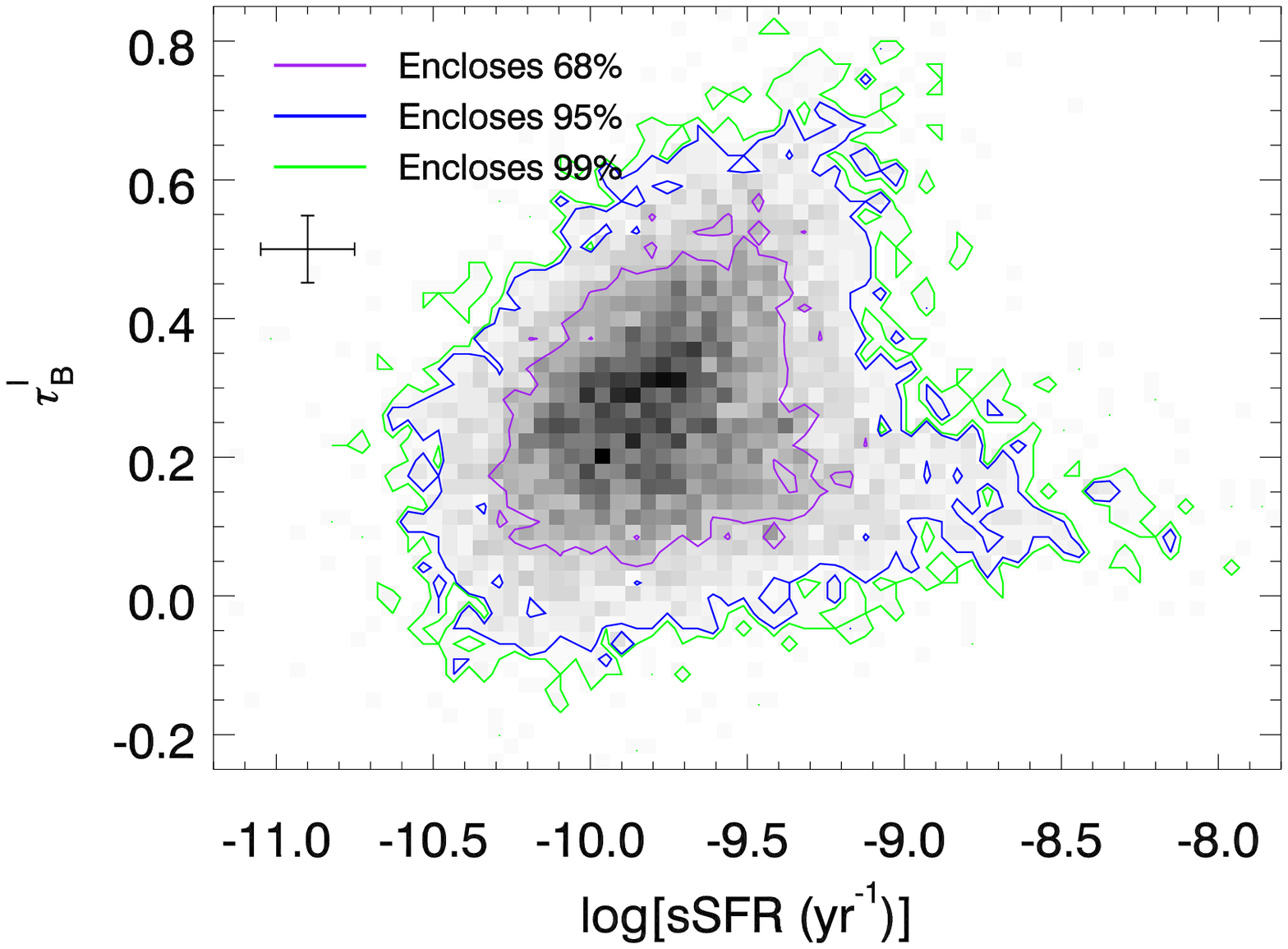} \\
\end{array}$
\end{center}
\caption{The UV power-law index, $\beta_{\rm{GLX}}$, and the Balmer optical depth, $\tau_B^l$, as a function of the $4000$~\AA\ break, ($D_n4000$; top) and also the sSFR ($\mathrm{sSFR}=\mathrm{SFR}/M_*$~yr$^{-1}$; bottom). A representative error bar of the median measurement uncertainties is shown in the top left of each panel. Slight trends appear to suggest that larger values of $\beta_{\rm{GLX}}$ or $\tau_B^l$ occur in galaxies with larger $D_n4000$, which roughly corresponds to older ages for the stellar population. A window of $1.1<D_n4000<1.3$ (dashed red lines) is used to achieve a more uniform mean stellar age as a function of $\tau_B^l$ when deriving the average attenuation curve (see \S~\ref{attenuation curve}). No clear trends are apparent with respect to sSFR. \label{fig:beta_tau_age_test}}
\end{figure*}

Before proceeding with the methodology described above, it is important to determine whether or not there are any systematic trends between our attenuation parameters and the stellar population age in order to ensure that the average age of each template for different bins in $\tau_B^l$ are consistent. One way to test this is by examining the value of the $4000$~\AA\ break and the sSFR, both of which are sensitive to the age of the stellar population, as a function of $\beta_{\rm{GLX}}$ and $\tau_B^l$ in our galaxy sample. We utilize the measurement of $D_n4000$ for the $4000$~\AA\ break and the galaxy SFR and stellar mass ($M_*$) within the spectroscopic fiber from the MPA/JHU catalog. The comparison between these two parameters and $\beta_{\rm{GLX}}$ and $\tau_B^l$ is shown in Figure~\ref{fig:beta_tau_age_test}. It can be seen that there are slight changes in the range of $\beta_{\rm{GLX}}$ or $\tau_B^l$ values spanned at a given $D_n4000$ value, suggesting that larger values of $\tau_B^l$ might correspond on average to galaxies with a slightly larger $D_n4000$ values (older stellar population). No obvious trends appear evident when comparing to sSFR. If the trend with $D_n4000$ is indeed significant, then we should see a noticeable difference among the inferred attenuation curves for the different bins of $\tau_B^l$ as a result of using the lower $\tau_B^l$ bins for comparison.

As a first attempt for deriving an attenuation curve, we work under the assumption that the average flux density of the spectra within all bins of $\tau_B^l$ create template spectra which represent galaxies with the same average stellar population age. The adopted bins are outlined in Table~\ref{Tab:tau_bin}. In order to reduce the large spread in flux density values within each bin, which result from the range of distances covered by our sample, we normalize the flux density to the rest-frame value at $\lambda=5500$~\AA. Since we are interested in understanding the attenuation of the stellar continuum, we make use of the available emission line subtracted optical spectra from the SDSS database for each galaxy. Each optical spectrum is smoothed by 50 channels ($\sim40-100$~\AA) to improve the $S/N$. The flux density in the UV region of $1250<\lambda<2600$~\AA\ is determined solely based on the two bands covered by \textit{GALEX} under the assumption that the entire region follows the $\beta_{\rm{GLX}}$ power-law behavior determined from the FUV and NUV, and we acknowledge this as a limitation of this study. Such an assumption would not distinguish possible features within the attenuation curve, such as a $2175$~\AA\ bump. Although if a feature is present, it would impact the values of $\beta_{\rm{GLX}}$ and we will discuss this in more detail in \S~\ref{influence_of_2175}. We also note that $\beta_{\rm{GLX}}$ is slightly redder than the actual UV spectrum (see \S~\ref{beta_GLX_vs_z}), but because we are using the ratio of flux densities for our analysis, the outcome for the attenuation curve is the same regardless of this effect.

For constructing our templates we choose not adopt a weighted-mean, as is done in \citet{calzetti94}, because we find that the reddest galaxies (largest $\beta_{\rm{GLX}}$) within each bin tend to be brighter and have lower uncertainties, which biases the average UV slope of the templates to larger values of $\beta_{\rm{GLX}}$ for all bins ($\Delta\beta_{\rm{GLX}}\sim0.2$). In contrast, adopting a simple average of the flux densities creates templates which have a UV slope nearly identical to the value found by taking the weighted-mean of all $\beta_{\rm{GLX}}$ within that bin ($\Delta\beta_{\rm{GLX}}\sim0.02$). For deriving the attenuation curve we are only considering the optical spectral region for which every galaxy has data. For $z<0.1$, this corresponds to $3793\le\lambda\le8325$~\AA. 

Given that the $\beta_{\rm{GLX}}$-$\tau_B^l$ relation of this sample demonstrates similarity to the results of \citet{calzetti94}, it would appear that the dust geometry is similar to that of starbursting systems (namely a foreground-like geometry). In this scenario, the optical depth is expected to follow a functional form similar to equation~(\ref{eq:intensity}) and we can use the flux density from lower bins of $\tau_B^l$ as a reference spectrum (i.e., representing lower attenuation cases).

Given a reference spectrum, $F_r(\lambda)$, we can determine 
\begin{equation}
\tau_{n,r}(\lambda) = -\ln \frac{F_n(\lambda)}{F_r(\lambda)} \,,
\end{equation}
where $\tau_{n,r}$ corresponds to the dust optical depth of template $n$ with flux density $F_n(\lambda)$, and it is required that $n>r$ for comparison. From this it is possible to determine the selective attenuation, $Q_{n,r}(\lambda)$, 
\begin{equation}\label{eq:Q_def}
Q_{n,r}(\lambda) = \frac{\tau_{n,r}(\lambda)}{\delta \tau_{Bn,r}^l} \,,
\end{equation}
where $\delta \tau_{Bn,r}^l=\tau_{Bn}^l-\tau_{Br}^l$ is the difference between the Balmer optical depth of template $n$ and $r$. We stress that the quantity $Q_{n,r}(\lambda)$ reflects the \textit{selective} attenuation, which is a difference in attenuation between two wavelengths, and not a total attenuation, and because of this the zero-point of $Q_{n,r}(\lambda)$ is arbitrary. Following \citet{calzetti94}, we select $Q_{n,r}(5500\mathrm{\AA})=0$ as the zero-point.  

To determine the influence of variation in stellar population age for our galaxies we compared the entire sample of galaxies spanning $0.8\lesssim D_n4000 \lesssim1.6$ to subsamples spanning smaller ranges in $D_n4000$. We find that using the entire sample clearly affects the inferred attenuation curve, $Q_{n,r}(\lambda)$, giving rise to artificially higher attenuation in the region of $\lambda<5500$~\AA\ as a result of the slightly older stellar population ages for galaxies with increasing $\tau_B^l$ (i.e., the intrinsic spectrum of these systems is redder relative to reference bins). In light of this, we adopt a window in average stellar population age of $1.1<D_n4000<1.3$ to derive our attenuation curve. This subsample still encompasses the majority of the sample (7265 galaxies), but works to restrict the majority of the observed age effects. 

\begin{figure}
\plotone{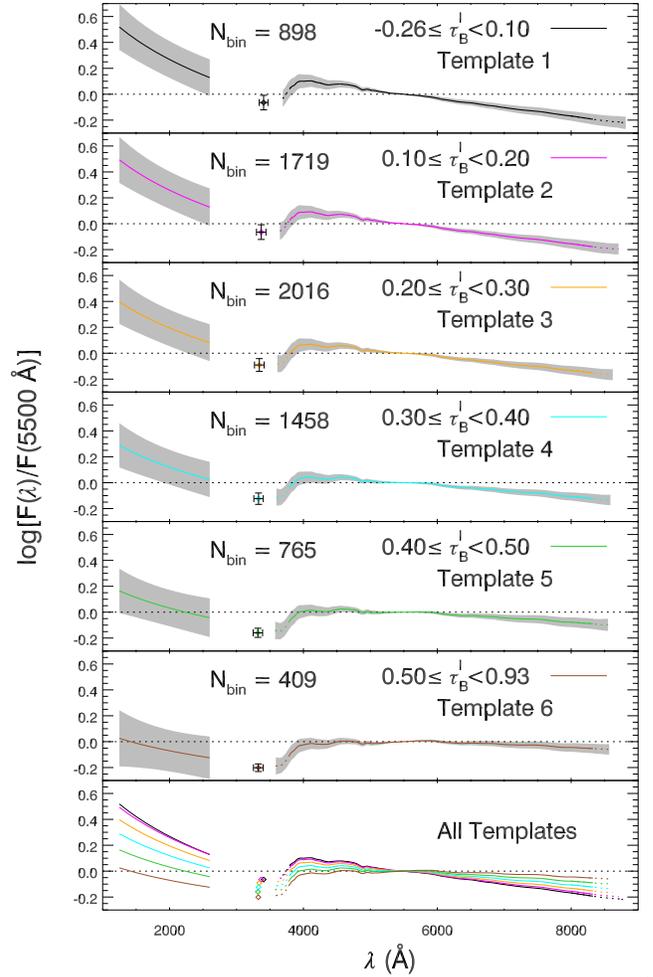}
\caption{Average flux density of galaxies, normalized at 5500\AA, within each bin of $\tau_B^l$ for the subsample of galaxies with $1.1<D_n4000<1.3$ ($N_{\mathrm{tot}}=7265$). The range in $\tau_B^l$ and the number of sources in each bin, $N_{\mathrm{bin}}$, are shown in each panel. The \textit{GALEX} FUV and NUV flux densities for each galaxy are used to determine the flux density over the region of $1250<\lambda<2600$~\AA\ by assuming it follows $\beta_{\rm{GLX}}$. The optical measurements are from SDSS spectroscopy. The gray regions denote the area enclosing approximately 68\% of the population. The dotted regions in the optical spectra indicate the average obtained from less than the full sample in that bin (due to varying redshifts), but still containing $>50\%$ of the bin sample. The symbols show the average $u$-band flux density. It can be seen that the UV and optical flux densities seem to agree within the scatter, indicating that the aperture corrections made are reasonable. For reference, the bottom panel shows a comparison of the average flux density of each bin without the dispersion included. \label{fig:F_vs_lam}}
\end{figure}

\begin{figure}
\plotone{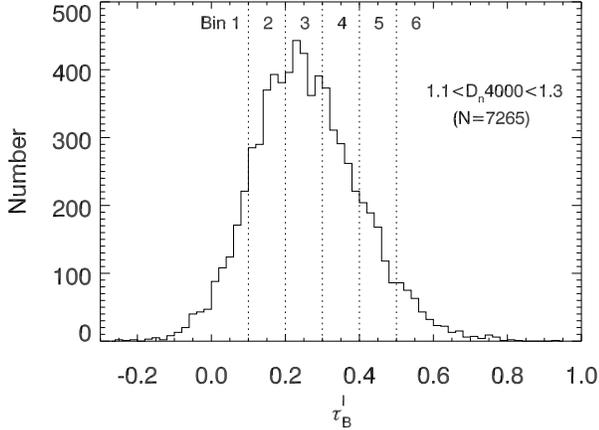}
\caption{Histogram of $\tau_B^l$ values for the sample of galaxies with $1.1<D_n4000<1.3$. Vertical dotted lines denote the boundaries for the bins adopted to construct the flux templates. The distribution of sources is peaked at $\tau_B^l\sim0.25$, resulting in the bins not being uniformly populated. This has a significant impact on the average template constructed for bin 1, which appears similar to template 2 as a result of the majority of sources in bin 1 lying at $\tau_B^l\sim0.1$. For this reason, we do not use bin 1 in deriving the selective attenuation curve. \label{fig:tau_hist_Dn4000}}
\end{figure}

The templates of the average flux density using these galaxies divided into the same 6 bins of $\tau_B^l$ shown in Table~\ref{Tab:tau_bin}, but only using the $1.1<D_n4000<1.3$ sample, can be seen in Figure \ref{fig:F_vs_lam}. It can be seen that the amplitudes of the UV and optical flux density appear to be in good agreement, which is more evident when we include the average flux density in wavelength regions that contain $>50\%$ of the bin sample (dotted lines in Figure \ref{fig:F_vs_lam}) and the average $u$-band flux density. The x-axis error bar of the average $u$-band flux density denotes the 1$\sigma$ range in rest-wavelength spanned in each bin. This suggests that on average, our aperture corrections for the UV and optical flux density appear to provide values consistent with each other. The similarity of template 1 and 2 is likely a result of the distribution of galaxies in bin 1 having a majority of cases towards $\tau_B^l\sim0.1$, which we show in Figure~\ref{fig:tau_hist_Dn4000} (also evident in the right panels of Figure~\ref{fig:beta_tau_age_test}), and this is skewing the average SED of the template to have similar attenuation to template 2.

\begin{figure}
\plotone{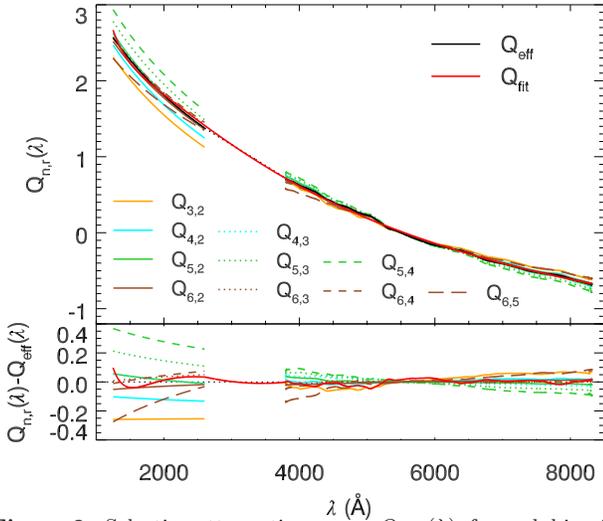}
\caption{Selective attenuation curve, $Q_{n,r}(\lambda)$, for each bin of $\tau_B^l$ (denoted by $n$) determined from comparing to a reference template (denoted by $r$) at lower $\tau_B^l$. Also shown is the effective curve, $Q_{\mathrm{eff}}(\lambda)$ (solid black line), which is the average value of $Q_{n,r}(\lambda)$ for all cases, but excludes use of template 1 due to it appearing nearly identical to template 2 (see \S~\ref{attenuation curve}). The gap region between 2600~\AA\ and 3800~\AA\ is denoted with a dotted line corresponding to a linear interpolation between the end points and is not used for constraining the fit. The solid red line is a single polynomial fit to $Q_{\mathrm{eff}}(\lambda)$. The lower panel shows the difference between each curve relative to $Q_{\mathrm{eff}}(\lambda)$. \label{fig:Q_eff}}
\end{figure}

We show the selective attenuation curve for each bin of Balmer optical depth for different reference templates in Figure~\ref{fig:Q_eff}. We exclude the use of template 1 in our analysis because its SED appears so similar to template 2 (but with a lower average $\tau_B^l$), a result which leads to significantly lower selective attenuation curves when it is used as a reference compared to those found using the other templates. It can be seen in Figure~\ref{fig:Q_eff} that templates 2-6 all give very similar selective attenuation curves. This implies that adopting a \textit{single} selective attenuation curve is appropriate to characterize the entire range of Balmer optical depths spanned by the majority of galaxies in this sample. We determine the effective attenuation curve, $Q_{\mathrm{eff}}(\lambda)$, by taking taking the average value of $Q_{n,r}(\lambda)$ found from templates 2-6. We have fit the value of $Q_{\mathrm{eff}}(\lambda)$ to a single third-order polynomial as a function of $x=1/\lambda$~($\micron^{-1}$):
\begin{multline}\label{eq:Q_fit}
Q_{\mathrm{fit}}(x) = -2.488 + 1.803x - 0.261x^2 + 0.0145x^3\,,\\ 0.125\micron\le\lambda<0.832\micron \,.
\end{multline}

\begin{figure}
\plotone{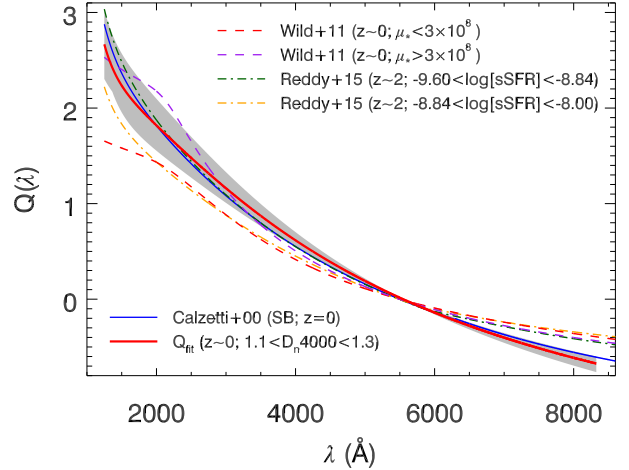}
\caption{Comparison of our selective attenuation curve (red solid line) compared to others in the literature. The gray region denotes the range of $Q_{n,r}(\lambda)$ values (i.e., region spanned by lines in Figure~\ref{fig:Q_eff}). The solid blue line is the starburst selective attenuation curve of \citet{calzetti00}, the dashed lines are the curves of local SDSS galaxies from \citet{wild11} divided according to stellar mass surface density, $\mu_*$, and the dash-dot lines are the curves of $z\sim2$ SFGs from \citet{reddy15} divided according to sSFR. Our selective attenuation curve appears most similar to that found by \citet{calzetti00}. \label{fig:Q_compare}}
\end{figure}

We compare our selective attenuation curve to other curves in the literature in Figure~\ref{fig:Q_compare}. To give a sense for the uncertainty, a gray region denoting the range of $Q_{n,r}(\lambda)$ is shown (i.e., region spanned by all lines shown in Figure~\ref{fig:Q_eff}). We include the curves of local starburst galaxies from \citet{calzetti00}, local SDSS galaxies from \citet{wild11}, and higher redshift ($z\sim2$) SFGs from \citet{reddy15}. The selective attenuation curves of \citet{wild11} are divided according to stellar mass surface density, $\mu_*$, with the break corresponding to the value which separates the bimodal local galaxy population into bulge-less ($\mu_*<3\times10^8~M_\odot~\mathrm{kpc}^{-2}$) and bulged ($\mu_*>3\times10^8~M_\odot~\mathrm{kpc}^{-2}$) galaxies \citep{kauffmann03b}, as well as sub-divided by the sSFR and the axial ratio ($b/a$). For clarity, we only reproduce the curves corresponding to $\log[\mathrm{sSFR~(yr}^{-1})]=-9.5$ and $b/a=0.6$ from that work.  The selective attenuation curves of \citet{reddy15} are divided according to sSFR. Our selective attenuation curve appears to be most similar to the \citet{calzetti00} curve. We find that the derived selective attenuation curve does not change much depending on the range of $D_n4000$ used, so long as it remains relatively narrow ($\Delta D_n4000\lesssim0.2$; see \S~\ref{curve_vs_param}).

The selective attenuation can be related to the total-to-select extinction, $k(\lambda)$, through the following relation
\begin{equation}\label{eq:k_def}
k(\lambda)= fQ(\lambda)+R_V \,,
\end{equation}
where $f$ acts to change the tilt of the curve and is necessary to make $k(B)-k(V)\equiv1$,
\begin{equation}\label{eq:f_def}
f= \frac{1}{Q_{\mathrm{eff}}(B)-Q_{\mathrm{eff}}(V)} \,,
\end{equation}
and where $R_V$ is the total-to-select extinction, which is the vertical offset from 5500~\AA. We assume $B$ and $V$ bands to be 4400~\AA\ and 5500~\AA, respectively. The term $f$ is necessary to account for differences in the reddening between the ionized gas, which is assumed to suffer from extiction, and the stellar continuum. Therefore, the quantity $fQ(\lambda)$ represents the true wavelength-dependent behavior of the attenuation curve on the stellar continuum, but does not represent a total attenuation curve without knowledge of the normalization (given by $R_V$). Determination of $R_V$ requires knowledge of either the NIR photometry (where the attenuation should approach zero; $k(\lambda\rightarrow\infty)=0$) or the total infrared data (to determine the total dust attenuation).   

The value of $f$ can be quantitatively expressed in terms of the differential reddening between the ionized gas and the stellar continuum by rewriting each term on the right side of equation~(\ref{eq:Q_def}). For the case where the reference source has $\tau_B^l=0$, we get for the numerator
\begin{equation}
\tau_n(\lambda) = -\ln\frac{F_n(\lambda)}{F_0(\lambda)}=0.921A_\lambda=0.921E(B-V)_{\mathrm{star}}k(\lambda) \,,
\end{equation}
where we have used the definition of total-to-selective extinction $k(\lambda)\equiv A_\lambda/E(B-V)_{\mathrm{star}}$, and for the denominator 
\begin{equation}
\delta \tau_{B}^l= \tau_{B}^l-\tau_{B0}^l=\frac{k(\mathrm{H}\beta)-k(\mathrm{H}\alpha)}{1.086}E(B-V)_{\mathrm{gas}} \,,
\end{equation}
where $k(H\beta)$ and $k(H\alpha)$ are the values for the intrinsic extinction curve of the galaxy and \textit{not} from the attenuation curve. Therefore, we can rewrite the equation as
\begin{equation}
Q(\lambda) = \frac{k(\lambda)-R_V}{k(H\beta)-k(H\alpha)}\frac{E(B-V)_{\mathrm{star}}}{E(B-V)_{\mathrm{gas}}} \,,
\end{equation}
where we have explicity added the term $R_V$ which corresponds to the zero-point normalization that was applied at 5500~\AA. Comparing this to equation~(\ref{eq:k_def}) implies that the term $f$ is equivalent to 
\begin{equation}
f = \frac{k(H\beta)-k(H\alpha)}{E(B-V)_{\mathrm{star}}/E(B-V)_{\mathrm{gas}}} \,.
\end{equation}
For extinction curves, such as the Milky Way, it is the case that $E(B-V)_{\mathrm{star}}=E(B-V)_{\mathrm{gas}}$, and $f$ is simply the difference in extinction between the Balmer emission lines. However, the same is not true of attenuation curves for which it is typically seen that $E(B-V)_{\mathrm{star}}<E(B-V)_{\mathrm{gas}}$ \citep[e.g.,][]{calzetti00,kreckel13,reddy15}. 

\begin{figure}
\plotone{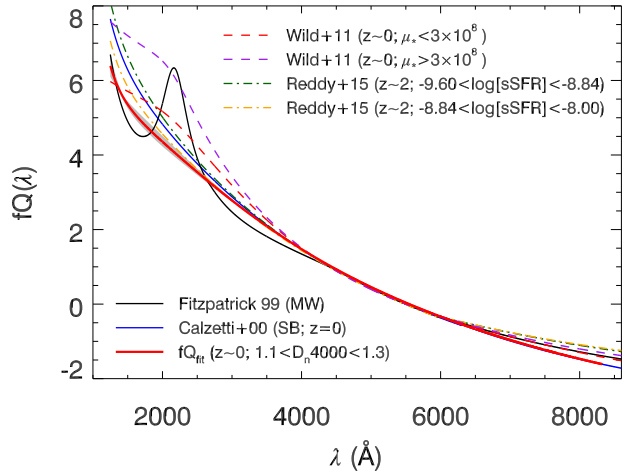}
\caption{Normalized selective attenuation curve $fQ(\lambda)$ derived from our sample of SFGs compared to values in the literature ($fQ(\lambda)=k(\lambda)-R_V$). The term $f$ is required to make the curve have $k(B)-k(V)\equiv1$. Lines are the same as in Figure~\ref{fig:Q_compare} but with the addition of the MW curve in solid black \citep{fitzpatrick99}. The gray region denoting the range of $fQ_{n,r}(\lambda)$ values (where $f$ varies in each case) is significantly reduced after this normalization. We find a slightly lower selective attenuation in the UV compared to previously determined attenuation curves, with a near-IR appearing similar to \citet{calzetti00}. \label{fig:fQ_compare}}
\end{figure}

The value of $f$ for our average selective attenuation curve, $Q_{\mathrm{fit}}(\lambda)$, is determined using equation~(\ref{eq:Q_fit}) to be 2.396$\substack{+0.33 \\ -0.29}$, where the uncertainty here reflects the maximum and minimum values from fits using individual $Q_{n,r}(\lambda)$. We compare our selective attenuation curve with this normalization term included to the attenuation curves mentioned before with their own $f$ values, in addition to the MW extinction curve in Figure~\ref{fig:fQ_compare}. It can be seen that the gray region denoting the range of $fQ_{n,r}(\lambda)$ values, where $f$ varies for each individual case, is significantly reduced after this normalization. For reference, the values of $f$ in other works are $f=2.659$ in \citet{calzetti00}, $f=3.609$ and 2.996 for the lower and higher $\mu_*$ sample, respectively, in \citet{wild11}, and $f=2.676$ and 3.178 for the lower and higher sSFR subsample, respectively, in \citet{reddy15}. In reference to the attenuation curve from \citet{calzetti00}, we find that the attenuation in SFGs is slightly lower in the UV by up to 20\% at 1250\AA. Out towards the near-IR, our curve appears similar to the starburst curve from \citet{calzetti00}. If this similarity were to hold out to longer wavelengths then we could expect the normalization term $R_V$ to be similar to the starburst curve value of $4.05$, since $k(\lambda\rightarrow\infty)=0$. An exact determination of the value of $R_V$ for this curve will be the subject of a future study. 

If we assume the underlying extinction curve for these galaxies to be the \citet{fitzpatrick99} MW extinction curve (for the values of $k(H\beta)$ and $k(H\alpha)$), then we find that $\langle E(B-V)_{\mathrm{star}}\rangle=0.52\langle E(B-V)_{\mathrm{gas}}\rangle$ for the average of the SFGs in our sample. It is worth noting that assuming different extinction curves for the ionized gas will result in subtle changes to this ratio (e.g., for a \citet{cardelli89} MW extinction curve, $\langle E(B-V)_{\mathrm{star}}\rangle=0.45\langle E(B-V)_{\mathrm{gas}}\rangle$) and that the value of $k(\lambda)$ is likely to be dependent on metallicity. This is important to consider when comparing values in the literature. This ratio is in agreement with previous results suggesting that the stellar continuum suffers roughly one-half of the reddening of the ionized gas \citep{calzetti00,kreckel13}.

\section{Sources of Concern in Adopting \texorpdfstring{$\beta_{\mathrm{GLX}}$}{beta_GLX} for the UV slope}

\subsection{Comparing \texorpdfstring{$\beta_{\mathrm{GLX}}$}{beta_GLX} to \texorpdfstring{$\beta$}{beta}} \label{beta_GLX_vs_z}
In order to compare the $\beta_{\rm{GLX}}$-$\tau_B^l$ relationship found in \S~\ref{beta_vs_tau} to similar studies in the literature it will be necessary to understand how $\beta_{\rm{GLX}}$ relates to the true UV slope $\beta$ estimated only from the UV continuum. The \textit{GALEX} filters have relatively wide passbands, which makes them susceptible to numerous stellar absorption features that appear in the UV. The influence of these features is redshift-dependent because various absorption features will pass in and out of each filter as each passband shifts. In this section we seek to address whether the differences in redshift for the galaxies in our sample is affecting the observed $\beta_{\rm{GLX}}$-$\tau_B^l$ relationship seen in \S~\ref{beta_vs_tau}. This will also allow us to transform $\beta_{\rm{GLX}}$ to $\beta$ and then make comparisons to previous studies.

Typically the conversion factor between $\beta_{\rm{GLX}}$ to $\beta$ is found using sample of galaxies for which both UV spectra and \textit{GALEX} observations have been obtained \citep[e.g.,][]{kong04,takeuchi12}. These results suggest that $\beta_{\rm{GLX}}$ is typically larger (i.e., redder) than $\beta$ by $\sim0.05-0.1$. However, because UV spectral data is not available for this sample, we utilize a Starburst99 \citep{leitherer99} spectrum of a continuously star forming galaxy with solar metallicity ($Z_\odot=0.02$) as a reference for an intrinsic galaxy spectrum. As will be shown, the exact age of the reference spectrum is not particularly important so long as the assumption of a continuous SFR is reasonable, as the shape of the UV slope remains relatively unaffected over a wide range of ages \citep[e.g., see Figure 2 of][]{leitherer99}. Working from this reference spectrum we can apply an attenuation curve to vary the shape of the UV slope and then shift the spectra to various redshifts. We will utilize the attenuation curve that we derived in \S~\ref{attenuation curve}. The lack of the curve normalization is not important because we are only examining the difference between $\beta_{\rm{GLX}}$ and $\beta$, and not the absolute values of the UV flux density (i.e., the shape of the UV SED after attenuation will remain the same regardless of this normalization). Using a color excess over the range $0.0<E(B-V)_{\mathrm{star}}<0.9$ reproduces the full range of $\beta_{\rm{GLX}}$ values seen in our sample. For each reddened and redshifted spectrum we can then determine what the corresponding value of $\beta_{\rm{GLX}}$ is relative to $\beta$. The value of $\beta$ is determined using the 10 rest-frame wavelength windows used by \citet{calzetti94} to measure the UV slope of starburst galaxies from the spectra observed by the \textit{International Ultraviolet Explorer (IUE)} ($\beta_{\rm{IUE}}$). These UV windows were designed to avoid strong stellar absorption features, including the 2175~\AA\ feature, and therefore represent an accurate measure of the UV slope. As the windows are taken in rest-frame wavelength, this measurement is not affected by redshift. 
 
\begin{figure}
\plotone{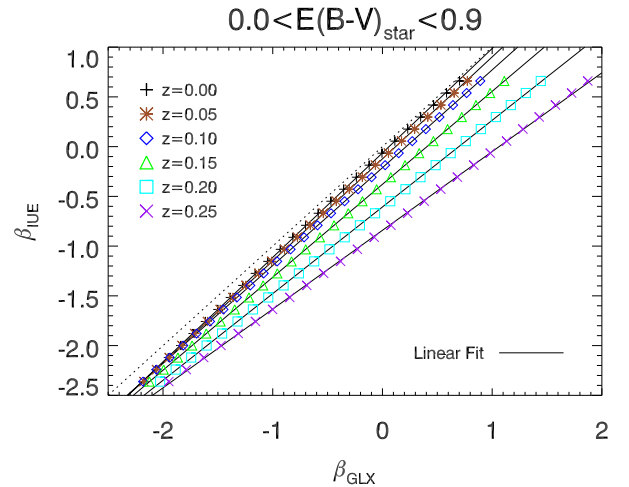}
\caption{The relationship between UV slope determined using $\beta_{\rm{GLX}}$ and $\beta_{\rm{IUE}}$ for several different redshifts when adopting a SB99 model of a 100~Myr galaxy with continuous star formation. Each symbol is a step in $E(B-V)_{\mathrm{star}}$ of 0.036 starting at 0 in the lower-left and increasing up to 0.9 in the upper-right, where we are assuming the attenuation curve derived in \S~\ref{attenuation curve}. For each redshift the behavior follows a simple linear relation, demonstrated by the linear least-squares fits (black lines). The dotted line shows the 1:1 relation. \label{fig:beta_compare}}
\end{figure}

The relationship between the two UV slope diagnostics for several different redshifts is shown in Figure~\ref{fig:beta_compare}. For each redshift, the steps in coverage correspond to changes $E(B-V)_{\mathrm{star}}$ of 0.036 starting at 0.0 in the lower-left and increasing up to 0.9 in the upper-right. It can be seen that for each redshift the relationship follows a simple linear relation, $\beta_{\rm{IUE}}=m\beta_{\rm{GLX}}+b$, but with varying values of slope and offset. As expected, the value of $\beta_{\rm{GLX}}$ becomes more discrepant with $\beta_{\rm{IUE}}$ as we move to higher redshifts where more absorption features below $1250$~\AA\ begin to come into the FUV filter passband. It is for this reason that we imposed a redshift cut in our initial sample selection of $z<0.1$, in order to limit significant deviations. It can be seen in Figure~\ref{fig:beta_compare} that the differences between the two estimators for $z<0.1$ is less drastic than compared to higher $z$.  

\begin{figure*}
\begin{center}$
\begin{array}{lr}
\includegraphics[width=3.5in]{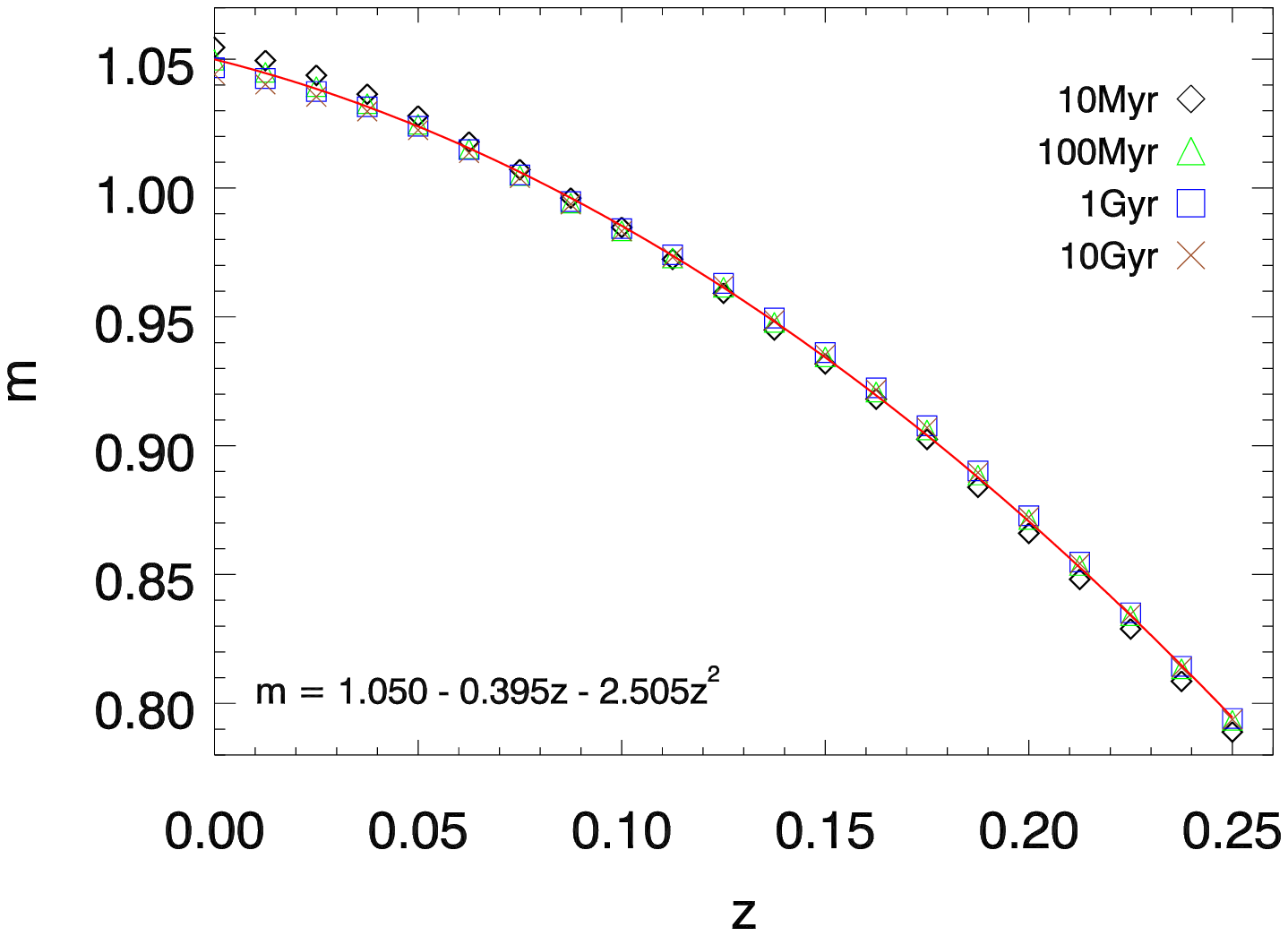}
\includegraphics[width=3.5in]{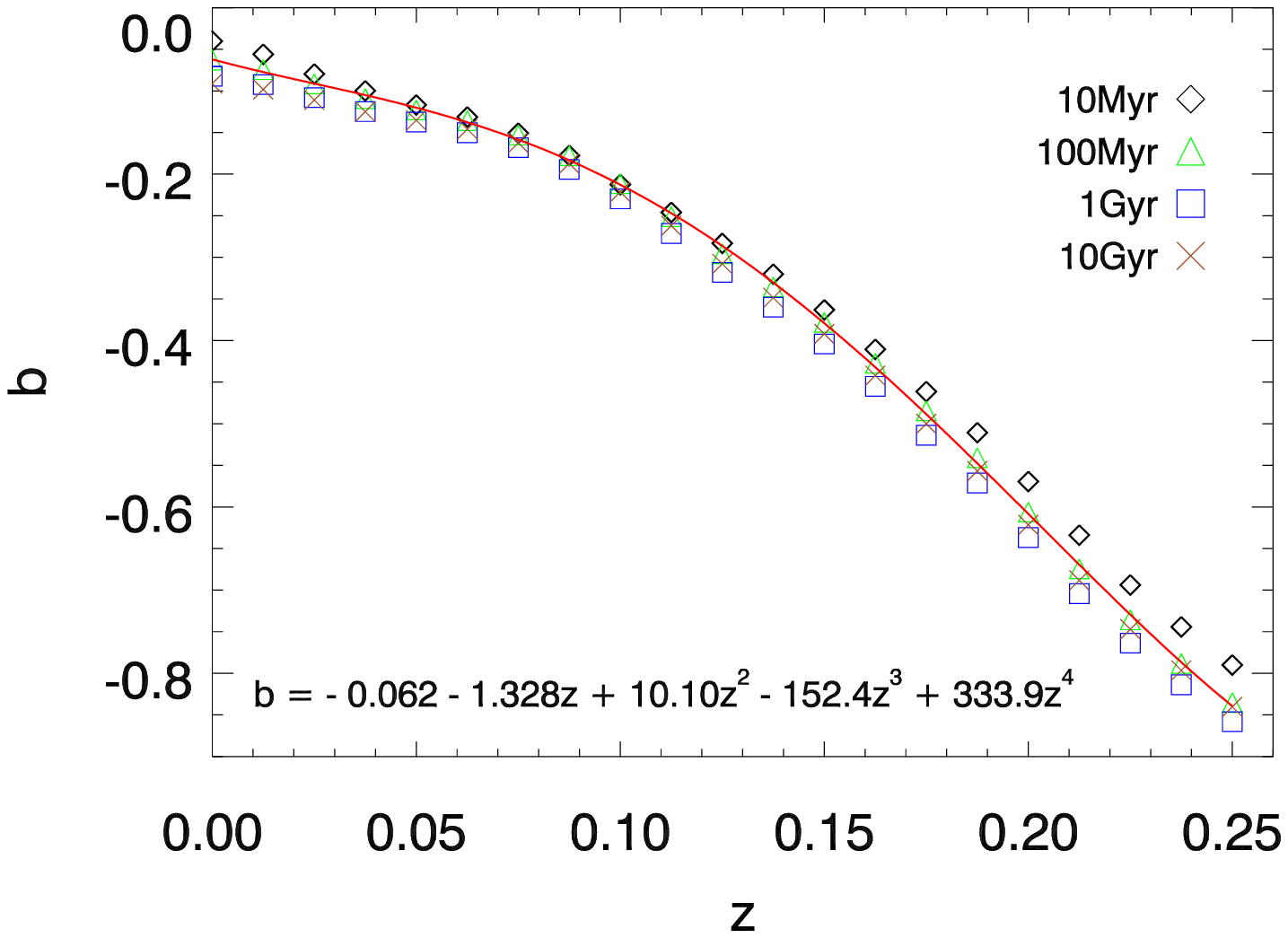} \\
\end{array}$
\end{center}
\caption{The relationship between the slope, $m$, and the offset, $b$, in the fit of $\beta_{\rm{IUE}}=m\beta_{\rm{GLX}}+b$ as a function of redshift for SB99 models with different ages of continuous star formation. There is relatively little variation in the behavior for the different age models (open symbols). Our sample was restricted to $z<0.1$, to limit the influence of absorption lines on the value of $\beta_{\rm{GLX}}$. We adopt the 100~Myr case (green triangles) for our conversion, as it is a fair representation of the average trend. The behavior of the slope and offset are well approximated by a second-order and fourth-order polynomials (red lines), respectively. The parameters of the fits to the 100~Myr values are shown. \label{fig:slope_offset}}
\end{figure*}

To determine the corrections, we take a linear least-squares fit to the relation between $\beta_{\rm{GLX}}$ and $\beta_{\rm{IUE}}$ for various redshifts. The behavior of the slope, $m$, and offset, $b$, as a function of redshift is shown in Figure~\ref{fig:slope_offset}. We show the value of these parameters for several models with different ages of continuous star formation. As expected, the variation among the parameters of the fit are relatively small between the different models. The 100~Myr case seems to be a fair representation of the average trend and so we will adopt it for determining corrections.

The value of $m$ behaves in a manner that can be well approximated by a second-order polynomial. Fitting the 100~Myr case to this functional form gives 
\begin{equation}
m(z) = 1.050-0.395z-2.505z^2 \,. 
\end{equation}
The value of $b$ behaves in a slightly more complicated manner and is better fit using a fourth-order polynomial. Fitting the 100~Myr case to this functional form gives 
\begin{equation}
b(z) = -0.062-1.328z+10.10z^2-152.4z^3+333.9z^4 \,.
\end{equation}
By adopting these fits to the relationship between $\beta_{\rm{GLX}}$ and $\beta_{\rm{IUE}}$, we can reproduce the observed trends in Figure~\ref{fig:beta_compare} very well. These conversion parameters are similar to those found in other studies using more robust techniques \citep[e.g.,][]{kong04,takeuchi12}.

We examined if the assumption of the attenuation curve is important in the conversion from $\beta_{\rm{GLX}}$ to $\beta$ by by also using the \citep{calzetti00} attenuation curve. We find that differences in the conversion are not significant over $0<z<0.1$ ($|\Delta m|\lesssim0.02$, $|\Delta b|\lesssim0.05$), and for this reason we do not expect this choice to have a significant impact on the results.

\begin{figure}
\plotone{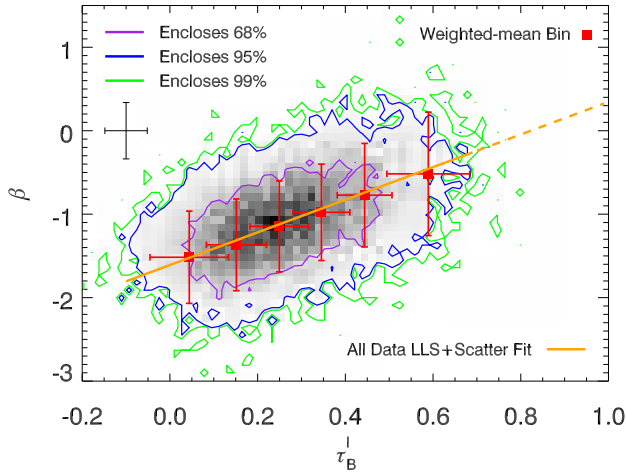}
\caption{The UV power-law index after correcting for stellar absorption features in the \textit{GALEX} passbands, $\beta$ (see \S~\ref{beta_GLX_vs_z} for conversion from $\beta_{\rm{GLX}}$), as a function of the Balmer optical depth, $\tau_B^l$, for our sample of SFGs. Symbols and lines have the same meaning as in Figure~\ref{fig:beta_tau}, however the values have changed. \label{fig:beta_corr_tau}}
\end{figure}

The relationship between the UV slope, $\beta$, as a function of $\tau_B^l$ is shown in Figure~\ref{fig:beta_corr_tau}, where we have used the average relation between $\beta_{\rm{GLX}}$ and $\beta_{\rm{IUE}}$. The Spearman and Kendall nonparametric correlation tests give $\rho_S=0.47$ and $\tau_K=0.32$, respectively, for this relation. Taking a linear least-squares fit gives
\begin{equation}
\beta = (1.95\pm0.03) \tau_B^l -(1.61\pm0.01) \,,
\end{equation}
with an intrinsic dispersion of $\sigma_{\mathrm{int}}=0.44$. Adopting the same six ranges in $\tau_B^l$ for the bins as before gives a similar trend, with the values also given in Table~\ref{Tab:tau_bin}. The behavior of the relationship is similar to before with nearly the same slope but with a lower offset by -0.15.

\begin{figure}
\plotone{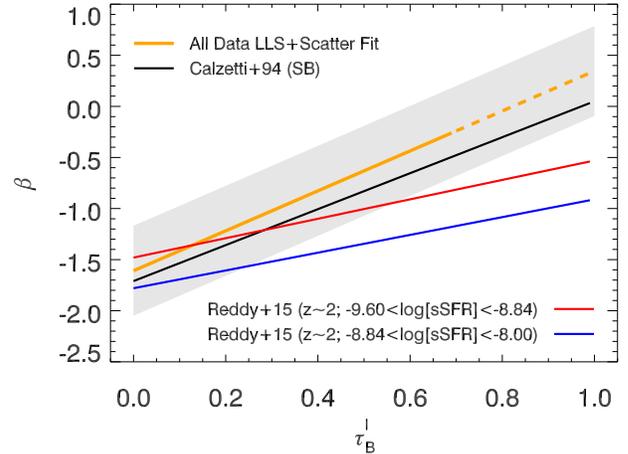}
\caption{Comparison of the UV power-law index, $\beta$,  as a function of the Balmer optical depth, $\tau_B^l$, for our sample of SFGs (orange line) to those in the literature. For each study $\beta$ is determined in a slightly different way, but in all cases it represents the UV slope inferred from the stellar continuum (i.e., avoiding regions with stellar absorption features, see \S~\ref{beta_GLX_vs_z} for details). We show the \citet{calzetti94} sample of starburst galaxies (black line) and the SFGs at $z\sim2$ from \citet{reddy15}, which are separated according to sSFR ($\mathrm{sSFR}=\mathrm{SFR}/M_*$~yr$^{-1}$; red and blue lines). Our fit at $\tau_B^l>0.7$ is shown with a dashed line to denote that there are limited data in this range. The intrinsic dispersion of our data is denoted by the gray filled region ($\sigma_{\mathrm{int}}=0.44$). The dispersion of the other samples are not shown for clarity. \label{fig:beta_tau_compare}}
\end{figure}

Now that the data has been expressed in terms of $\beta$, our results can be directly compared to previous studies. We compare our $\beta-\tau_B^l$ relationship to the sample of local starburst (SB) galaxies from \citet{calzetti94} and also the sample of $z\sim2$ SFGs from \citet{reddy15} in Figure~\ref{fig:beta_tau_compare}. The relationship from \citet{calzetti94} is $\beta_{\rm{IUE}} = (1.76\pm 0.25) \tau_B^l -(1.71\pm 0.12)$, and has a dispersion of $\sigma_{\mathrm{int}}\sim0.4$. For the sample of $z\sim2$ SFGs in \citet{reddy15}, the authors found significant variation in the $\beta-\tau_B^l$ relationship with sSFR ($\mathrm{sSFR}=\mathrm{SFR}/M_*$~yr$^{-1}$). As a result of this they separate their sample into two bins, $\beta_{\rm{SED}} = (0.95\pm 0.14) \tau_B^l -(1.48\pm 0.02)$ for $-9.60\le\log[\mathrm{sSFR}]<-8.84$ ($\sigma_{\mathrm{int}}=0.31$), and $\beta_{\rm{SED}} = (0.87\pm 0.09) \tau_B^l -(1.78\pm 0.03)$ for $-8.84\le\log[\mathrm{sSFR}]<-8.00$ ($\sigma_{\mathrm{int}}=0.20$), where the SED subscript denotes that this measurement is based on using the 10 UV windows of \citet{calzetti94} on best-fit stellar population models of the photometric data.   

Given the similarities with our attenuation curve with \citet{calzetti94}, it is not so surprising that Figure~\ref{fig:beta_tau_compare} shows that the $\beta-\tau_B^l$ relationship of our sample of SFGs is similar to their starburst sample. The overall vertical offset could be linked to the intrinsic spectrum of the starburst galaxies being a little bluer. In contrast, we see a significantly different relation from the $z\sim2$ SFGs of \citet{reddy15}. Such differences likely reflect variations in the geometry of the dust relative to the ionized regions with redshift, but it could also result from changes in the properties of dust grains (e.g., chemical composition, absorption/scattering cross-sections, size distribution) in galaxies with redshift. Although, the range in sSFR values spanned by the sample in \citet[][$-9.6<\log(\mathrm{sSFR})<-8.0$]{reddy15} is different from that of our sample ($-10.5\lesssim\log(\mathrm{sSFR})\lesssim-8.9$). We explore the role that sSFR has in our local $\beta-\tau_B^l$ relation in more detail in \S~\ref{atten_vs_param}.

An interesting feature of the $\beta-\tau_B^l$ relation is that galaxies at the lowest dust attenuation ($\tau_B^l\sim0$) have UV slopes which are still reddened ($\beta\sim-1.6$) relative to what is expected for a nearly dust-free system undergoing moderate star formation ($\beta\sim-2.2$). This can arise if the dust attenuation in star forming regions acts in a different manner than the surrounding interstellar medium \citep[e.g.,][]{charlot&fall00,calzetti01}. Given the observations, we expect a scenario in which the most active star forming regions of the galaxy can be dust obscured but not ``seen'' in the Balmer decrement. This is possible if some dust is homogeneously mixed with stars in the star forming regions, as such a geometry will result in overall attenuation which is gray \citep[e.g.,][]{calzetti13}. More specifically, the flux density of the star forming regions will be significantly reduced relative to the older population, such that it does not contribute to the observed UV slope, but which still provides an observed Balmer decrement which is similar to the intrinsic value. Such a scenario has been seen in star clusters \citep{calzetti15}, and can explain the appearance of UV reddening despite the optical diagnostics that suggest minimal attenuation from the interstellar medium.

\subsection{Disentangling the Influence of a 2175\AA\ Feature}\label{influence_of_2175}
A characteristic feature of the MW extinction curve is the dust feature at 2175~\AA\ (see Figure~\ref{fig:fQ_compare}). It is well known that the presence of scattering for an extended source by a foreground dust screen has the effect of reducing the overall optical depth, flattening the attenuation curve, and diminishing the strength of the 2175~\AA\ feature \citep{natta84,calzetti94}. In addition, it has been suggested that the strength of the UV field may affect the dust that produces this feature \citep[e.g.,][]{gordon03}. Together, these effects can explain the absence of this feature from the starburst attenuation curve \citep{calzetti01}. However, a 2175~\AA\ feature has been seen to some extent in the attenuation curve of other local \citep[e.g.,][]{conroy10,wild11} and high-redshift galaxies \citep[e.g.,][]{noll09,buat11,buat12,kriek&conroy13,scoville15}. In this section we will address the effects that such a feature would have on our observations if it persisted to some degree.

\begin{figure}
\plotone{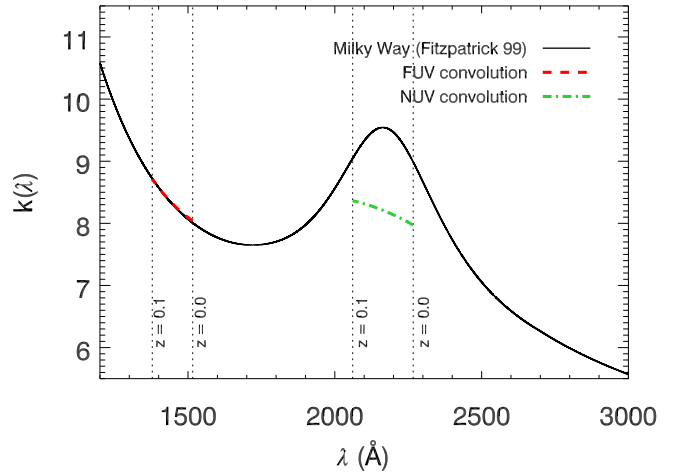}
\caption{The Milky Way extinction curve and the corresponding value inferred from a convolution of GALEX bands over the redshift range of our sample ($0<z<0.1$). The wide passband of the NUV acts to suppress the strength of a 2175~\AA\ feature, thus making it more difficult to detect the presence of a feature. \label{fig:MW_Extinct_GALEX}}
\end{figure}

In Figure~\ref{fig:MW_Extinct_GALEX}, we show the location of the FUV and NUV bands as a function of redshift relative to the MW extinction curve. It can be seen that for the redshift range of our sample ($0<z<0.1$), the 2175~\AA\ feature always lies within the \textit{GALEX} NUV filter. The limited nature of our UV coverage also implies that we cannot separate our sample into redshift bins to determine underlying differences in the shape of the UV continuum. Therefore, there is a legitimate concern whether such a feature can be influencing the derived values of the UV slope in this work. Despite our inability to directly measure the presence of the 2175~\AA\ feature, we provide several considerations below which suggest that \textit{on average} it is quite weak in our sample. 

\begin{figure}
\plotone{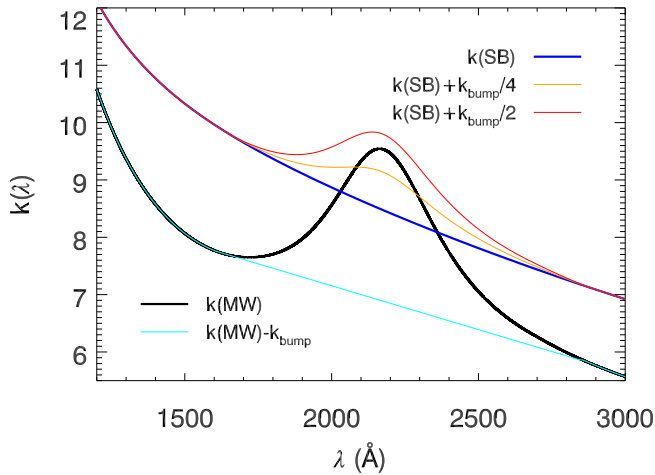}
\caption{As a test of the presence of a Milky Way-like 2175~\AA\ feature in our attenuation we can determine if adding a similar feature to the \citet{calzetti00} attenuation can reproduce the differences between that attenuation curve and the one we are finding. To achieve this we assume a linear relation underlying the 2175~\AA\ feature and subtract off the bump feature. The residual bump, $k_{\mathrm{bump}}$, can then be added to the \citet{calzetti00} at various strengths. \label{fig:k_custom}}
\end{figure}

\begin{figure}
\plotone{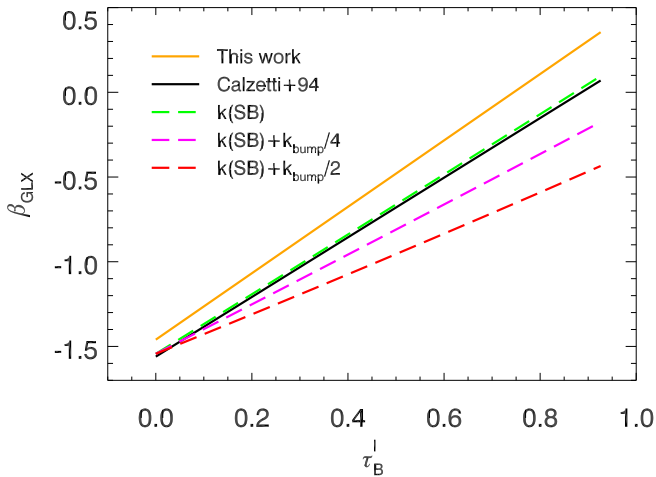}
\caption{The $\beta_{\rm{GLX}}$-$\tau_B^l$ relation found using the \citet{calzetti00} attenuation curve, $k(\mathrm{SB})$, modified with an added 2175~\AA\ feature at various strength relative to the MW (dashed lines). These curves are applied to a smooth UV flux density profile with $\beta=-1.56$ (chosen because it has an intrinsic UV slope similar to the zero-point of the observed relations). It can be seen that adding this feature acts to keep the observed UV slope bluer as a function of $\tau_B^l$, a result of the extra attenuation in the NUV band. Since such a trend is not seen in our sample relative to the \citet{calzetti94} relation, this would suggest that such a feature is not significant in a majority of these galaxies. \label{fig:beta_tau_2175_eff}}
\end{figure}

First, the presence of a 2175~\AA\ feature can only make a UV spectrum, as measured by $\beta_{\rm{GLX}}$, appear bluer. We demonstrate this effect by showing the results of adding a MW-like 2175~\AA\ feature of varying strength to the \citet{calzetti00} starburst attenuation curve on a template with a fixed UV slope. The strength of the 2175~\AA\ feature is crudely determined by subtracting off the excess extinction in the MW curve between $1600-2850$~\AA\,assuming an underling linear relation. The nature of this method, along with the modified SB attenuation curves, are shown in Figure~\ref{fig:k_custom}. For the purpose of comparing effects on $\beta_{\rm{GLX}}$, we offset the $\beta$-$\tau_B^l$ relation from \citet{calzetti94} by the average difference found in our sample due to stellar absorption lines ($\beta_{\rm{GLX}}\sim\beta+0.15$; see \S~\ref{beta_GLX_vs_z}). For simplicity, the template assumed is a smooth UV flux density profile with $\beta=-1.56$ ($F=\lambda^\beta$), which is chosen because it is similar to the zero-point of the offset of the \citet{calzetti94} relation. To determine $\beta_{\rm{GLX}}$ as a function of $\tau_B^l$ for each attenuation curve, we vary $E(B-V)_{\mathrm{gas}}$, which relates to $\tau_B^l$ through equation~\ref{eq:EBV_gas}, and assume $\langle E(B-V)_{\mathrm{star}}\rangle=0.44\langle E(B-V)_{\mathrm{gas}}\rangle$ \citep{calzetti00} to attenuate the template ($F_{\mathrm{obs}}=F_{\mathrm{int}}10^{-0.4E(B-V)_{\mathrm{star}}k(\lambda)}$). Looking at Figure~\ref{fig:beta_tau_2175_eff}, it is apparent that the $\beta_{\rm{GLX}}$-$\tau_B^l$ relation is flatter as the strength of a 2175~\AA\ feature is increased. This trend occurs because the added absorption within the NUV passband from the feature results in the observed UV slope appearing bluer. Since we do not observe a flatter relation in comparison to that of \citet{calzetti94}, this would argue against a significant feature in the attenuation curve. In addition, if the strength of the feature were to vary significantly over the sample we would expect to see a larger dispersion in the observed $\beta_{\rm{GLX}}$-$\tau_B^l$ relation at larger $\tau_B^l$, which does not appear to occur in Figure~\ref{fig:beta_tau}. However, given the large scatter of the observed $\beta_{\rm{GLX}}$-$\tau_B^l$ relation, a more subtle influence of such a feature cannot be ruled out.

Second, if an absorption feature is affecting the NUV flux density, then we would expect to see discrepancies in the appearance of the UV-optical spectra seen in Figure~\ref{fig:F_vs_lam}. Any absorption in the NUV band would lead to a bluer looking slope, which in turn would cause large offsets between the estimated flux density at 2600~\AA\ and the start of the SDSS spectra at $\sim$4000~\AA. Following the previous argument, the presence of a feature would additionally manifest itself in the derived selective attenuation curve (Figure~\ref{fig:Q_eff}), resulting in an offset between the UV and optical regions. Such a feature would cause larger values of $Q(\lambda)$ for the NUV portion, which does not appear to be the case because they are all well approximated by a single third-order polynomial over the entire UV-optical wavelength region.

Since we do not see the influence of a 2175~\AA\ feature in our average attenuation curve, we can rule out that it is playing a significant role in the average SFG. However, more detailed analysis of the UV spectrum, ideally with spectroscopy, should be pursued to determine this conclusively. Such analysis is beyond the immediate scope of this work.

\section{Discussion}
\subsection{Influence of Galaxy Properties on Dust Attenuation}
Since the dust in these galaxies is giving rise to the observed attenuation, one would naturally expect correlations to exist between parameters that are strong indicators of the presence of dust. The two ingredients needed for significant dust content in a galaxy are high metal content and high gas content \citep[e.g.,][]{calzetti99}. Previous studies have found a positive correlation between the total dust mass in galaxies with their gas-phase metallicity \citep{draine07,galametz11}, however the functional form of this relation is still under considerable debate. In addition, there are the well-known positive correlations between stellar mass and metallicity \citep[e.g.,][]{tremonti04} and stellar mass and SFR, also called the star forming main-sequence \citep[e.g.,][]{brinchmann04,cook14}. Taken together, these relations suggest a scenario in which the most massive galaxies and/or actively star forming galaxies accumulate a larger amount of dust that can lead to elevated levels of attenuation compared to their low mass and/or weakly star forming counterparts. A relation between SFR and amount of dust attenuation is observed, for instance, in local galaxies \citep[e.g.,][]{wang&heckman96,calzetti01}. However, the requirement of high gas content implies that there will eventually be a turnover in this behavior as one moves to the most massive galaxies, which are dominated by elliptical galaxies. This is a consequence of the baryon efficiency of galaxies experiencing a turnover at around $M_*$, implying a gas-deficiency in the most massive galaxies \citep[e.g.,][]{guo10}. Such a scenario naturally explains why massive elliptical galaxies typically have negligible dust content. Unfortunately, the complex nature of the many physical mechanisms giving rise to the correlations mentioned above have made quantifying this picture difficult. 

\begin{figure*}
\begin{center}$
\begin{array}{lr}
\includegraphics[width=3.in]{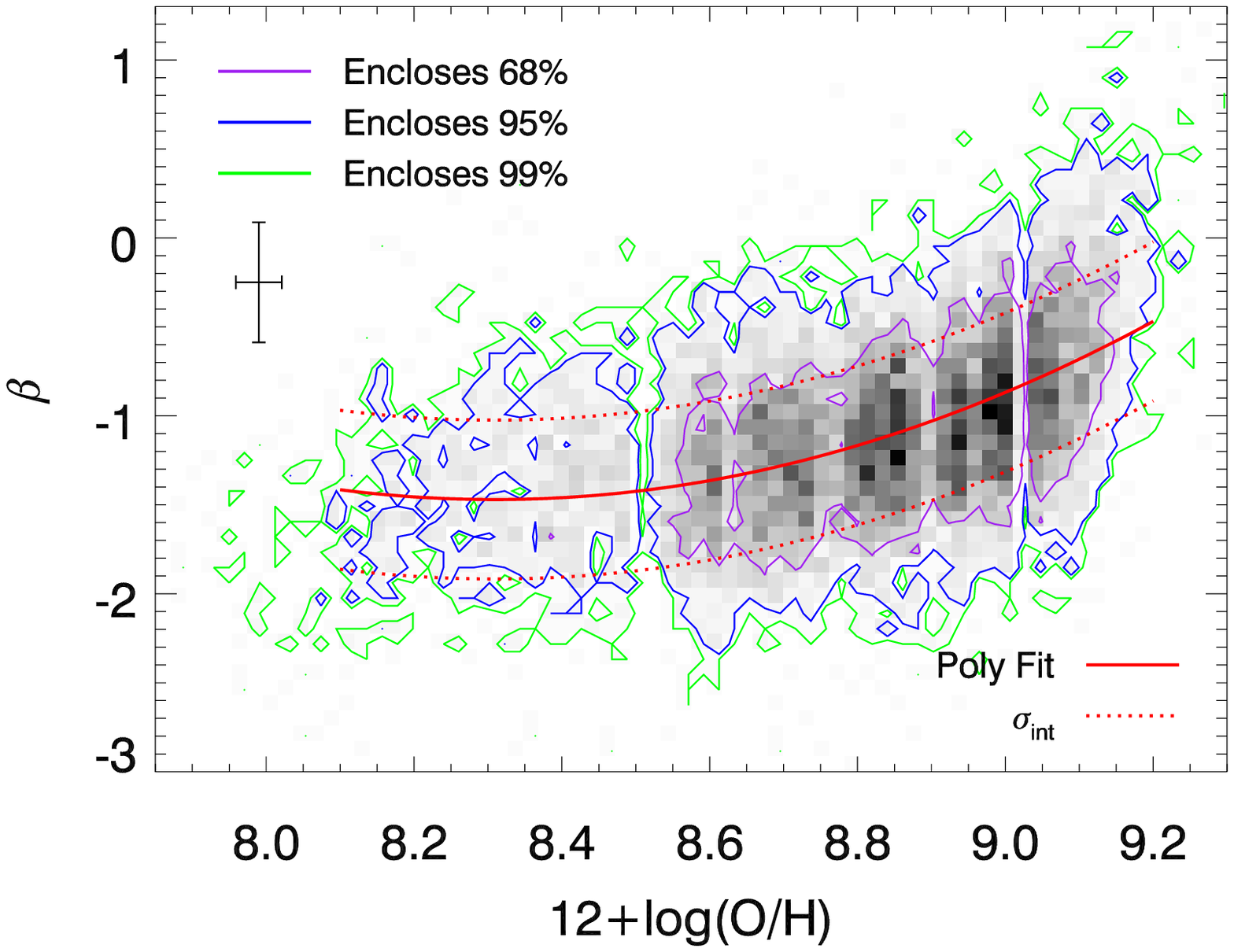}
\includegraphics[width=3.in]{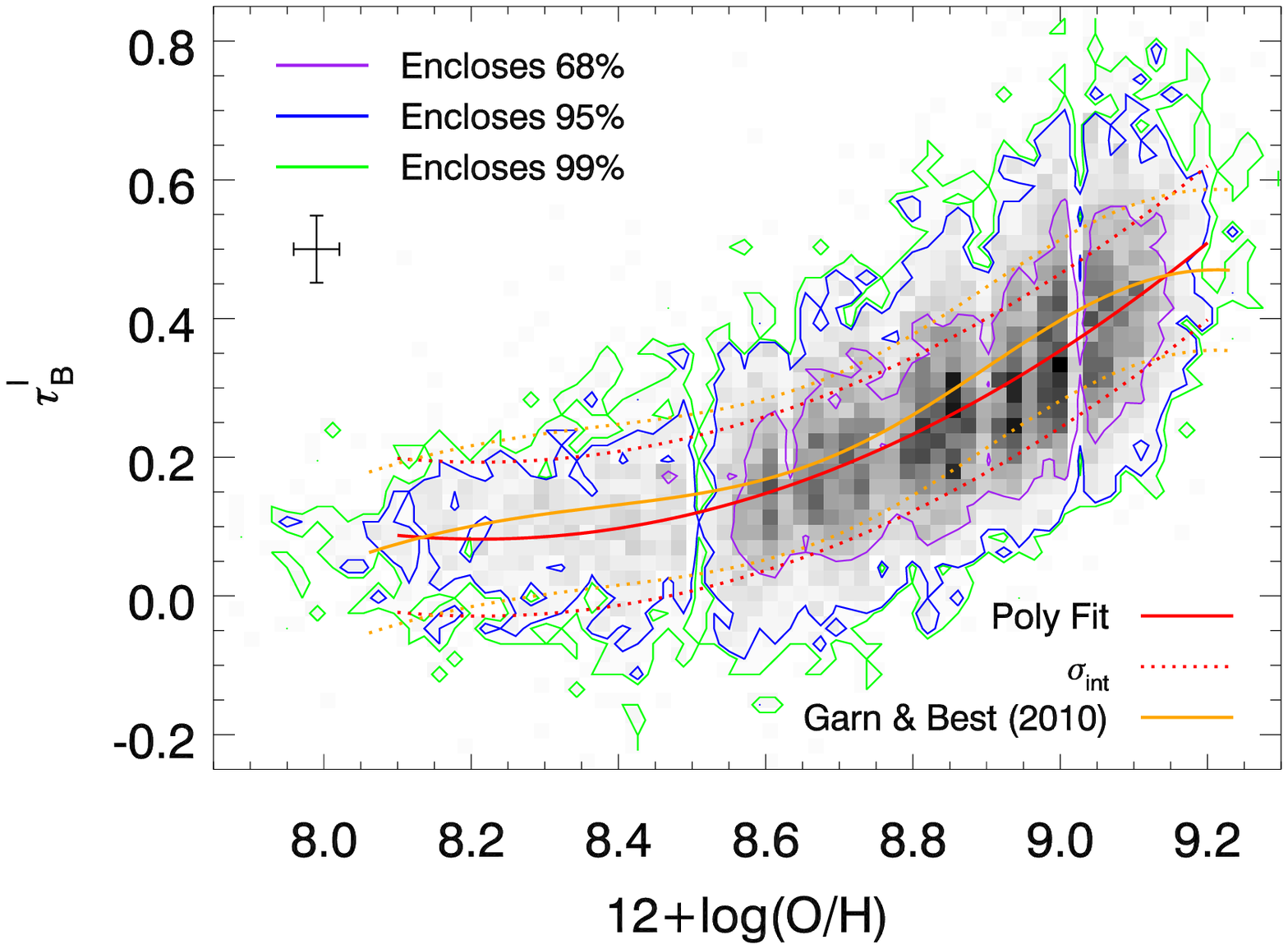} \\
\includegraphics[width=3.in]{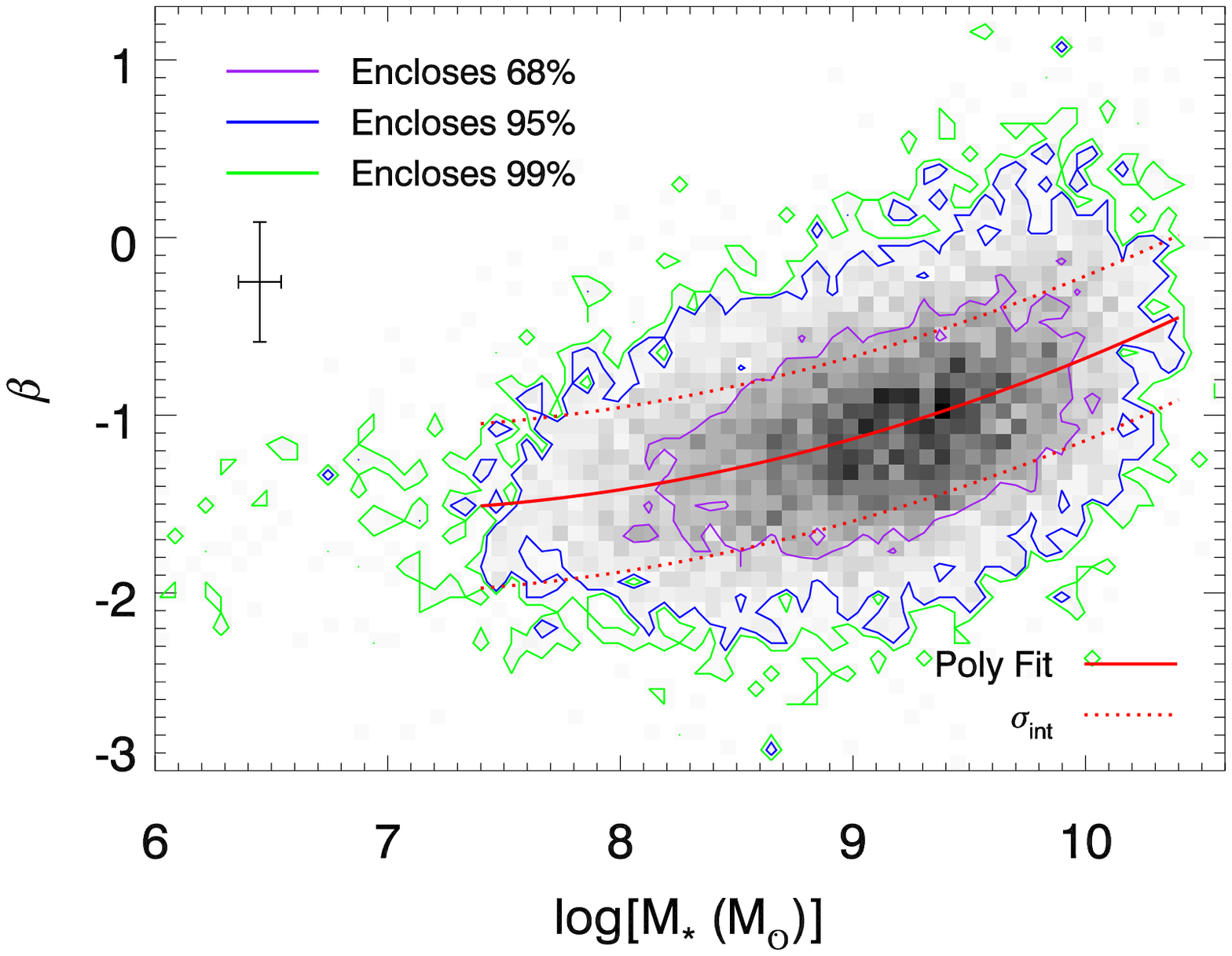}
\includegraphics[width=3.in]{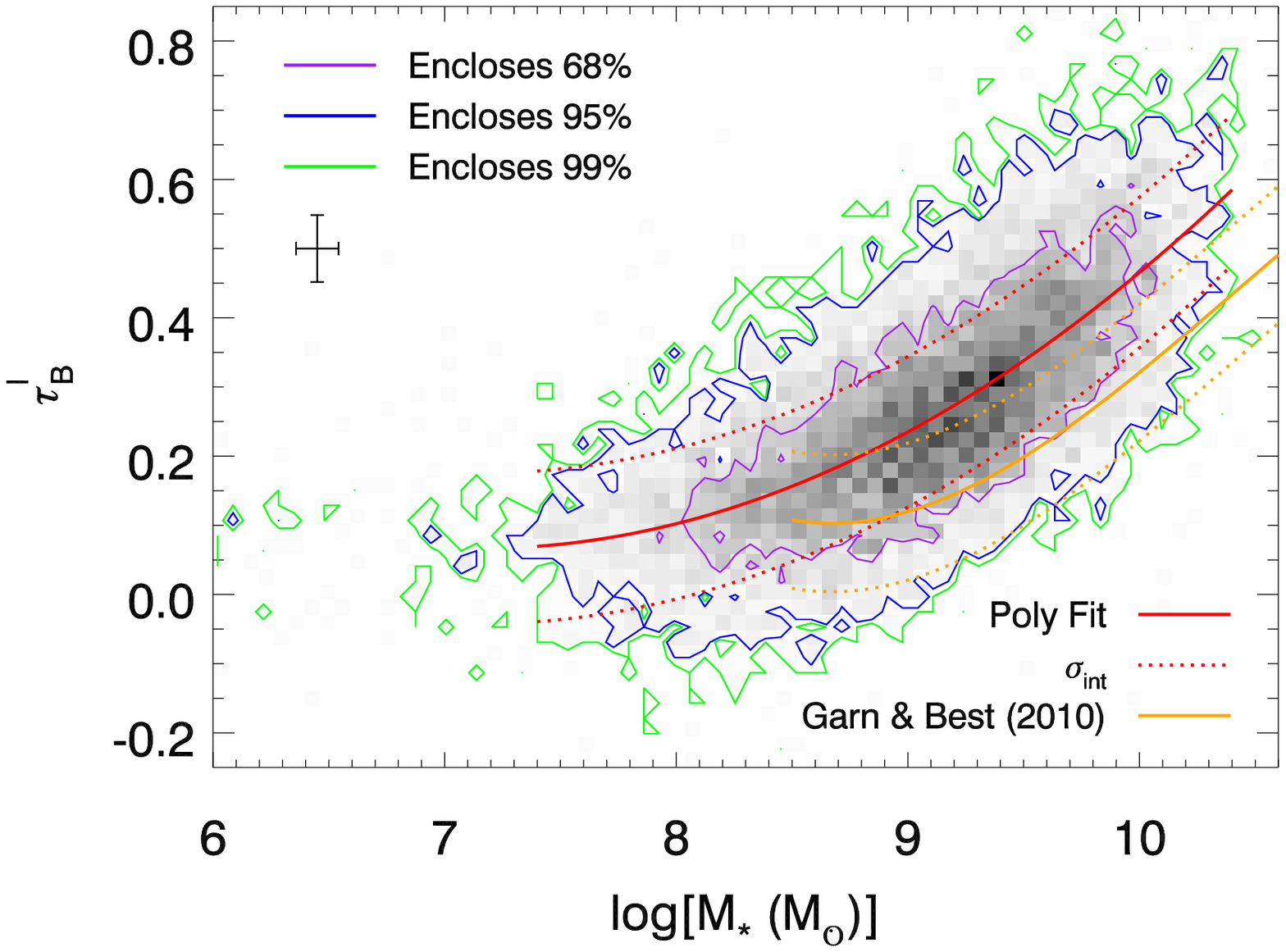} \\
\includegraphics[width=3.in]{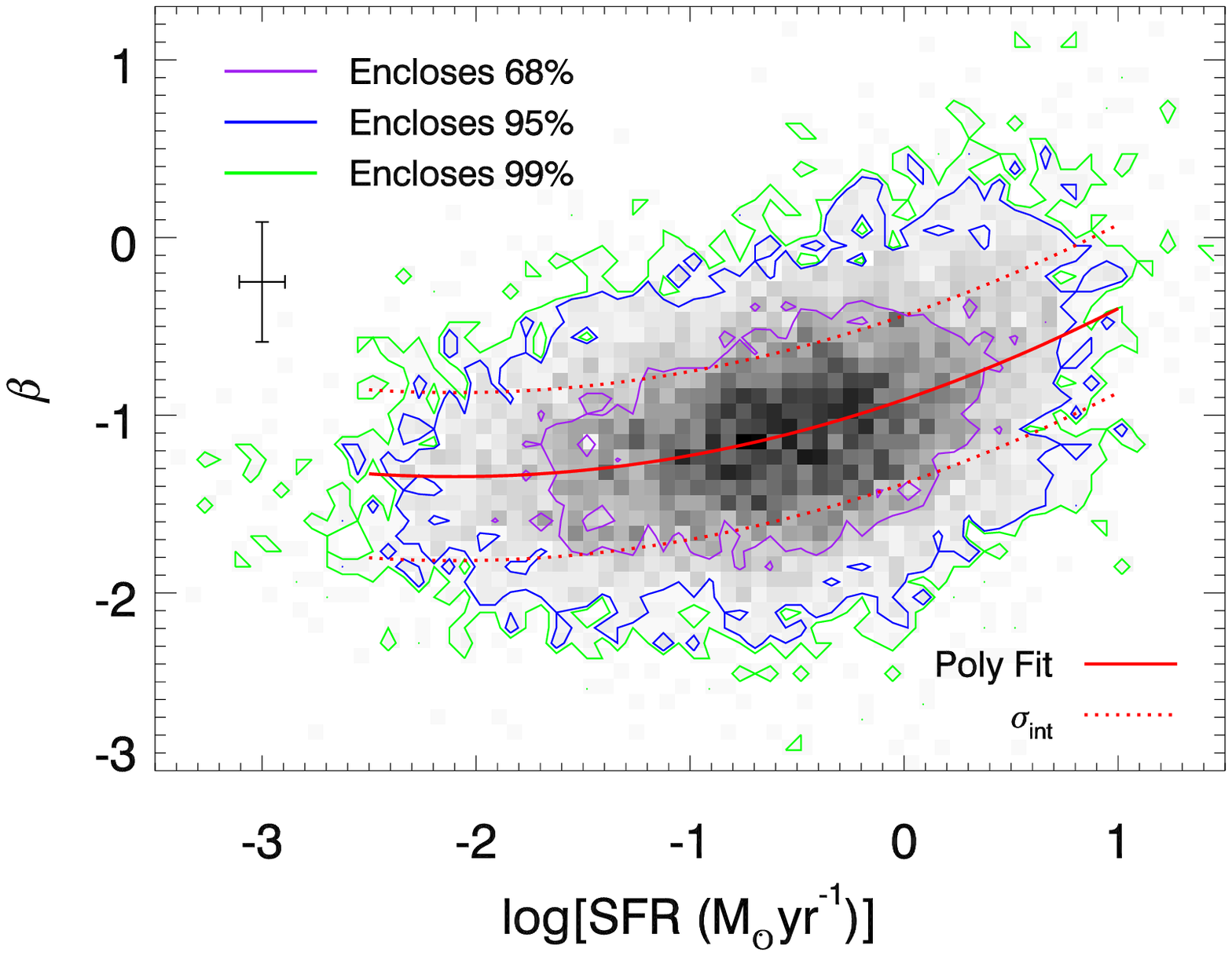}
\includegraphics[width=3.in]{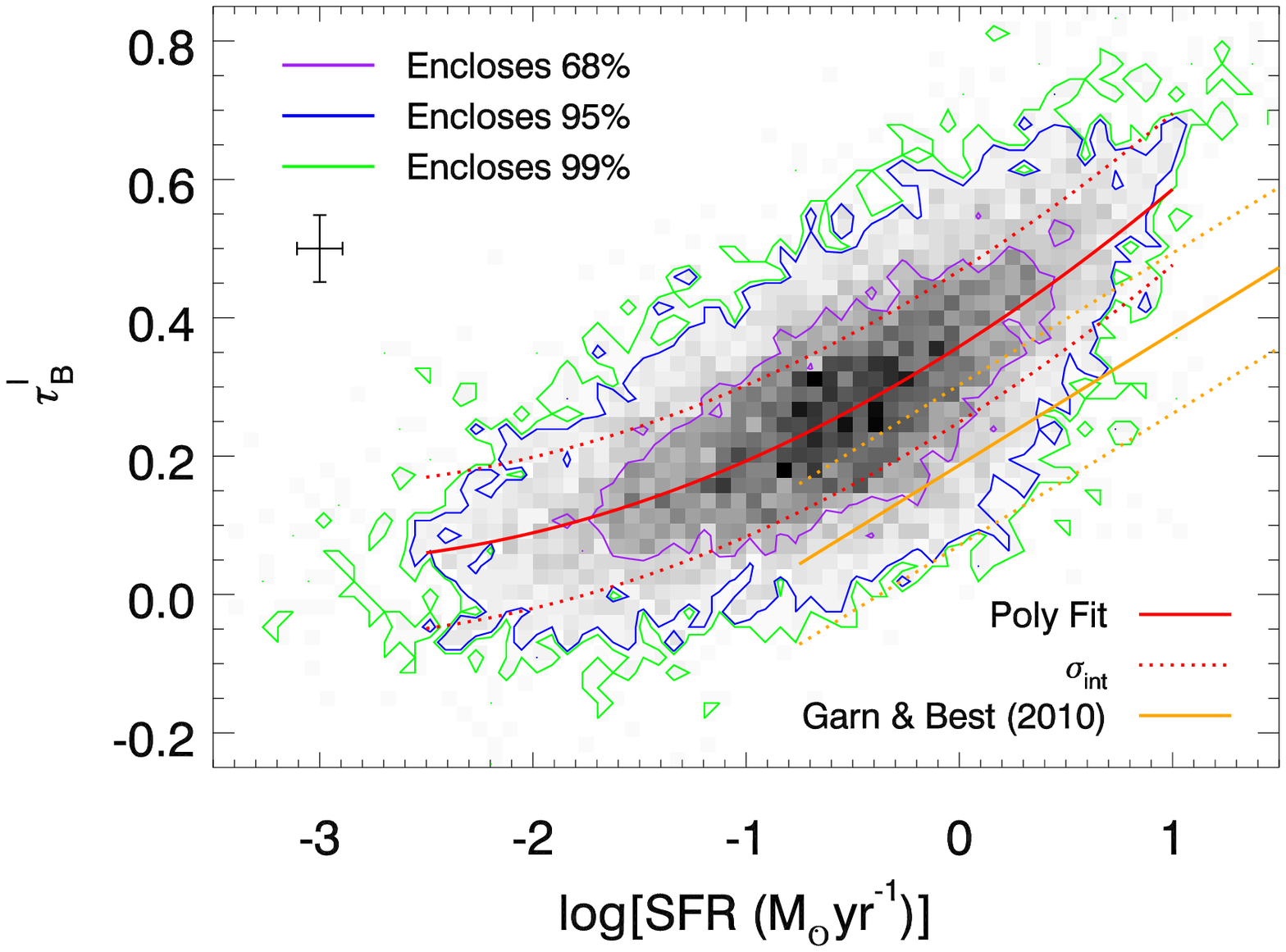} \\
\includegraphics[width=3.in]{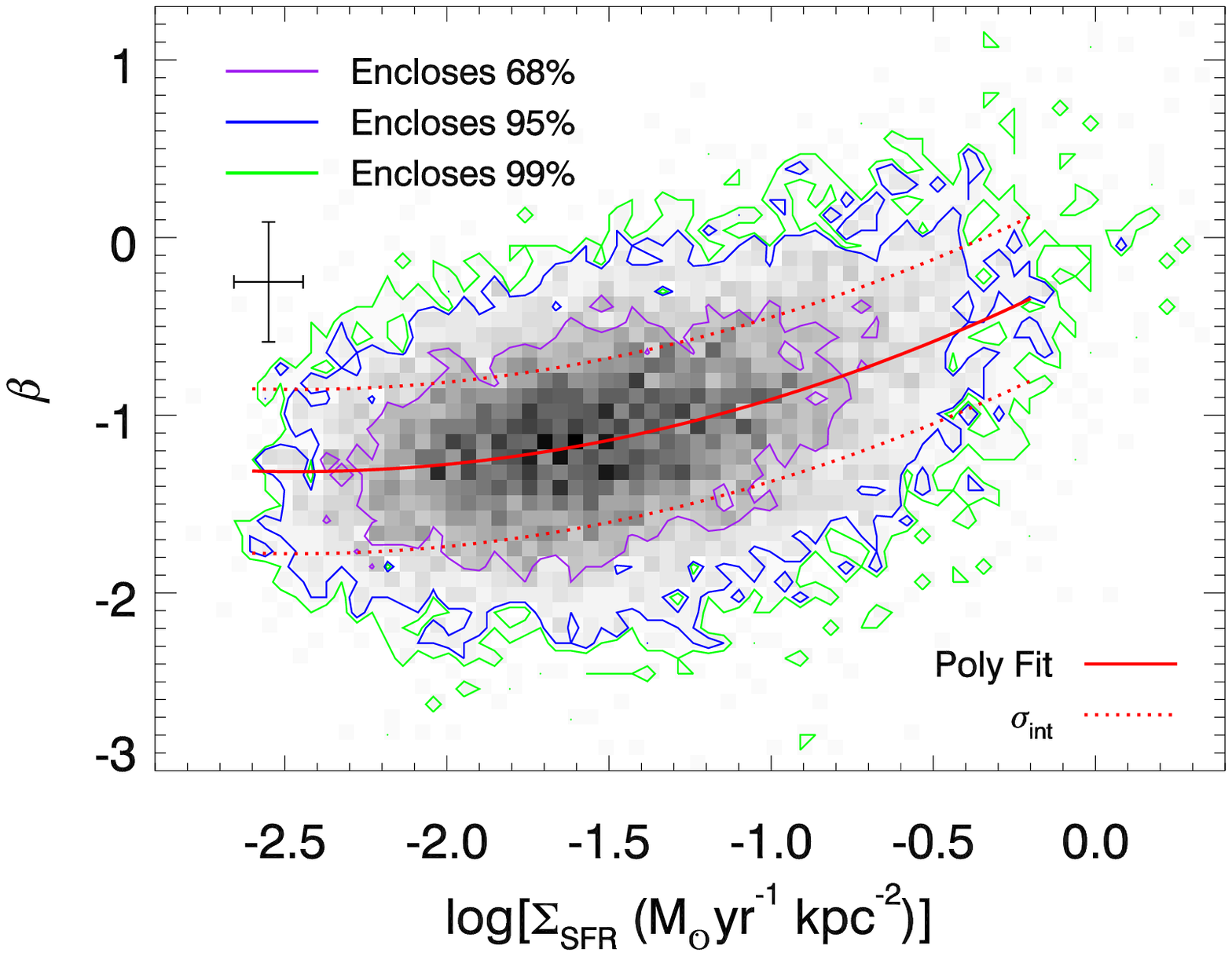}
\includegraphics[width=3.in]{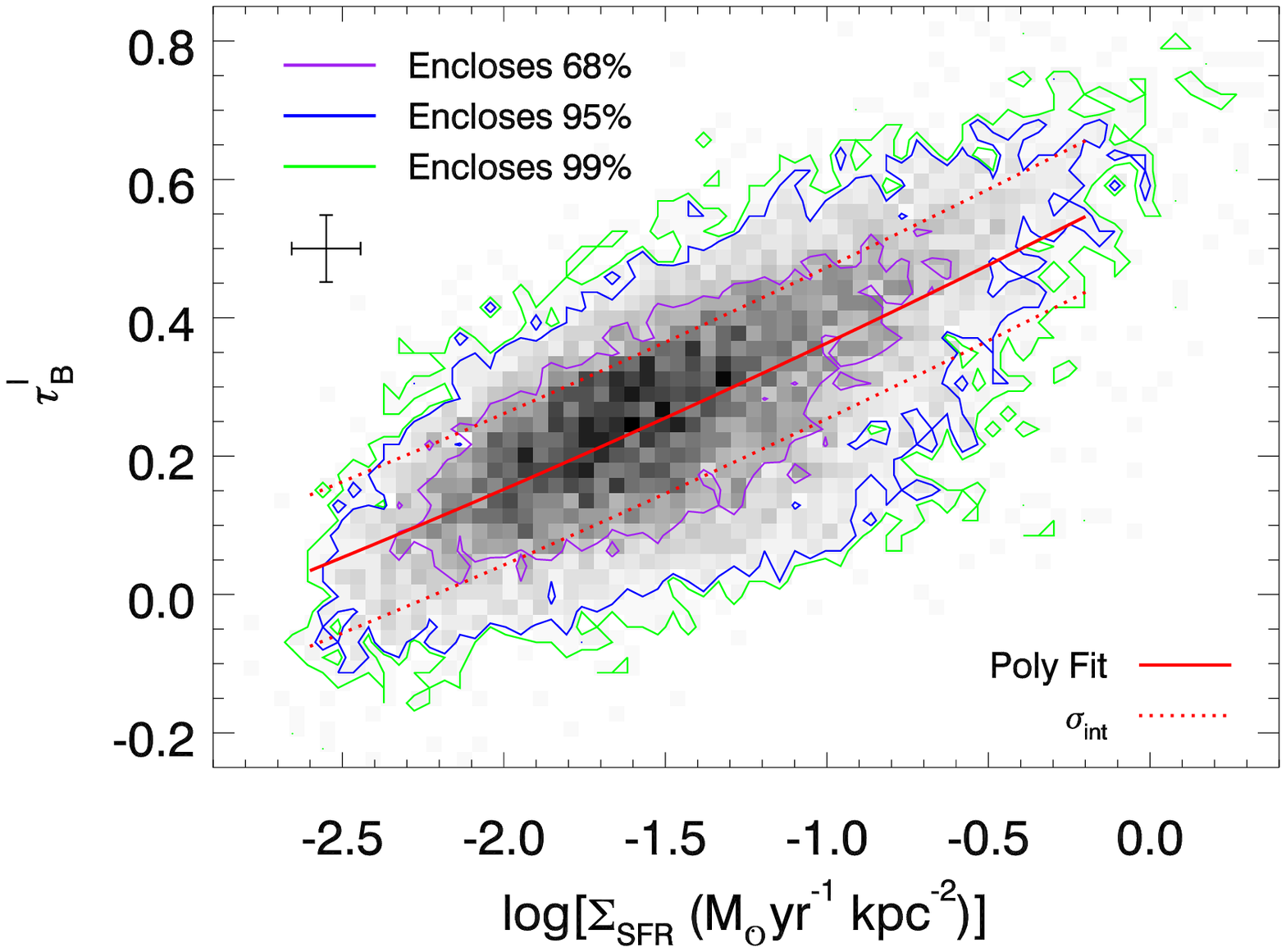} \\
\end{array}$
\end{center}
\caption{The UV power-law index after correcting for stellar absorption features in the \textit{GALEX} passbands, $\beta$ (see \S~\ref{beta_GLX_vs_z}),, and the Balmer optical depth, $\tau_B^l$, as a function of various galaxy properties. Shown from top to bottom are the gas-phase metallicity, stellar masses, SFRs, and SFR surface density ($\Sigma_{\mathrm{SFR}}$), respectively (all from $3\arcsec$ SDSS fiber). A representative error bar of the median measurement uncertainties is shown in the top left of each panel. All quantities show a positive correlation with the amount of UV and optical attenuation. Second-order polynomial fits to the data are shown as the solid red lines, with the dispersion shown as dotted red lines ($\pm\sigma_{\mathrm{int}}$). When possible, we compare to \citet{garn&best10}, which are offset in $M_*$ and SFR due to their use of total quantities in contrast to our fiber-only values (see \S~\ref{atten_vs_param}). \label{fig:beta_tau_prop1}}
\end{figure*}

Here we explore the presence of correlations between observational parameters associated with dust attenuation and various galaxy properties in order to help understand which properties are important for the presence of dust. In this section we will make use of $\beta$ for the UV slope in order to limit systematic effects with redshift. We examine the relationships between $\beta$ and $\tau_B^l$ with the metallicity, stellar mass, SFR, and SFR surface density ($\Sigma_{\mathrm{SFR}}$) of the galaxies. We plan to examine the role of galaxy inclination on attenuation in a future paper. As a reminder, we utilize the measurements of these quantities within the 3\arcsec\ SDSS fiber, along with their $1\sigma$ uncertainties, derived by the MPA/JHU group. The resulting comparisons are shown in Figure~\ref{fig:beta_tau_prop1}. Looking at this Figure, it is evident that tighter relationships arise for $\tau_B^l$ than for $\beta$. Part of this is likely due to the smaller uncertainties in former quantity, but it is also the case that differences in the underlying SFH of these galaxies could have a larger influence on the UV luminosity, and hence the UV slope. As has been found in many previous studies, the trends here suggest that star forming galaxies with larger metallicities, stellar masses, and star formation rates tend to have higher dust attenuation \citep[e.g.,][]{wang&heckman96,hopkins01,garn&best10,reddy15}. We perform a simple second-order polynomial fit to the data shown in Figure~\ref{fig:beta_tau_prop1}, to illustrate the general trends. We stress that these fits do not formally account for the uncertainties in the parameters and should only be used as a guideline. We present each fit, along with the intrisic dispersion from this relation, in Table~\ref{Tab:beta_vs_param}. In this table, we also show the coefficients for the Spearman, $\rho_S$, and Kendall, $\tau_K$, nonparametric correlation tests, which act as a gauge of the correlation strength. Similar to before, the large sample size makes the probability of no correlation be very close to zero and no longer meaningful to report.

For comparison, we plot the relationships found in \citet{garn&best10} using the full SDSS SFG dataset ($\sim$100000 galaxies) for H$\alpha$ attenuation, $A_{H\alpha}$ as a function of metallicity, stellar mass, and SFR in Figure~\ref{fig:beta_tau_prop1}. We convert $A_{H\alpha}$ back into a Balmer decrement following their assumption of a \citet{calzetti00} attenuation curve (see their equation~(1)), and then convert that into $\tau_B^l$ using equation~(\ref{eq:tau}). For consistency, we convert their metallicity values, derived using the O3N2 indicator \citep{pettini&pagel04}, into \citet{tremonti04} values using the relation provided in \citet{kewley&ellison08}. As expected, we find that the relationship we see with metallicity is nearly identical to theirs, although we do not see any evidence of a turnover at high metallicities as they suggest. However, significant differences appear between our relationships and those of \citet{garn&best10} for $\tau_B^l$ as a function of stellar mass and SFR. This is a result of their use of \textit{total} galaxy values for stellar mass and SFR, whereas we utilize fiber measurements, because the Balmer decrement measurement in both studies are coming from only the fiber region (same $\tau_B^l$ values) but the enclosed stellar mass and SFR are aperture dependent. The aperture-dependence of these values gives rise to the horizontal offset between the relations. In their analysis \citet{garn&best10} assume that the Balmer decrement is not dependent on the fiber aperture, which has been found to hold in some studies albeit with a large dispersion among individual cases \citep[e.g.,][]{kewley05,zahid13}. However, other studies have found radial dependencies which suggest the Balmer decrement decreases with increasing radius \citep[e.g.,][]{munoz-mateos09,iglesias-paramo13}, with more massive galaxies having larger gradients \citep{nelson15}. Given the inconclusive nature of this effect, we choose not to make assumptions regarding the aperture corrections for the attenuation, as we want all comparisons to be as self-consistent as possible. 

Understanding the general relationships between the amount of attenuation and these parameters offers a potential avenue for determining appropriate dust corrections at higher redshifts. However, several studies have found that the relationship between dust attenuation and SFR appears to evolve with redshift \citep[e.g.,][]{reddy06,reddy10,sobral12,dominguez13}. The studies of \citet{sobral12} and \citet{dominguez13} also examined the relation between dust attenuation and total stellar mass and find that it does not show significant evolution from redshift $z=0.1$ to 1.5 when comparing to \citet{garn&best10}. As a consequence, they state that total stellar mass might be a fundamental predictor of dust attenuation corrections. However, since we find an offset relative to the \citet{garn&best10} relation when using fiber-only measurements (due to less mass being enclosed), this result suggests that the relation between dust attenuation and total stellar mass is not fundamental. We suspect that the correlations between the attenuation and stellar mass or SFR is a byproduct of their strong correlation with metallicity and/or gas content, which are more fundamental predictors for the presence of dust. Additional studies are needed to determine if the relationship between attenuation and metallicity is redshift-dependent.

\begin{sidewaystable*}
\begin{center}
\caption{Fit Parameters of $\beta$ and $\tau_B^l$ as a Function of Galaxy Properties \label{Tab:beta_vs_param}}
\begin{tabular}{cc|cccccc|cccccc}
\hline\hline 
             &              &   \multicolumn{6}{c|}{$\beta$} & \multicolumn{6}{c}{$\tau_B^l$}   \\
      $x$    &     range    &  $p_0$    &  $p_1$    & $p_2$   & $\sigma_{\mathrm{int}}$ & $\rho_S$ & $\tau_K$ & $p_0$    &  $p_1$    & $p_2$ & $\sigma_{\mathrm{int}}$ & $\rho_S$ & $\tau_K$  \\ \hline

12+log(O/H) &  $8.1<x<9.2$ &  85.55  & -20.95 & 1.261 & 0.45 & 0.44 & 0.33 & 29.55  & -7.177  & 4.370$\times$10$^{-1}$ & 0.11 & 0.72 & 0.52 \\

log[$M_*$ ($M_\odot$)] & $7.4<x<10.4$ &  2.334  & -1.140 & 8.392$\times$10$^{-2}$ & 0.46 & 0.43 & 0.30 & 2.561  & -6.984$\times$10$^{-1}$ & 4.889$\times$10$^{-2}$ & 0.11 & 0.72 & 0.53 \\

log[SFR ($M_\odot$yr$^{-1}$)] & $-2.5<x<1.0$ & -9.115$\times$10$^{-1}$ &  4.131$\times$10$^{-1}$ & 9.833$\times$10$^{-2}$ & 0.47 & 0.36 & 0.25 & 3.585$\times$10$^{-1}$ &  1.962$\times$10$^{-1}$ & 3.078$\times$10$^{-2}$ & 0.11 & 0.68 & 0.50 \\

log[$\Sigma_{\mathrm{SFR}}$ ($M_\odot$yr$^{-1}$kpc$^{-2}$)] & $-2.6<x<-0.2$ & -1.651$\times$10$^{-1}$ & 9.349$\times$10$^{-1}$ & 1.897$\times$10$^{-1}$ & 0.46 & 0.37 & 0.26 & 5.940$\times$10$^{-1}$ & 2.402$\times$10$^{-1}$ & 9.625$\times$10$^{-3}$ & 0.11 & 0.67 & 0.49 \\ \hline
\end{tabular}
\end{center}
\textbf{Notes.} The functional form of these fits are $y = p_0+p_1x+p_2x^2$. We also report the intrinsic dispersion, $\sigma_{\mathrm{int}}$, which is taken as the standard deviation from the fitted relation. These fits are performed without accounting for the uncertainties in the variables and are only intended to illustrate general trends. The coefficients for the Spearman, $\rho_S$, and Kendall, $\tau_K$, nonparametric correlation tests are also given for each case.
\end{sidewaystable*}

\subsection{Variation in \texorpdfstring{$\beta$}{beta_GLX}-\texorpdfstring{$\tau_B^l$}{tau}}\label{atten_vs_param}
\begin{figure*}
\begin{center}$
\begin{array}{c}
\includegraphics[width=7.0in]{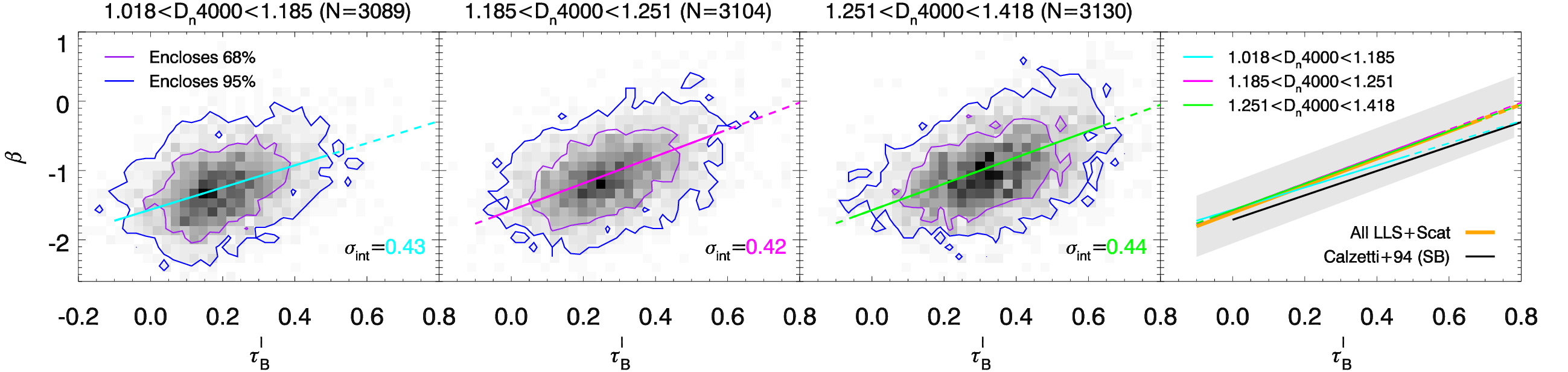} \\
\includegraphics[width=7.0in]{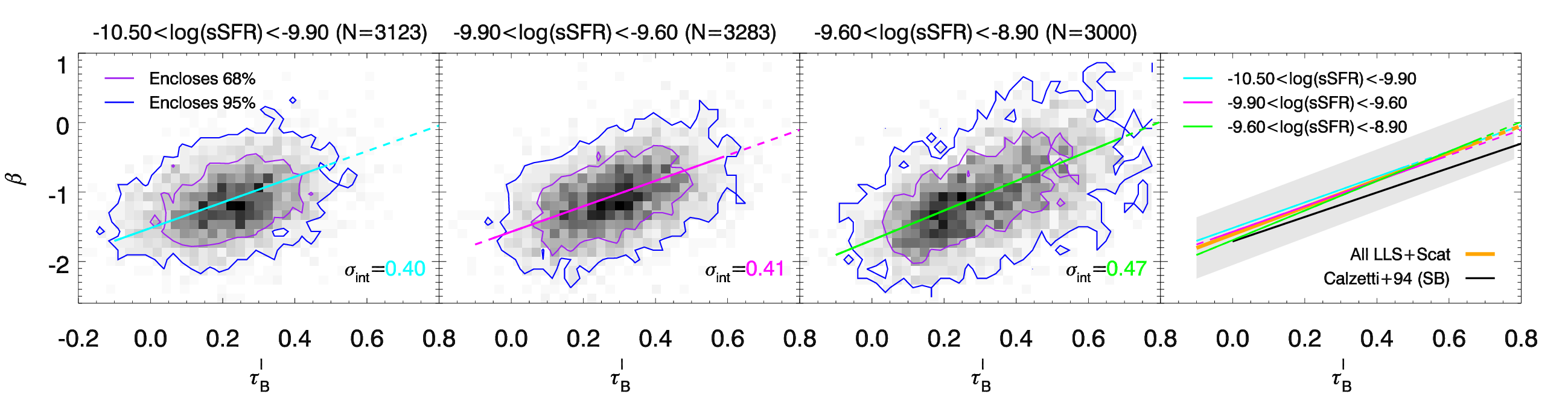} \\
\includegraphics[width=7.0in]{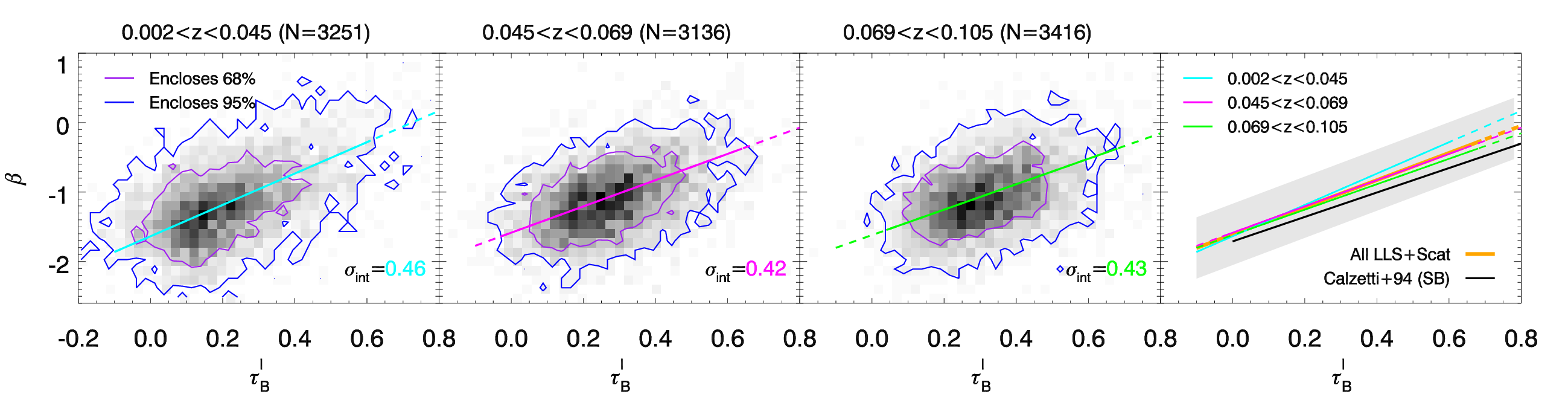} \\
\includegraphics[width=7.0in]{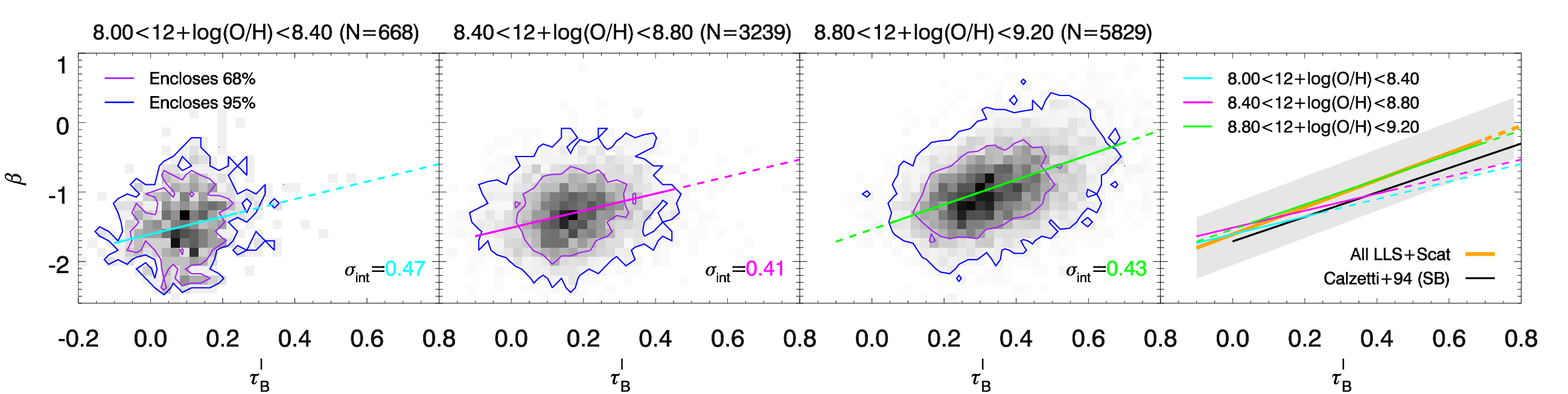} \\
\end{array}$
\end{center}
\caption{The $\beta$-$\tau_B^l$ relation for subsamples of galaxies with different properties ($\beta$ is the UV power-law index after correcting for stellar absorption features, see \S~\ref{beta_GLX_vs_z}). The parameters considered here are $D_n4000$, sSFR, $z$, and gas-phase metallicity. Only small differences appear among the different subsamples, which are not significant given the scatter in the data. In addition, the intrinsic dispersion is not seen to decrease among these subsamples. The lower panel acts to illustrate that the strong correlation of $\beta$ and $\tau_B^l$ with metallicity isolates different regions of parameter space. \label{fig:beta_tau_subsample}}
\end{figure*}

Given the large number of sources in our sample, we can also examine the variation in the behavior of the $\beta$-$\tau_B^l$ relation for galaxies with specific properties and identify key drivers of the large intrinsic scatter. Recent findings by \citet{reddy15} examining SFGs at $z\sim2$, found that galaxies appear to show significant differences in the $\beta$-$\tau_B^l$ relation as a function of galaxy sSFR. We separate our sample into three subsamples according to $D_n4000$ and sSFR, as these may be expected to correlate to the intrinsic UV slope. We also divide the sample by $z$ in order to test whether a fixed aperture with redshift influences the measurements. The results are shown in Figure~\ref{fig:beta_tau_subsample}. We chose the subsamples to consist of roughly one-third of the sample ($\sim3000$ galaxies). We note regions with low sampling of galaxies with dashed lines. Looking at Figure~\ref{fig:beta_tau_subsample}, it can be seen that no significant differences are evident in the $\beta$ vs $\tau_B^l$ relation with $D_n4000$, sSFR, or $z$, which indicates that the intrinsic scatter is not driven primarily by these parameters. We do not observe large variations in the offset of $\beta$ with sSFR, as is seen in \citet{reddy15}, but we note these studies probe different regimes, with the bin of highest sSFR ($-9.6<\log[\mathrm{sSFR}]<-8.9$) in our sample corresponding to the bin of lowest sSFR their sample ($-9.6<\log[\mathrm{sSFR}]<-8.84$), which makes direct comparison difficult.  

We find that it is not particularly informative to divide the sample according to galaxy metallicity, mass, SFR, or $\Sigma_{SFR}$ because these show significant correlation with the attenuation parameters, which implies that these parameters segregate galaxies in $\beta$-$\tau_B^l$ parameter space. As a result, this makes it difficult to differentiate between variation in attenuation relations because low and high attenuation galaxies would be separated. As an example, we illustrate the sample separated into three equally spaced bins in metallicity, which is argueably more informative here instead of equal number bins, in Figure~\ref{fig:beta_tau_subsample}. It can be seen that the higher metallicity galaxies ($8.8<12+\log(O/H)<9.2$), which correspond to the majority of our sample, are driving much of the observed trend between $\beta$ and $\tau_B^l$. 

\subsection{Variation in the Attenuation Curve}\label{curve_vs_param}
In a similar manner to how the sample was divided to see the effect on the observed values of $\beta$ and $\tau_B^l$, we can now examine how the attenuation curve changes with these properties. We utilize the same subsamples presented in \S~\ref{atten_vs_param} of $D_n4000$, sSFR, and $z$. For each parameter except $D_n4000$, we add an additional constraint that $1.1< D_n4000 <1.3$ in order to limit the stellar population age effects. We follow the same methodology presented in \S~\ref{attenuation curve}, dividing each sample into 6 bins of $\tau_B^l$ and constructing average flux templates, to derive the attenuation curve for these subsamples. We do not consider bins with less than 100 galaxies in determination of the effective attenuation curve. For nearly all cases the effective attenuation curve in the UV (derived from $\beta_{\rm{GLX}}$) and optical (derived from SDSS spectra) appear to be in agreement and well approximated by a single third-order polynomial. All of the fits to these subsamples are presented in Table~\ref{Tab:Q_vs_param}. 

\begin{figure*}
\begin{center}$
\begin{array}{lr}
\includegraphics[width=3.3in]{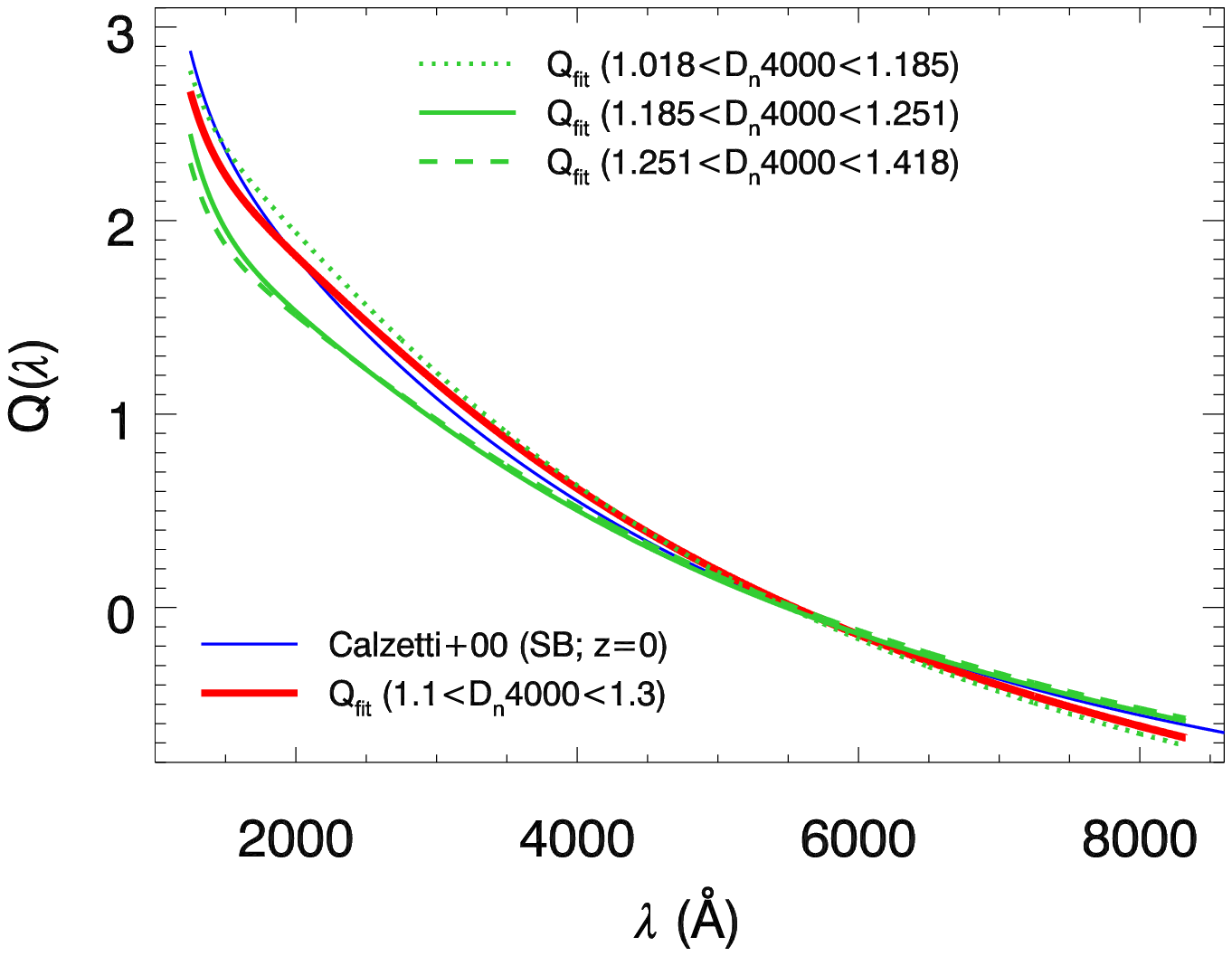}
\includegraphics[width=3.3in]{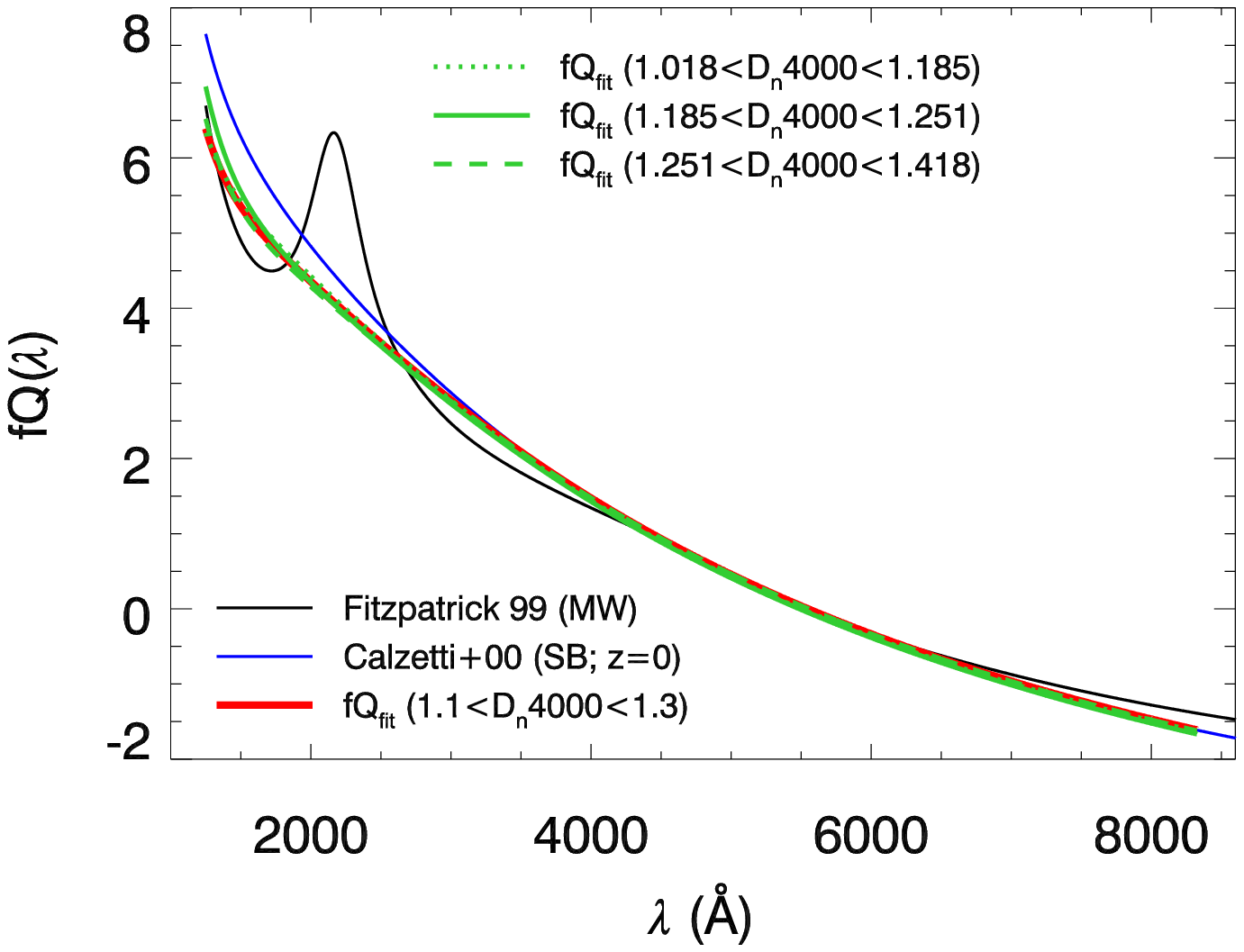} \\
\includegraphics[width=3.3in]{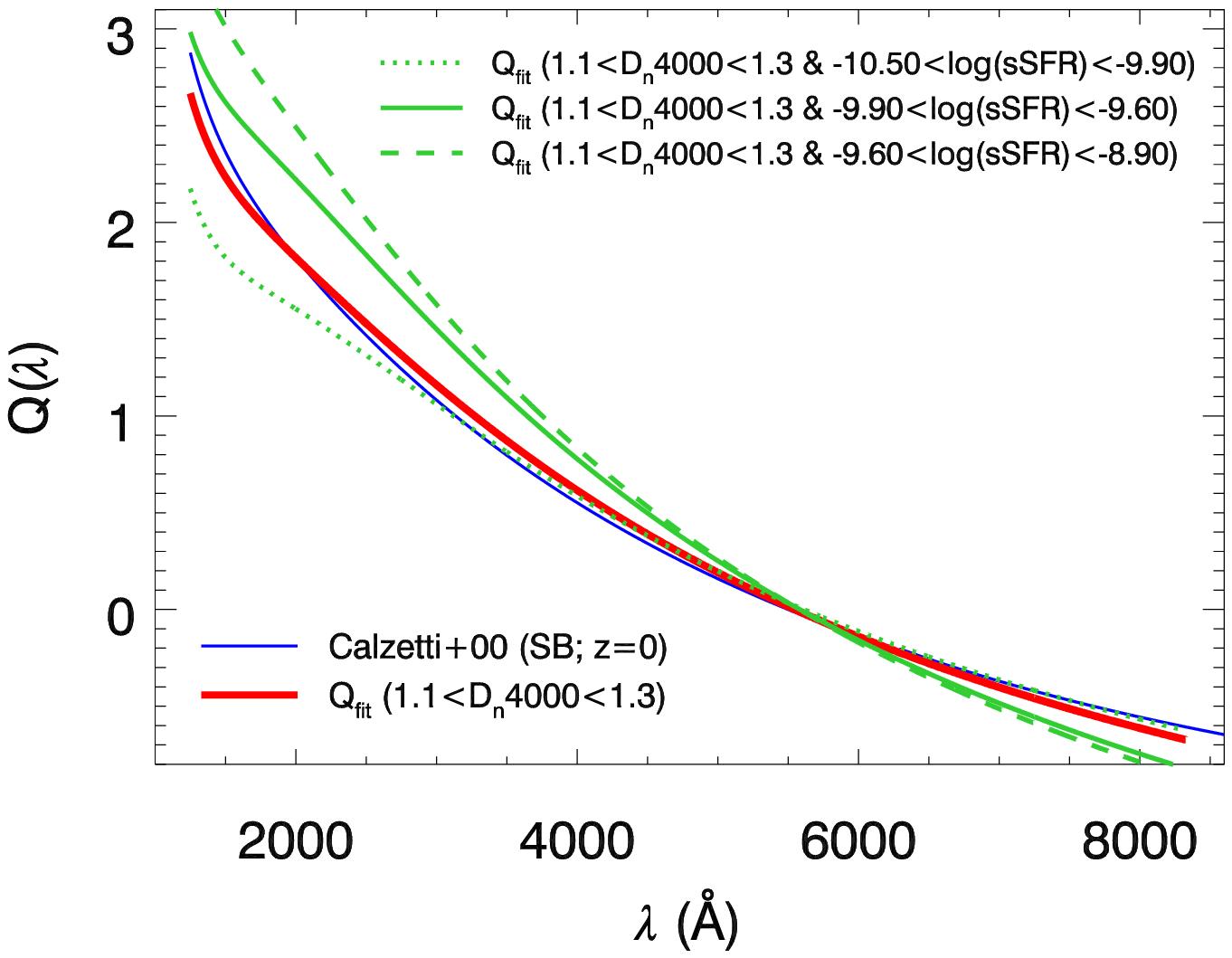}
\includegraphics[width=3.3in]{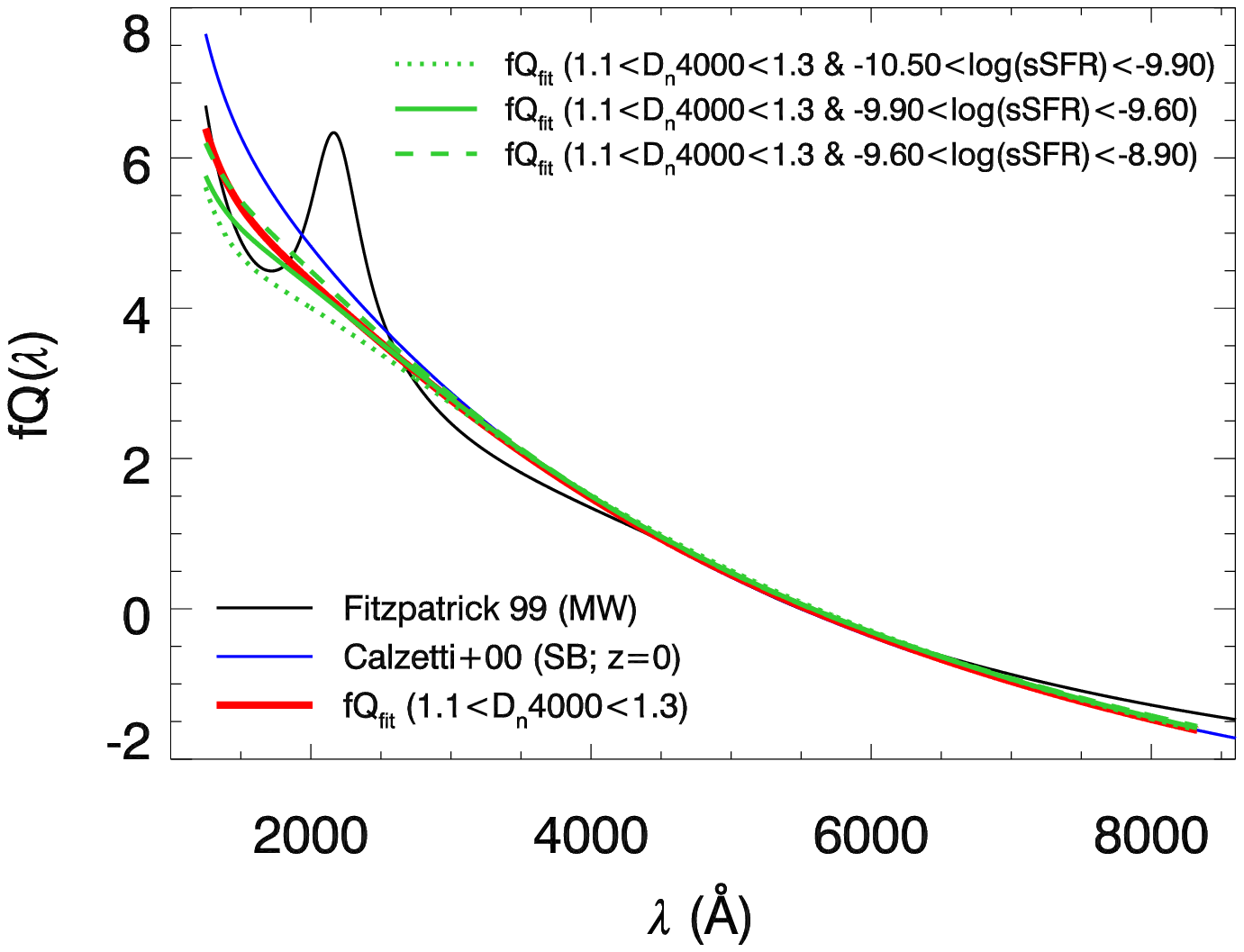} \\
\includegraphics[width=3.3in]{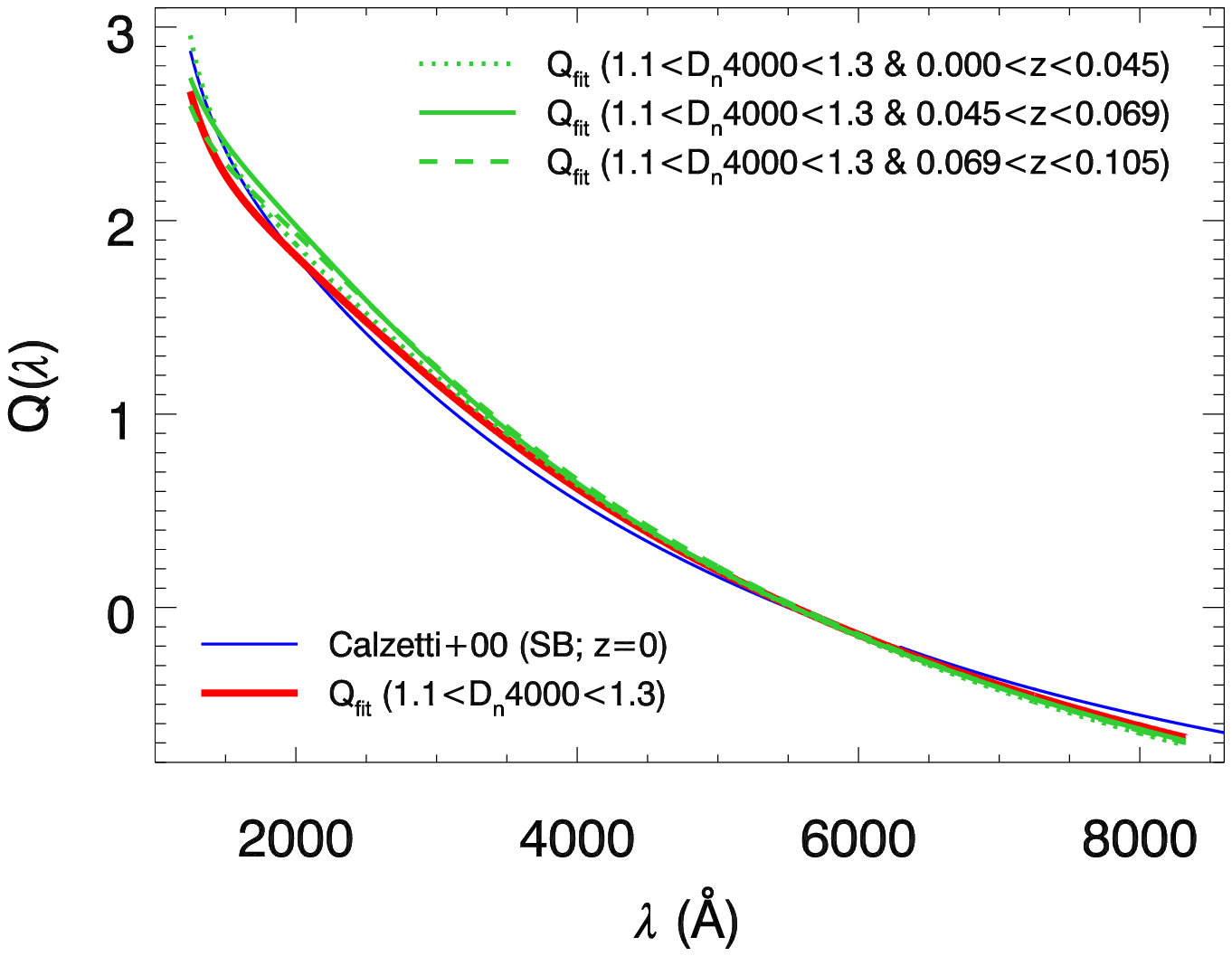}
\includegraphics[width=3.3in]{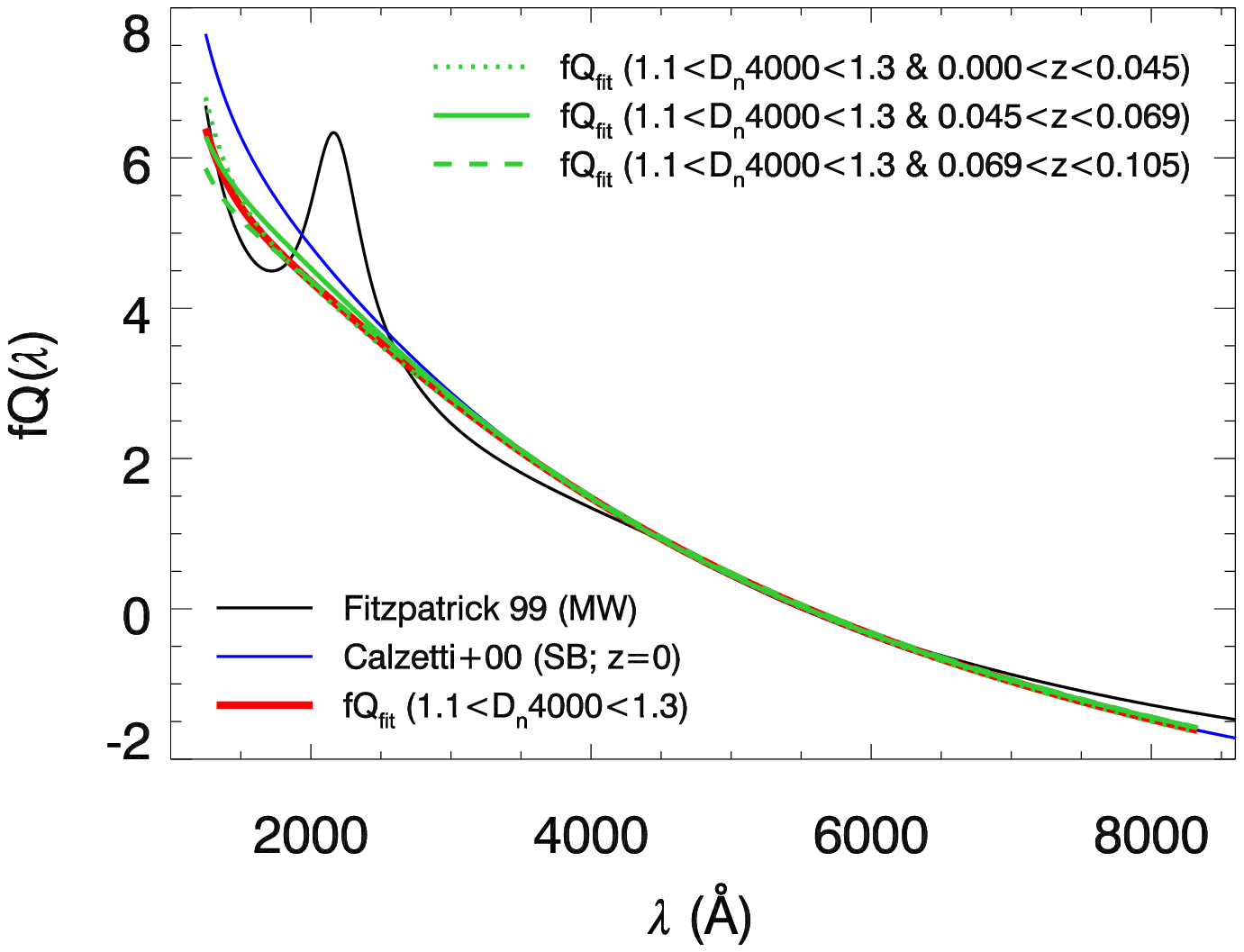} \\
\end{array}$
\end{center}
\caption{The effective attenuation curve for subsamples of galaxies with different properties. The parameters considered here are $D_n4000$, sSFR, and $z$. Slight differences in $Q(\lambda)$ appear with changes in $D_n4000$ and sSFRs, with the latter being more significant (uncertainty not shown for clarity). This indicates variation in $f$ (or equivalently in $E(B-V)_{\mathrm{star}}/E(B-V)_{\mathrm{gas}}$), which can result from changes in the relative contribution to the global flux density from massive stars (see \S~\ref{curve_vs_param}). However, after normalizing the curves as $fQ(\lambda)$, all cases are consistent with the same attenuation curve found from our total sample average (red line). The lack of variation in $Q(\lambda)$ with redshift indicates that the choice of fixed aperture (corresponding to different physical scales with $z$) does not seem to influence the derived curve. \label{fig:Q_compare_subsample}}
\end{figure*}

We plot the subsample curves of $Q(\lambda)$ alongside our average curve derived earlier and the \citet{calzetti00} curve (left panels), as well as the curves normalized by $f$ and alongside the \citet{fitzpatrick99} MW curve (right panels), in Figure~\ref{fig:Q_compare_subsample}. Slight differences in $Q(\lambda)$ appear in the $D_n4000$ and sSFR subsamples, with the latter case being more significant. If we make the crude assumption that these galaxies have similar extinction curves, then this would indicate variation in $\langle E(B-V)_{\mathrm{star}}\rangle/\langle E(B-V)_{\mathrm{gas}}\rangle$. These differences suggest that galaxies with lower $D_n4000$ or higher sSFR have slightly higher ratios of $\langle E(B-V)_{\mathrm{star}}\rangle/\langle E(B-V)_{\mathrm{gas}}\rangle$ (lower $f$). These changes are quite interesting because they indicate differences in the relative reddening of the ionized gas and the stellar continuum. One possible explanation for this is that in galaxies with elevated SFRs the UV reddening is more heavily weighted towards the same regions that dominate the Balmer line emission, thus increasing $\langle E(B-V)_{\mathrm{star}}\rangle/\langle E(B-V)_{\mathrm{gas}}\rangle$. One can imagine that for the extreme scenario in which nearly all of the global flux density is exclusively produced in HII regions, that this ratio would approach unity. However, this behavior is dependent on the optical depth of these star forming regions. In their study of $z\sim2$ SFGs, \citet{reddy15} find a lower ratio with increasing sSFR which they attribute to a larger fraction of the star formation in the galaxy becoming obscured in optically thick regions as the SFR increases. These obscured regions do not contribute significantly to the UV emission, but they continue to contribute to the Balmer line emission. Thus, the UV slope underestimates the dust attenuation relative to the Balmer line inferred attenuation for these galaxies such that $\langle E(B-V)_{\mathrm{star}}\rangle/\langle E(B-V)_{\mathrm{gas}}\rangle$ decreases toward larger SFRs. It is important to note that a lower ratio of $\langle E(B-V)_{\mathrm{star}}\rangle/\langle E(B-V)_{\mathrm{gas}}\rangle$ is also possible if higher star formation activity gives rise to significant outflows which reduce the overall optical depth affecting the stellar continuum. It is also important to state again that the sample of \citet{reddy15} probes a higher range of sSFR than this work and this may lead to differences in the underlying physical mechanisms at work between the two samples. Given the complex dependence of this ratio, we can not make any clear statement as to the cause of the differences that we see in our sample. 

Another notable result that can be seen in Figure~\ref{fig:Q_compare_subsample} is that there is virtually no dependence on the behavior of the attenuation curve as a function of the redshift spanned by our sample. We take this to indicate that the different physical aperture scales being probed as a result of our choice of fixed angular aperture does not seem to significantly alter the resulting curve. This also indicates that the average attenuation properties over smaller galaxy regions do not significantly deviate from the total values.  

Despite the changes seen in the behavior of $Q(\lambda)$ with proxies for stellar age, after normalizing the curves as $fQ(\lambda)$ the difference in the curves is significantly reduced in all cases (see right panels of Figure~\ref{fig:Q_compare_subsample}). A similar result was found by \citet{reddy15} for their $z\sim2$ sample of galaxies separated by sSFR, albeit with a different overall shape than we find for local galaxies. This remarkable result indicates that despite the differences in physical properties and SFHs that are spanned by SFGs in the local universe, on average they appear to suffer from a \textit{similar attenuation curve}. However, the large scatter in the $\beta$-$\tau_B^l$ relation and the flux density SEDs likely implies that there are variations in the attenuation on a case-by-case basis, part of which can stem from differences in the star-dust geometry. 

\begin{table*}
\begin{center}
\caption{Fit Parameters of $Q(\lambda)$ as a Function of Galaxy Properties \label{Tab:Q_vs_param}}
\begin{tabular}{ccccccc}
\hline\hline 
      $x$    &     range   & $f$ &  $p_0$    &  $p_1$    & $p_2$ &  $p_3$   \\ \hline
$D_n4000$ & $1.1<x<1.3$  & 2.396$\substack{+0.33 \\ -0.29}$ & -2.488 & 1.803 & -2.609$\times$10$^{-1}$ & 1.452$\times$10$^{-2}$  \\

          & $1.016<x<1.185$ & 2.283$\substack{+0.31 \\ -0.26}$ & -2.565 & 1.823 & -2.510$\times$10$^{-1}$ & 1.332$\times$10$^{-2}$ \\ 
          & $1.185<x<1.251$ & 2.840$\substack{+1.18 \\ -0.35}$ & -2.139 & 1.551 & -2.320$\times$10$^{-1}$ & 1.373$\times$10$^{-2}$ \\ 
          & $1.251<x<1.418$ & 2.840$\substack{+1.16 \\ -0.58}$ &-2.138 & 1.567 & -2.358$\times$10$^{-1}$ & 1.364$\times$10$^{-2}$  \\ \\ 

log[sSFR (yr$^{-1}$)] & $-10.50<x<-9.90$$^*$ & 2.580$\substack{+0.22 \\ -0.26}$ & -2.423 & 1.817 & -2.862$\times$10$^{-1}$ & 1.636$\times$10$^{-2}$  \\
          & $-9.90<x<-9.60$$^*$ & 1.931$\substack{+0.58 \\ -0.26}$ & -3.065 & 2.225 & -3.168$\times$10$^{-1}$ & 1.664$\times$10$^{-2}$ \\
          & $-9.60<x<-8.90$$^*$ & 1.804$\substack{+0.45 \\ -0.27}$ & -3.185 & 2.275 & -3.063$\times$10$^{-1}$ & 1.567$\times$10$^{-2}$ \\ \\

$z$       & $0.002<x<0.045$$^*$ & 2.302$\substack{+0.36 \\ -0.38}$ & -2.642 & 1.931 & -2.913$\times$10$^{-1}$ & 1.717$\times$10$^{-2}$  \\ 
          & $0.045<x<0.069$$^*$ & 2.296$\substack{+0.36 \\ -0.41}$ & -2.488 & 1.759 & -2.317$\times$10$^{-1}$ & 1.169$\times$10$^{-2}$  \\ 
          & $0.069<x\le0.105$$^*$ & 2.260$\substack{+0.32 \\ -0.16}$ & -2.584 & 1.859 & -2.572$\times$10$^{-1}$ & 1.321$\times$10$^{-2}$  \\ \hline 
\end{tabular}
\end{center}
\textbf{Notes.} The uncertainty in $f$ denotes the maximum and minimum values from fits using individual $Q_{n,r}(\lambda)$ for each subsample (see \S~\ref{attenuation curve}). The functional form of these fits are $Q = p_0+p_1x+p_2x^2+p_3x^3$. $^*$These cases also have the constraint that $1.1<D_n4000<1.3$.
\end{table*}

\section{Conclusions}
We use a sample of $\sim$10000 local ($z\lesssim0.1$) star forming galaxies to constrain the nature of dust attenuation in galaxies as a function of their physical properties. Utilizing aperture-matched UV and optical data, we find a linear relationship between the UV power-law index, $\beta$, and the Balmer line optical depth, $\tau_B^l$, which is similar to the local starburst relation. The large scatter ($\sigma_{\mathrm{int}}=0.44$) of this relation suggests there is significant variation in the attenuation of individual galaxies local Universe. Using this large sample, we are able to quantify how the attenuation is influenced by varying galaxy parameters. We observe significant correlations between the amount of UV and ionized gas reddening with galaxy metallicity, $M_*$, SFR, and $\Sigma_{\mathrm{SFR}}$. A weaker negative correlation is seen with the mean stellar age, traced by the 4000~\AA\ break ($D_n4000$). These trends are consistent with a scenario in which the total dust content increases with star formation activity and also builds up slowly with age. These relationships can provide a way for determining attenuation in other studies if these parameters are available. However, we stress that the redshift evolution of some of these relationships \citep[e.g.,][]{reddy06,reddy10,sobral12,dominguez13} poses a problem in the application to higher redshift studies.

Using our sample we derive an attenuation curve over the wavelength range $1250\mathrm{\AA}<\lambda<8320\mathrm{\AA}$. We find a lower selective attenuation in the UV compared to previously determined attenuation curves and which is about 20\% lower than the starburst curve from \citet{calzetti00} at 1250\AA. However, given that the normalization of our curve is still unknown, it is not clear whether this also corresponds to lower \textit{total} attenuation. Such an analysis will be the subject of a future study. We see no evidence to suggest that a significant 2175~\AA\ feature is present in this curve, although this cannot be conclusively determined without available UV spectroscopy. The relative reddening of the stellar continuum is roughly one-half of the amount suffered by the ionized gas, with $\langle E(B-V)_{\mathrm{star}}\rangle=0.52\langle E(B-V)_{\mathrm{gas}}\rangle$ (assuming \citet{fitzpatrick99} MW extinction for the ionized gas), in good agreement with previous studies \citep{calzetti94,wild11,kreckel13,reddy15}. We emphasize that this is the average relation and that individual cases will vary from this ratio depending on their properties.

When dividing the sample according to different galaxy properties we find that galaxies with larger sSFRs have smaller $f$ values (larger ratios of $\langle E(B-V)_{\mathrm{star}}\rangle/\langle E(B-V)_{\mathrm{gas}}\rangle$). However, after normalizing the curves to remove the effects of differential reddening, the variation in the curves is significantly reduced in all cases. This result indicates that despite differences in physical properties and SFHs spanned by SFGs in the local universe, on average they appear to suffer from a \textit{similar underlying attenuation curve}. This single attenuation curve is well suited for application to large statistical studies of SFGs, but should be used with caution on a case-by-case basis.

\section*{Acknowledments} The authors thank the anonymous referee whose suggestions helped to clarify and improve the content of this work. AJB also thanks K. Grasha for comments that improved the clarity of this paper.

Part of this work has been supported by NASA, via the Jet Propulsion Laboratory Euclid Project Office, as part of the ``Science Investigations as Members of the Euclid Consortium and Euclid Science Team'' program.

This work is based on observations made with the NASA Galaxy Evolution Explorer. \textit{GALEX} is operated for NASA by the California Institute of Technology under NASA contract NAS5-98034.
This work has made use of SDSS data. Funding for the SDSS and SDSS-II has been provided by the Alfred P. Sloan Foundation, the Participating Institutions, the National Science Foundation, the US Department of Energy, the National Aeronautics and Space Administration, the Japanese Monbukagakusho, the Max Planck Society and the Higher Education Funding Council for England. The SDSS website is http://www.sdss.org/. The SDSS is managed by the Astrophysical Research Consortium for the Participating Institutions.

\appendix
\section{Viability of Using the Optical Slope Instead of the UV Slope}\label{opt_slope}
Future large area IR surveys, such as those planned with \textit{Euclid} and the \textit{Wide-Field Infrared Survey Telescope} (\textit{WFIRST}), will image vast numbers of galaxies. Given the shortest wavelengths available to these missions, $\sim5500$~\AA\ and $\sim7600$~\AA\ for \textit{Euclid} and \textit{WFIRST}, respectively, the shortest rest-frame wavelengths available for galaxies with $z<1$ this will correspond to the optical portion of the spectrum. Therefore, a proper utilization of the $\beta$-$\tau_B^l$ relation would require separate measurements from another facility to determine UV slope. Here we investigate the possibility of using the optical slope, from the observed SDSS $u$ ($\lambda_{\mathrm{eff}}=3543$~\AA) and $g$ ($\lambda_{\mathrm{eff}}=4770$~\AA) fiber photometry, as an indicator for reddening of the continuum instead of $\beta$. This is calculated using the same method as for the UV slope, 
\begin{equation}
\beta_{\rm{opt}} = \frac{\log[F_\lambda(\mathrm{u})/F_\lambda(\mathrm{g})]}{\log(\lambda_{\mathrm{u}}/\lambda_{\mathrm{g}})}\,.
\end{equation}
We expect that the correlation between the optical slope and $\tau_B^l$ to be weaker than the UV slope, given that this region is less sensitive to the effects of dust and more sensitive to older stellar populations.  

\begin{figure}
\plotone{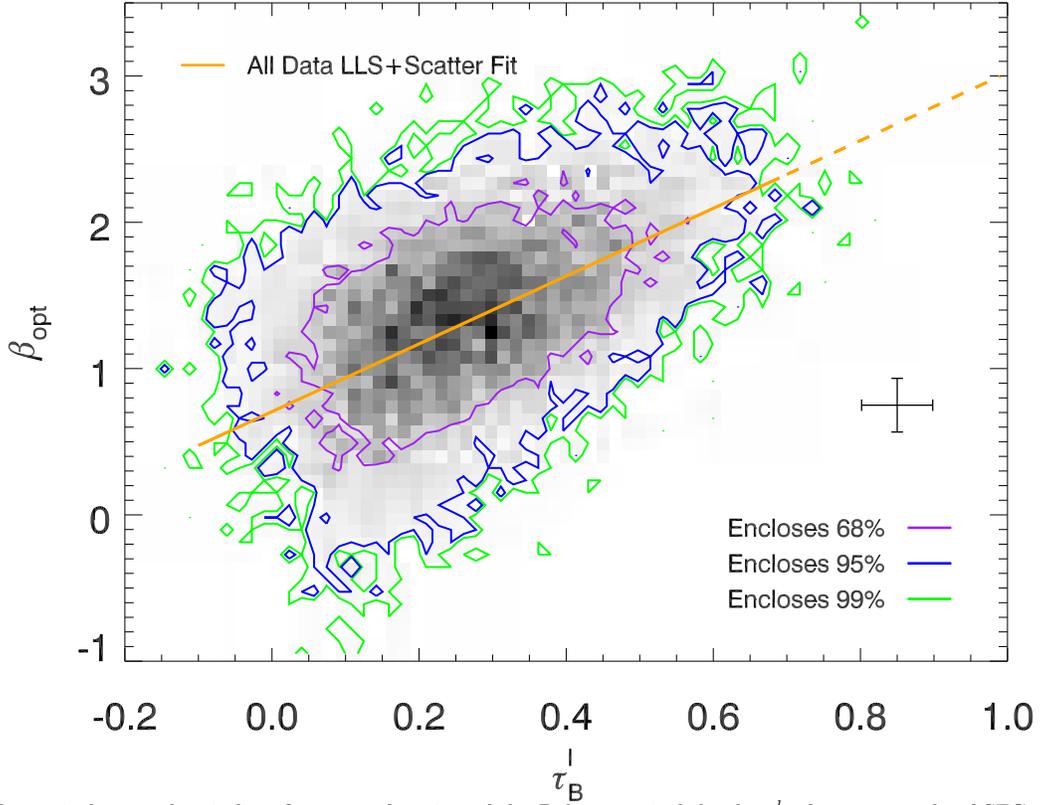}
\caption{The optical power-law index, $\beta_{\rm{opt}}$, as a function of the Balmer optical depth, $\tau_B^l$, for our sample of SFGs (orange line). A representative error bar for the SFG sample is shown in the bottom right. Our fit at $\tau_B^l>0.7$ is shown with a dashed line to denote that there are limited data in this range. \label{fig:ug_beta_tau}}
\end{figure}

\begin{figure}
\plotone{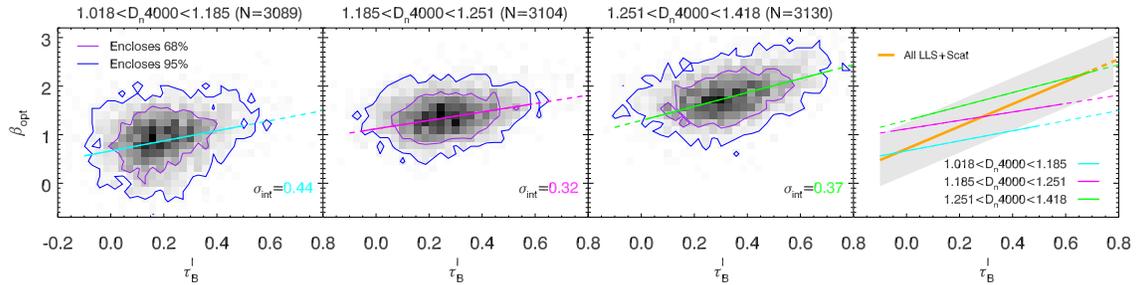}
\caption{The $\beta_{\rm{opt}}$-$\tau_B^l$ relation for subsamples of galaxies with different $D_n4000$. The subsamples show significant offsets relative to each other and also have a lower dispersion relative to the total sample. This indicates that $D_n4000$ acts as a good diagnostic of the intrinsic optical slope (i.e., vertical normalization). Galaxies with an older stellar population (larger $D_n4000$) have a redder intrinsic values of $\beta_{\rm{opt}}$. \label{fig:ug_beta_tau_subsample}}
\end{figure}

In Figure~\ref{fig:ug_beta_tau} we show the $\beta_{\rm{opt}}$ and $\tau_B^l$ values for our sample of 9813 SFGs. A linear fit to the data using the MPFITEXY routine \citep{williams10} gives 
\begin{equation}
\beta_{\rm{opt}} = (2.32\pm0.04) \tau_B^l +(0.71\pm0.01) \,,
\end{equation}
with an intrinsic dispersion of $\sigma_{\mathrm{int}}=0.53$.  As expected, this dispersion is larger than that found using the UV slope, but not by a very large amount. Similar to earlier anaylsis, we divide the sample by $D_n4000$ to determine its role in the dispersion. We plot the sample divided into three ranges of $D_n4000$ in Figure~\ref{fig:ug_beta_tau_subsample} and the fitted relationships are
\begin{equation}
\beta_{\rm{opt}}(1.018<D_n4000<1.185) = (1.04\pm0.06) \tau_B^l +(0.67\pm0.01) \,,
\end{equation}
\begin{equation}
\beta_{\rm{opt}}(1.185<D_n4000<1.251) = (0.87\pm0.04) \tau_B^l +(1.12\pm0.01) \,,
\end{equation}
\begin{equation}
\beta_{\rm{opt}}(1.251<D_n4000<1.418) = (1.43\pm0.04) \tau_B^l +(1.29\pm0.02) \,.
\end{equation}
It is evident that the dispersion is reduced by $\Delta\sigma_{\mathrm{int}}\sim-0.15$. This indicates that $D_n4000$ acts as an indicator of the intrinsic optical slope (i.e., the normalization of the $\beta_{\rm{opt}}-\tau_B^l$ relation). From this we believe that the optical slope might be viable for determining appropriate corrections in a large statistical sample, especially if information on the 4000\AA\ break is available to correct for stellar age effects.

\end{document}